\documentclass[prb,aps,floatfix,floats,superscriptaddress,notitlepage]{revtex4-1}
\usepackage{epsfig}
\usepackage{enumerate}
\usepackage{amsmath}
\usepackage{amssymb}
\usepackage{hyperref}
\usepackage{color}
\usepackage{theorem}
\usepackage{braket}
\usepackage{mathtools}
\usepackage{comment}
\usepackage{tensor}
\usepackage{tikz}
\usetikzlibrary{decorations.markings}

\newcommand{\be}{\begin{equation}}
\newcommand{\ee}{\end{equation}}
\newcommand{\bea}{\begin{eqnarray}}
\newcommand{\eea}{\end{eqnarray}}

\definecolor{violet}{rgb}{0.62,0,1}
\definecolor{lightblue}{rgb}{0.12,0.56,1}
\definecolor{green}{rgb}{0.13,0.55,0.13}

\def\fr#1{(\ref{#1})}

\def\xt{\xi}

\definecolor{violet}{rgb}{0.62,0,1}
\definecolor{lightblue}{rgb}{0.12,0.56,1}
\definecolor{green}{rgb}{0.13,0.55,0.13}

\def\doi{http://dx.doi.org/}

\theoremstyle{plain}

\allowdisplaybreaks[4]
\begin{document}
\title{Approximate light cone effects 
in a non-relativistic quantum field theory\\ after a local quench}
\author{Bruno Bertini}
\affiliation{SISSA and INFN, via Bonomea 265, 34136, Trieste, Italy}
\begin{abstract}
We study the spreading of correlations after a local quench in a non-relativistic quantum field theory. We focus on noninteracting non-relativistic fermions and study the time evolution after two identical systems in their ground states are suddenly joined together with a localized impurity at the junction. We find that, even if the quasi-particles of the system have unbounded dispersion, the correlations show light cone effects. We carry out a detailed study of these effects by developing an accurate asymptotic expansion of the two-point function and determining exactly the density of particles at any time after the quench. In particular, we find that the width of the light cone region is $\propto t^{1/2}$. The structure of correlations, however, does not show a pure light cone form -- ``superluminal corrections'' are much larger than in the bounded-dispersion case. These findings can be explained by inspecting the structure of excitations generated by the initial state. We show that a similar picture also emerges in the presence of a harmonic trapping potential and when more than two systems are suddenly joined at a single point.
\end{abstract}
\maketitle
\section{Introduction}
During the last decade, pioneering experiments in cold atomic gases and trapped ions allowed, for the first time, the observation of many-body systems undergoing (nearly) unitary time evolution\cite{exp, explc}. This breakthrough has given fresh vigour to the study of non-equilibrium dynamics in closed quantum many-body systems, and  a strenuous theoretical effort initiated -- see, \emph{e.g.}, the reviews~[\onlinecite{CC:review, C16, CaCh16, EF:review, DeLucaArxiv16, BD:review, VM:review, LangenReview, VR:review, ProsenReview}]. In most of these studies non-equilibrium dynamics is generated using a \emph{quantum quench} protocol -- the system is taken in the ground state of some Hamiltonian $H_0$ and time evolved with a different Hamiltonian $H$, related to the first by the change of some parameters. It was understood that expectation values of local-in-space observables \emph{can relax to time independent values} and these values can be computed without solving the entire dynamics -- they are determined by the local and quasi-local\cite{ProsenReview} conservation laws of the system\cite{EF:review, VR:review, RigolPRL07}. In particular,  the two extreme cases are those of \emph{integrable models} -- possessing a macroscopic number of local and quasilocal conservation laws -- and \emph{generic models} -- where the set of local and quasilocal conservation laws is reduced to the only Hamiltonian. Stationary values of local observables are believed to be described by a standard Gibbs ensemble (GE)\cite{D91, S94, R08, RS12} in the latter case, and by a generalized Gibbs ensemble (GGE)\cite{RigolPRL07} in the former. These conjectures have been verified in many highly non-trivial examples \cite{RigolPRA06,CazalillaPRL06,CalabreseJStatMech07,
CramerPRL08,BarthelPRL08,FiorettoNJP10,P:meanvalues,CEF:TFIC,DBZ:work,FE,CauxPRL12,EsslerPRL12,ColluraPRL13, QAPRL,
MussardoPRL13,PozsgayJStatMech13,FagottiJStatMech13,KSCCI,FagottiPRB14,WoutersPRL14,
PozsgayPRL14,BKCexcited, KCC:freetohardbosons,DeNardisPRA14,SotiriadisJStatMech14,GoldsteinArxiv14,
EsslerArxiv14, Betal:LL, qbosons, MPTW:XXZ, IlievskiPRL15, S:memory, Ilievskietal, Pirolibound, BPC:sinhG, PVC:XXZ, BS:qlGGE, SGcharges, RigolPRL16}.   

The finite-time dynamics revealed itself to be an harder problem to tackle; in numerous cases, however, the time evolution of non trivial observables has been determined\cite{CazalillaPRL06,CEF:TFIC, EsslerPRL12, ColluraPRL13, BKCexcited, KCC:freetohardbosons, SotiriadisJStatMech14, RCK:QS, RCK:2componentLL, KZ:semiclassicalSG, S:memory, ES:IFT, RS-14, MCKC-14, RS:long, B:sG, NC-14, D:nogo, C:planarQQ, NIC14,AC:entropy, MK:prethermalization, RoschPRL08, KollarPRB11, worm13, MarcuzziPRL13, EsslerPRB14, Fagotti14, konik14, BF15, CTGM:pret, knap15, SmacchiaPRB15, FC15, BEGR:PRL, MenegozJStatMech15, KaminishiNatPhys15, BEGR:long, MKZ:uSC, CC,CalabreseJStatMech07, DPC:relaxationdynamics, CF:EE,CEF:TFIC,DMCF:entropyheisenberg,LK:lcbosehubbard,MWNM:fermioncorrelation,CBSSF:lcbosons, BEL:PRL, ABGM:magneticimpurityXY, KSZ:spinchargeseparation,CC07,SD:localquenches,GREE:boundstateslocalquenches, F:LocalSwitch, BF:defect, CADY:hydro, BCDF:transport, C:furtherresults, confinementnoneq, RMCKT:confinementnoneq, CCE:XXZquench,  EP:latticeversion,EP:localquench, VSDH:XX, F:Currents, CP:localquenchXXZ}. In this context a number of fascinating features emerged; one example worth mentioning is the phenomenon of \emph{prethermalization}\cite{MK:prethermalization, RoschPRL08, KollarPRB11, worm13, MarcuzziPRL13, EsslerPRB14, Fagotti14, konik14, BF15, CTGM:pret, knap15, SmacchiaPRB15, FC15, BEGR:PRL, MenegozJStatMech15, KaminishiNatPhys15, BEGR:long, NIC14} in weakly non integrable models. There, at intermediate times observables approach quasi-stationary values close to the unperturbed GGE prediction before drifting to their final GE values. Another notable phenomenon is the appearance of ``light cone effects'' in correlation functions. They show abrupt changes as functions of time when they receive some information produced by the quench and propagating at finite velocity. These effects have been theoretically predicted in many different settings\cite{CC:review,EF:review,BD:review, VM:review,CC,CalabreseJStatMech07,CF:EE,CEF:TFIC,DMCF:entropyheisenberg,LK:lcbosehubbard,MWNM:fermioncorrelation,CBSSF:lcbosons, BEL:PRL, ABGM:magneticimpurityXY, KSZ:spinchargeseparation,CC07,SD:localquenches,GREE:boundstateslocalquenches, F:LocalSwitch, BF:defect, CADY:hydro, BCDF:transport, C:furtherresults, confinementnoneq, RMCKT:confinementnoneq, VSDH:XX,  F:Currents, CP:localquenchXXZ}  and have been observed in experiments \cite{explc}. 

The physical mechanism behind this behaviour can be qualitatively described by the semi-classical interpretation  introduced in Ref.~[\onlinecite{CC}]. The dynamics are characterized in terms of correlated pairs of quasi-particles emitted at the time of the quench and propagating freely throughout the system. These quasi-particles are stable in integrable models, while they acquire a finite life-time (depending on the structure of the initial state) in the presence of integrability breaking interactions. The semi-classical interpretation allows one to understand the dynamics generated by both local quenches, where the Hamiltonian is changed only in a finite region of space, and global ones, where the Hamiltonian is changed over a macroscopic region. In both cases, one has to imagine that the quasi-particles are generated only in the region where the Hamiltonian is changed. According to this interpretation,  light cone effects are due to the finiteness of quasi-particles' maximal group velocity -- in quantum spin chains, this is generically implied by the Lieb-Robinson bounds\cite{LR72}, and in relativistic field theories by their causality structure. This is, in other words, an inherent feature of these systems, and does not depend on the quench procedure or the initial state. It is worth mentioning that, in the case of global quenches, determining the velocities of quasi-particle excitations in the presence of interactions is a highly non-trivial task. Since the quench creates a finite density of excitations the velocities get non-trivially renormalized by the interactions, in a state-dependent way\cite{BEL:PRL,CADY:hydro, BCDF:transport,AC:entropy}. Relativistic invariance or Lieb-Robinson bounds, however, ensure that a maximal velocity always exists.

An immediate question arising from the above discussion is the following: what happens in cases where the quasi-particles have unbounded spectra as in, \emph{e.g.}, non-relativistic quantum field theories? Do light cone effects also persist there? If so, what is the mechanism preserving them? In the context of global quenches in non-relativistic field theories some modifications to the light cone effects have been observed.\cite{KCC:freetohardbosons} Here we focus on local quenches in a non-interacting non-relativistic fermionic field theory. In particular, we consider the so called ``cut and glue" quench protocol\cite{CC07,CC:review}. Two copies of the system are initially separated, each in its own ground state, and at time $t=t_0=0$ they are instantaneously glued together. By investigating the time dependence of the correlation functions we find that approximate light cone effects survive but there are tangible corrections; we motivate this by computing the distribution of the quasi-particle excitations produced by the quench and showing that the vast majority of them move at the initial Fermi velocity. These excitations are the ones responsible for the light cone effects. The corrections are generated by faster quasi-particles, which are also produced by the quench. 

The paper is laid out as follows. In Section~\ref{Sec:model} we introduce the non-relativistic fermionic quantum field theory considered in this work, while in Section~\ref{Sec:quench} we describe the quench protocol adopted. Section \ref{Sec:twopoint} is devoted to the calculation of the two-point function of the fermionic operators and the density of particles in the thermodynamic limit; we also perform a detailed analysis of the results, computing the distribution of excitations created by the quench. In Section \ref{sec:EE} we study the time evolution of the entanglement entropy, while in Section~\ref{Sec:generalizations} we discuss two generalisations of the problem examined: the inclusion of a harmonic trapping potential and the sudden joining of more then two systems at one point. Section~\ref{Sec:conclusions} contains our conclusions. A number of technical points and further details are reported in the appendices.

\section{The model}
\label{Sec:model}
\begin{figure}
\begin{tikzpicture}[xscale=4, yscale=4]
\draw[-, line width= 0.75mm] (0,0) -- (1,0) ;
\draw[-,  line width= 0.75mm] (-1,0) -- (0,0) ;
\draw[-, line width= 0.75mm] (1,-0.025) -- (1,0.025) ;
\draw[-, line width= 0.75mm] (-1,-0.025) -- (-1,0.025) ;
\draw node [above] at (0,0.05) {$\mathbb{S}(k)$};
\draw node[below] at (0.5,-0.01) {$L$};
\draw node[below] at (-0.5,-0.01) {$L$};
\draw node[above] at (0.75,0.01) {$E_1$};
\draw node[above] at (-0.75,0.01) {$E_2$};
\filldraw (0,0) circle (1.5pt);
\end{tikzpicture}
\caption{Our system; non-relativistic spinless fermions of mass $m_0$ live on two edges $E_1, E_2$ of length $L$, joined through the localized defect $\mathbb S(k)$.}\label{Fig:system}
\end{figure}
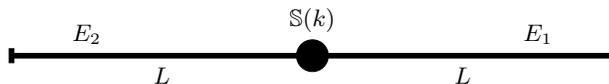

We consider non-relativistic spinless fermions of mass $m_0$ living on two edges ($E_1$ and $E_2$) of length $L$, which have a common endpoint -- ``the junction" -- featuring a localized defect. A graphical representation of our system is given in Fig.~\ref{Fig:system}.  It is convenient to parametrize a point in the system with the pair $(i,x)$, where $i=1,2$ specifies the edge and $x\in[0,L]$ the position on the edge -- measured from the junction.  The particles are described by the field $\psi_i(x,t)$, such that $\psi_i^\dag(x,0)$ creates a fermion at the point $(i,x)$. The field satisfies the canonical anti-commutation relations
\be
\label{Eq:CARxy}
\left\{\psi^{\phantom{\dag}}_i(x,t),\psi^{\dag}_j(y,t)\right\}=\delta_{ij}\delta(x-y)\,,\qquad\qquad\left\{\psi^{\phantom{\dag}}_i(x,t),\psi^{\phantom{\dag}}_j(y,t)\right\}=0=\left\{\psi^{\dag}_i(x,t),\psi^{\dag}_j(y,t)\right\}\,,
\ee
where $\delta_{ij}$ is a Kronecker delta and $\delta(x)$ is a Dirac delta.
To keep things simple, we assume that in the bulk of every edge ($0<x<L$) there is no interaction between the particles, \emph{i.e.} the equation of motion (EOM) for the field $\psi_i(x,t)$ is a free Schr\" odinger equation
\be
\label{Eq:EOM}
\left(i\partial_t  +\frac{1}{2m_0}\partial^2_x\right)\psi_i(x,t)=0\,, \qquad\qquad\qquad i=1,2\,.
\ee  
At the boundary $x=L$ we impose Dirichlet (open) boundary conditions
\be
\label{Eq:Cboundary}
\psi_i(L,t)=0\,,\qquad\qquad\qquad\qquad\qquad \forall i\,,
\ee 
we expect this choice not to affect the results presented  -- the $x=L$ conditions remain fixed in our quench protocol and we will be interested in the the infinite $L$ limit.

The defect at the junction ($x=0$) generates a point-like interaction. We model this interaction in the simplest non-trivial way\cite{H:graph, KS:graph,SGQFT,SGbos,SGQFTreview,AnyonsSG,SGNESS, MSS, SGEE} -- we assume it conserves the number of particles and generates non-trivial scattering only at the one-body level. In other words, we assume that the scattering matrix can be decomposed as a direct sum of scattering matrices $\mathbb S^{(n)}$ acting on sectors of fixed particle-number $n$ and that all the $\mathbb S^{(n)}$ are written in terms of a non trivial one-body scattering matrix $\mathbb S^{(1)}=\mathbb S$.  

The one body scattering is equivalent to some linear boundary conditions on the field $\psi_i(x,t)$ at the junction. The form of these boundary conditions is fixed by requiring the time evolution to be unitary\cite{H:graph, KS:graph, SGQFT,SGQFTreview} and reads as  
\be
\sum_{j=1}^{2}\lambda\left[\mathbb{I}-\mathbb{U}\right]_{ij}\psi_j(0,t)-i\left[\mathbb{I}+\mathbb{U}\right]_{ij}\partial_x\psi_j(0,t)=0\,,\label{Eq:Cvertex}
\ee
where $\lambda$ is a real positive parameter with dimension of mass, $\mathbb U$ is a unitary matrix; these quantities specify the physical properties of the scattering at the junction. The conditions \fr{Eq:Cvertex} correspond to the following momentum-space scattering matrix\cite{H:graph, KS:graph, SGQFT,SGbos,SGQFTreview}  
\be
\mathbb{S}(k)=-\frac{\lambda\left(\mathbb{I}-\mathbb{U}\right)-k\left(\mathbb{I}+\mathbb{U}\right)}{\lambda\left(\mathbb{I}-\mathbb{U}\right)+k\left(\mathbb{I}+\mathbb{U}\right)}\,,\qquad\qquad k\in\mathbb{R}\,,
\ee
where we used that $\lambda\left(\mathbb{I}-\mathbb{U}\right)+k\left(\mathbb{I}+\mathbb{U}\right)$ and $\left(\lambda\left(\mathbb{I}-\mathbb{U}\right)-k\left(\mathbb{I}+\mathbb{U}\right)\right)^{-1}$ commute to write $\mathbb{S}(k)$ in ``fraction form''. The diagonal element $[\mathbb{S}(k)]_{ii}$ of the scattering matrix is the reflection amplitude at the junction of a fermion with momentum $k$ on the $i$-th edge. The off diagonal elements, $[\mathbb{S}(k)]_{ij}$ with $i\neq j$, are the transmission amplitude of the fermion from the $i$-th to the $j$-th edge. Accordingly, the scattering matrix fulfils the unitarity condition\cite{SGQFT,SGQFTreview}  
\be
\label{Eq:unitarity}
\mathbb{S}^{\dag}(k)\mathbb{S}(k)=\mathbb{S}(k)\mathbb{S}^{\dag}(k)=\mathbb{I}\,.
\ee
In addition, $\mathbb{S}(k)$ also satisfies\cite{SGQFT,SGQFTreview} 
\be
\mathbb{S}^{\dag}(k)=\mathbb{S}(-k)\qquad\qquad\mathbb{S}(\lambda)=\mathbb{U}\,.
\ee
The matrix $\mathbb{U}$ is then nothing but the scattering matrix for particles with momentum $\lambda$.
The matrix $\mathbb S(k)$ can be simultaneously diagonalised for all $k$ by the unitary matrix $\mathcal{U}$ which diagonalises $\mathbb U$
\be
\mathcal{U}^{\dag} \mathbb S(k)  \mathcal{U} = \textrm{diag}\left(\frac{k-i\lambda \tan(\alpha_1/2)}{k+i\lambda \tan(\alpha_1/2)},\frac{k-i\lambda \tan(\alpha_2/2)}{k+i\lambda \tan(\alpha_2/2)}\right)\equiv \mathbb S_d(k)\,,
\ee
where $\alpha_i\in[0,2\pi[$ are the phases of the eigenvalues of $\mathbb U$. Consequently, defining the new fields
\be
\varphi_i(x,t)\equiv\sum_{j=1}^2[\mathcal{U}^\dag]_{ij}\psi_j(x,t)\,, 
\ee  
we see that they ``diagonalize'' the conditions at the junction
\be
\label{Eq:vertexdiag}
\partial_x \varphi_i(0,t)=\left(\lambda\tan(\alpha_i/2)\right) \varphi_i(0,t)\,,
\ee
while they satisfy the Dirichlet conditions \fr{Eq:Cboundary} at $x=L$. We stress that the new fields are \emph{non-local} in space, because they are linear  combinations of $\psi_i(x,t)$ on different edges of the graph. For this reason we shall address them as ``{unphysical fields}'' and use them only as a convenient tool for computations, while all the physical questions are asked in terms of the ``{physical fields}'' $\psi_i(x,t)$ \cite{SGEE}. 

\subsection{Scale-invariant scattering matrices}
Before proceeding we introduce a further simplification: we will restrict our attention to junction conditions on the unphysical fields $\varphi_i(x,t)$ which are  of Dirichlet (open) type
\be
\varphi_i(0,t)=0\,,
\label{Eq:dirichlet}
\ee
or Neumann (free) type
\be
\partial_x\varphi_i(0,t)=0\,.
\label{Eq:neumann}
\ee
This can be achieved by taking $\alpha_i=0,\pi$ (\emph{cf}. Eq.~\fr{Eq:vertexdiag}). 
In this limit the scattering matrix becomes $k$-independent, and in particular its eigenvalues can be either equal to $-1$ or $+1$. We note that the restriction \fr{Eq:dirichlet} -- \fr{Eq:neumann} captures all the possible \emph{scale invariant} scattering matrices (and in turn junction conditions), which are the only relevant ones when a CFT description\cite{BD:review} applies.

Let us briefly review the classification all the possible scale-invariant $2\times 2$ scattering matrices\cite{SGEE}, as the explicit expressions found shall be useful in the following. The classification is based upon the number $p$ of negative eigenvalues of $\mathbb S$, there are three possible cases:
\begin{itemize}
\item[i.] $p=0$, \emph{i.e.}, $\mathbb S(k)=\mathbb S=\mathbb{I}$. This corresponds to Neumann conditions at the junction for physical fields on both edges. 
\item[ii.] $p=1$, \emph{i.e.}, $\mathbb S(k)=\mathbb S = \mathcal{U}\mathbb{S}_d\mathcal{U}^{\dag}$ where $\mathbb{S}_d=\textrm{diag}({1,-1})$ and $\mathcal{U}$ is a unitary $2\times2$ matrix. Imposing that $\mathbb S$ is unitary itself one obtains its most general form  
\be
\label{Eq:transmissive}
\mathbb S \equiv \mathbb S(\varepsilon,\theta) =\frac{1}{\varepsilon^2+1}
\begin{pmatrix}
\varepsilon^2-1 & 2\varepsilon e^{i\theta}\\
2\varepsilon e^{-i\theta} & 1-\varepsilon^2
\end{pmatrix}\,,\qquad\varepsilon\in\mathbb{R}\,,\quad\theta\in[0,2\pi[\,.
\ee
The matrix $\mathbb S(\varepsilon,\theta)$ is diagonalised by 
\be
\label{Eq:matrixU}
\mathcal{U}\equiv\mathcal{U}(\varepsilon,\theta)=\frac{1}{\sqrt{\varepsilon^2+1}}
\begin{pmatrix}
\varepsilon & e^{i\theta}\\
e^{-i\theta} & -\varepsilon
\end{pmatrix}\,.
\ee
The junction conditions for the physical fields read as 
 \begin{align}
 \varepsilon \partial_x\psi_1(0,t)&=-e^{i\theta} \partial_x\psi_2(0,t)\\
 \psi_1(0,t)&=\varepsilon e^{i\theta} \psi_2(0,t)\,.
 \end{align}
\item[iii.] $p=2$, \emph{i.e.}, $\mathbb S(k)=\mathbb S=-\mathbb{I}$. This corresponds to Dirichlet conditions at the junction for physical fields on both edges.
\end{itemize}
In the cases $p=0$ and $p=2$ the two edges are completely disconnected: there is no transmission of information between the two halves of the system. The only case corresponding to a connected system is realized for $p=1$, the transmission and reflection amplitudes read as 
\be
\label{Eq:TransRef}
T_{12}=[\mathbb S]_{12}=\frac{2\varepsilon e^{i \theta}}{\varepsilon^2+1}=T_{21}^*\,,\qquad R_{11}=[\mathbb S]_{11}= \frac{\varepsilon^2-1}{\varepsilon^2+1}=-R_{22}\,.
\ee
We see that the transmission amplitude has maximal absolute value for $\varepsilon=1$, which corresponds to zero reflection amplitude at the defect in $x=0$. The limits ${\varepsilon\rightarrow 0,\pm\infty}$ correspond again to zero transmission, in these limits one of the two disjoint edges is subject to Dirichlet conditions and the other to Neumann ones. 

\subsection{Mode expansion}
\label{subsec:mode}

Let us consider a scale-invariant scattering matrix ${\mathbb{S}}$ with $p\in[0,2]$ negative eigenvalues. The unphysical field fulfilling the junction conditions determined by ${\mathbb{S}}$ can be represented through the following mode-expansion\cite{SGQFT,SGQFTreview}
\be
\label{Eq:modeunphys}
{\varphi}_i(x,t)=\sum_{m=1}^{\infty}\phi^{i}(x,m)\,e^{-i \omega_{i}(m)t}a_i(m)\,.
\ee
Here all the quantities depend on ${\mathbb{S}}$, however, we suppress their explicit dependence to lighten notations. The operators $\{a^\dag_i(m),a^{\phantom{\dag}}_i(m)\}$ satisfy the ``momentum space'' canonical anti-commutation relations  
\begin{align}
&\{a^\dag_i(n),a^{\phantom{\dag}}_j(m)\}=\delta_{nm}\delta_{ij}\qquad\qquad \{a^\dag_i(n),a^\dag_j(m)\}=0=\{a^{\phantom{\dag}}_i(n),a^{\phantom{\dag}}_j(m)\}\,,\label{Eq:CARaj}
\end{align}
and we introduced the following functions 
\be
\phi^{i}(x,m)=
\begin{cases}
\phi^{\textsc{dd}}(x,m)\qquad \textrm{if}\quad 1\leq i\leq p\,,\\
\phi^{\textsc{nd}}(x,m)\qquad \textrm{if}\quad p< i \leq 2\,,
\end{cases}
\qquad
\omega_{i}(m)=
\begin{cases}
\omega_{\textsc{dd}}(m)\equiv\frac{1}{2m_0}\left(\frac{\pi m}{L}\right)^2\qquad\qquad\,\,\,\, \textrm{if}\quad 1\leq i\leq p\,,\\
\omega_{\textsc{nd}}(m)\equiv\frac{1}{2m_0}\left(\frac{\pi (m-1/2)}{L}\right)^2\qquad \textrm{if}\quad p< i \leq 2\,.
\end{cases}
\ee
Finally the ``elementary'' single-particle wave-functions are given by  
\begin{align}
&\phi^{\textsc{dd}}(x,m)=\sqrt{\frac{2}{L}}\sin\left(m\frac{\pi x}{L}\right)\quad\qquad\qquad m=1,\ldots\,,\\
&\phi^{\textsc{nd}}(x,m)=\sqrt{\frac{2}{L}}\cos\left((m-\frac{1}{2})\frac{\pi x}{L}\right)\qquad m=1,\ldots\,.
\end{align}
These are complete orthonormal sets of functions in $\mathbb{L}^2([0,L])$. The labels ${\textsc{dd}}$ and ${\textsc{nd}}$ emphasize that these functions respectively fulfil Dirichlet and Neumann conditions at $x=0$, while both fulfil Dirichlet conditions at $x=L$.

It is straightforward to show that \fr{Eq:modeunphys} satisfies the EOM \fr{Eq:EOM}, the commutation relations \fr{Eq:CARxy}, the boundary conditions \fr{Eq:Cboundary} and the junction conditions \fr{Eq:vertexdiag}. The physical field is easily obtained from \fr{Eq:modeunphys} using the matrix $\mathcal{U}$
\be
\label{Eq:mode}
\psi_i(x,t)=\sum_{j=1}^{2}[\mathcal{U}]_{ij}\sum_{m=1}^{\infty}\phi^{j}(x,m)\, e^{-i \omega_{j}(m) t}a_j(m)\,.
\ee 
This expression satisfies the EOM \fr{Eq:EOM}, the commutation relations \fr{Eq:CARxy}, the boundary conditions \fr{Eq:Cboundary} and the junction conditions \fr{Eq:Cvertex}.
 
Using the modes $\{a^\dag_i(m),a^{\phantom{\dag}}_i(m)\}$ the Hamiltonian $H_{\mathbb{S}}$ can be written in second quantization as\cite{SGQFT}
\be
H_{\mathbb{S}}\equiv\sum_{i=1}^2 \sum_{m=1}^\infty \omega_i(m) a^\dag_i(m) a^{\phantom{\dag}}_i(m)\,,
\ee
and, consistently, the time evolved physical fields are 
\be
\psi_i(x,t) =e^{i H_{\mathbb{S}} t} \psi_i(x,0) e^{-i H_{\mathbb{S}} t}\,.
\ee
{The domain of the Hamiltonian is composed by all finite linear combinations of Fock states constructed using the modes $\{a^\dag_i(m),a^{\phantom{\dag}}_i(m)\}$; on its domain $H_{\mathbb{S}}$ is self-adjoint.}

\subsubsection{Modes on a fixed edge}

It is useful to define mode operators that describe the propagation of a particle on a given edge in the infinite volume limit. These are indeed very helpful in developing a correct physical intuition. The modes $\{a_i(m),a^\dag_i(m)\}$ do not satisfy this requirement, they are the modes of the unphysical field and describe excitations propagating in different edges at the same time. This can be seen by considering the state $\ket{a_j(m)}\equiv a_j^\dag(m)\ket{0}$, where $\ket{0}$ is the vacuum state such that $a_i(m)\ket{0}=0$ for any $i$ and $m$. Computing the wave function of $\ket{a_j(m)}$ by taking the overlap with an eigestate of the position operator, in the infinite volume limit we find\cite{note}
\be
\lim_{L\rightarrow\infty} \sqrt{L}\braket{0|\psi_i(x,0)a_j^\dag(m)|0}=\frac{1}{\sqrt{2}}\left([\mathcal {U V}]_{i j} e^{i k x} + [\mathcal {U V}^*]_{ij} e^{-i k x}\right)\,,\qquad\qquad k = \frac{\pi m}{L}\,,
\ee
where $\mathcal V= \textrm{diag}(1,-i)$. We see that this wave-function describes a particle asymptotically propagating (for $t\rightarrow \infty$) on any edge $i$ with probability $|\mathcal U_{ij}|^2$.  The modes fulfilling the ``on-edge'' requirement are given by the following linear combinations 
\be
\label{Eq:fixededgemodes}
b_i(m) \equiv \sum_{j=1}^2[\mathcal U]_{ij} [\mathcal V]_{jj} a_{j}(m)\,.
\ee
They satisfy  
\be
\lim_{L\rightarrow\infty} \sqrt{L}\braket{0|\psi_i(x,0)b_j^\dag(m)|0}=\frac{1}{\sqrt{2}}\left(\delta_{i j} e^{i k x} + [\mathbb S]_{ij} e^{-i k x}\right)\,,\qquad\qquad k = \frac{\pi m}{L}\,.
\ee
This implies that $b_j^\dag(m)\ket{0}$ describes a particle asymptotically propagating only on the edge $j$, as required. In the following we will use the ``unphysical modes" $\{a_i(m),a^\dag_i(m)\}$ in the calculations, as they allow to simplify the problem. However, we will see that the modes $\{b_i(m),b^\dag_i(m)\}$ are necessary for a semi-classical interpretation of the results.

\section{Local Quench}%
\label{Sec:quench}
Our goal is to study the time evolution generated by a sudden change of defect at $x=0$. Namely, we take the system in the ground state $\ket{\Psi}_{\mathbb{S}_0}$ of the Hamiltonian $H_{{\mathbb{S}_0}}$ with $2\mathcal N$ particles, and time evolve it for $t>0$ by means of $H_{{\mathbb{S}_1}}$. Pictorially, we denote the local quench by
\be
\mathbb{S}_0\longrightarrow\mathbb{S}_1\,.
\label{eq:genericlocalquench}
\ee
We note that ground states of Hamiltonians with different scattering matrices at the junction are orthogonal because of Anderson's orthogonality catastrophe.~\cite{orthogonalitycatastrophe} In most of the cases we consider, we focus on the scenario where the two edges are initially disjoint with open boundary conditions and they are suddenly joined with some imperfection at the junction. This corresponds to the local quench 
\be
-\mathbb{I}\longrightarrow\mathbb S(\varepsilon,\theta)\,,
\label{eq:cutandglue}
\ee 
where $\mathbb S(\varepsilon,\theta)$ is given in Eq.~\fr{Eq:transmissive}. Eq.~\fr{eq:cutandglue} is an example of a \emph{cut and glue quench}. This kind of local quench has already been considered in conformal field theory\cite{CC07, SD:localquenches} and on the lattice.\cite{EP:localquench,EP:latticeversion, CP:localquenchXXZ} Here we consider the time evolution it induces in our non-relativistic quantum field theory; we will take advantage of our simple continuum theory in order to obtain some exact results. 

Note that the closely related quench 
\be
\mathbb S(1,0)\longrightarrow \mathbb S(\varepsilon,\theta)
\label{eq:xrayedge}
\ee
describes instead the sudden switching on of a localised potential in a previously homogeneous system. This is exactly the setting of the x-ray edge singularity problem,\cite{gogolinbook,ND:solution, SS:bos} where the appearance of a core-hole in a metal after the absorption of an x-ray generates a localised potential in the conduction band. More precisely, the quench \fr{eq:xrayedge} describes the x-ray edge setting when the core hole is created at $t_i=0$ and destroyed at $t_f=\infty$. The main difference between the quench and the x-ray edge point of view is in the observables of interest. In the quench context one normally focuses on one-time two point functions away from the junction, while in the x-ray problem on two-time two point functions at the junction.

For the reader interested in the mathematical aspects of the local quench \fr{eq:genericlocalquench}, we note that the two unbounded operators $H_{{\mathbb S}_0}$ and $H_{{\mathbb S}_1}$ (we are dealing with a continuous theory) have generically different domains. So, in general, $H_{{\mathbb S}_1}$ can not be applied to the eigenstates of $H_{{\mathbb S}_0}$ and \emph{vice versa}. The two Hamiltonians are, however, both self-adjoint because they encode the conditions \fr{Eq:Cvertex}. As a consequence, the time evolution operator is unitary and its domain can be extended to the whole Hilbert space.

\subsection{Mapping between the modes}

Let us start by considering a generic local quench \fr{eq:genericlocalquench}. Since we are dealing with a quadratic theory both before and after the quench, the easiest way to determine the time evolution of observables is by using the mode expansion \fr{Eq:mode}. In order to do that, we need to find the expectation values of the mode operators $\textstyle \{a^{\phantom{\dag}}_{j}(m),a^{\dag}_{j}(m)\}$ of the final Hamiltonian $H_{\mathbb{S}_1}$ in the initial state $\ket{\Psi}_{\mathbb{S}_0}$. The standard procedure is to find a mapping between $\textstyle \{a^{\phantom{\dag}}_{j}(m),a^{\dag}_{j}(m)\}$ and $\textstyle \{a^{\phantom{\dag}}_{0j}(m),a^{\dag}_{0j}(m)\}$, the modes of the initial Hamiltonian $H_{{\mathbb{S}_0}}$ -- this mapping is determined in Appendix~\ref{app:modeopocc} and reads as
\be
\label{Eq:mapping}
a_i(n)=\sum_{k=1}^{2}\sum_{m=1}^{\infty} [\mathcal{U}^\dag\mathcal{U}_0]_{ik} B^{ik}_{nm}a_{0 k}(m)\,.
\ee
Here $\mathcal{U}_0$ and $\mathcal U_1$ diagonalize respectively $\mathbb S_0$ and $\mathbb S_1$, while the ``overlap matrix'' $B^{ij}_{nm}$ is given by 
\be
B^{ik}_{nm}=
\begin{cases}
\delta_{nm}\qquad &\textrm{if}\quad i\leq p_1\,,\quad k\leq p_0\\
B_{nm}\qquad &\textrm{if}\quad i> p_1\,,\quad k\leq p_0\\
B^*_{mn}\qquad &\textrm{if}\quad i\leq p_1\,,\quad k> p_0\\
\delta_{nm}\qquad &\textrm{if}\quad i> p_1\,,\quad k> p_0
\end{cases}\,,
\ee
where $p_1$ and $p_0$ are respectively the number of negative eigenvalues of $\mathbb S_1$ and of $\mathbb S_0$, and we defined the ``elementary overlap matrix" $B_{nm}$ as
\be
\label{Eq:B}
B_{nm}\equiv\int_0^L{\rm d}x\, \phi^{\textsc{nd}}(x,n)^* \phi^{\textsc{dd}}(x,m)=\frac{8}{\pi}\frac{m}{4m^2-(2n-{1})^2}\,.
\ee
Using the mapping \fr{Eq:mapping}, expectation values of the modes $\{a_i^{\phantom{\dag}}(n),a_i^\dag(n)\}$ are then expressed in terms of expectation values of the initial modes $\textstyle \{a^{\phantom{\dag}}_{0j}(m),a^{\dag}_{0j}(m)\}$. Since $\ket{\Psi}_{\mathbb{S}_0}$ is the $2\mathcal N$ particle ground state of $H_{\mathbb{S}_0}$, the expectation values of the initial modes are easily written -- the expectation value of a string of initial modes is zero if the number of creation and annihilation operators appearing is odd, while it can be computed by Wick's theorem if their number is even. The building blocks are then the expectation values of bilinears  
\begin{align}
\label{Eq:thermalEVs}
& \tensor*[_{\mathbb{S}_0}]{ \braket{\Psi|a_{0i}^\dag(n) a_{0j}^{\phantom{\dag}}(m)|\Psi}}{_{\mathbb{S}_0}}=\delta_{nm}\delta_{ij}\theta_{\textsc{h}}(\mathcal{N}-n)=\delta_{nm}\delta_{ij}-\tensor*[_{\mathbb{S}_0}]{ \braket{\Psi| a_{0j}^{\phantom{\dag}}(m)a_{0i}^\dag(n)|\Psi}}{_{\mathbb{S}_0}}\,,\\
&\tensor*[_{\mathbb{S}_0}]{ \braket{\Psi|a_{0i}^{\phantom{\dag}}(n)a_{0j}^{\phantom{\dag}}(m)|\Psi}}{_{\mathbb{S}_0}}=0=\tensor*[_{\mathbb{S}_0}]{ \braket{\Psi|a_{0i}^{\dag}(n)a_{0j}^{\dag}(m)|\Psi}}{_{\mathbb{S}_0}}\,,\notag
\end{align}
where $\theta_\textsc{h}(x)$ is the step function, which is non zero only for $x\geq0$ where is equal to $1$. Using the mapping \fr{Eq:mapping} we then conclude that only strings with equal number of creation and annihilation operators have non zero expectation value; these can be computed again by Wick's theorem with the building blocks 
\begin{align}
\label{Eq:thermalEVsas}
& \tensor*[_{\mathbb{S}_0}]{ \braket{\Psi|a_{i}^\dag(n) a_{j}^{\phantom{\dag}}(m)|\Psi}}{_{\mathbb{S}_0}}=\sum_{k=1}^{2}\sum_{q=1}^{\mathcal{N}} [\mathcal{U}^\dag_0 \mathcal{U}]_{k i}   [\mathcal{U}^\dag\mathcal{U}_0]_{j k}{B^{i k}_{nq}}^* {B^{j k}_{mq}} =\delta_{nm}\delta_{ij}-\tensor*[_{\mathbb{S}_0}]{ \braket{\Psi| a_{j}^{\phantom{\dag}}(m)a_{i}^\dag(n)|\Psi}}{_{\mathbb{S}_0}}\,,\\
&\tensor*[_{\mathbb{S}_0}]{ \braket{\Psi|a_{i}^{\phantom{\dag}}(n)a_{j}^{\phantom{\dag}}(m)|\Psi}}{_{\mathbb{S}_0}}=0=\tensor*[_{\mathbb{S}_0}]{ \braket{\Psi|a_{i}^{\dag}(n)a_{j}^{\dag}(m)|\Psi}}{_{\mathbb{S}_0}}\,.\notag
\end{align}

We note that the unitary mapping \fr{Eq:mapping} does not mix creation and annihilation operators, so it conserves the number of particles -- namely
\be
\label{Eq:number}
\hat N \equiv \sum_{i=1}^2\sum_{m=1}^\infty a^\dag_{0i}(m)a^{\phantom{\dag}}_{0i}(m)= \sum_{i=1}^2\sum_{m=1}^\infty a^\dag_{i}(m)a^{\phantom{\dag}}_{i}(m)\,.
\ee

\section{Two-point function}%
\label{Sec:twopoint}
We now turn to the calculation of our main object of interest, the equal-time two-point function of fermionic operators. It is defined as  
\be
\label{Eq:corrfun}
C_{ij}(x,y,t)\equiv \tensor*[_{\mathbb{S}_0}]{ \braket{\Psi|\psi_i^\dag(x,t)\psi_j(y,t)|\Psi}}{_{\mathbb{S}_0}}\,.
\ee
In particular, we will be often concerned with a particular limit of $C_{ij}(x,y,t)$
\be
\label{Eq:n}
n_{i}(x,t)\equiv \lim_{y\rightarrow x}C_{ii}(x,y,t)\,,
\ee
describing the density of particles at the point $(i,x)$. For the local quench considered, \fr{Eq:corrfun} is the only independent non zero two-point function. Since both the initial state and final theories are Gaussian (in the bulk), the time evolution of all observables can be reconstructed from $C_{ij}(x,y,t)$ by means of the Wick's theorem.

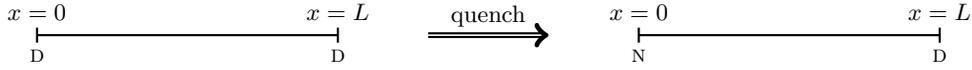
\begin{figure}
\begin{tikzpicture}[xscale=4, yscale=4]
\draw[-, thick] (0.5,0) -- (1.5,0) ;
\draw[-, thick] (-1.5,0) -- (-0.5,0) ;
\draw[-, thick] (1.5,-0.025) -- (1.5,0.025) ;
\draw[-, thick] (-1.5,-0.025) -- (-1.5,0.025) ;
\draw[-, thick] (0.5,-0.025) -- (0.5,0.025) ;
\draw[-, thick] (-0.5,-0.025) -- (-0.5,0.025) ;
\draw[->,double,thick] (-0.2,0) -- (0.2,0) ;
\draw node[above] at (0,0) {quench};
\draw node[below] at (0.5,-0.02) {$\textsc{n}$};
\draw node[below] at (-0.5,-0.02) {$\textsc{d}$};
\draw node[below] at (1.5,-0.02) {$\textsc{d}$};
\draw node[below] at (-1.5,-0.02) {$\textsc{d}$};
\draw node[above] at (0.5,0.02) {$x=0$};
\draw node[above] at (-0.5,0.02) {$x=L$};
\draw node[above] at (1.5,0.02) {$x=L$};
\draw node[above] at (-1.5,0.02) {$x=0$};
\end{tikzpicture}
\caption{Single-edge problem. The label $\textsc{d}$ denotes Dirichlet boundary conditions (\emph{cf}.~\fr{Eq:dirichlet}) while $\textsc{n}$ Neumann boundary conditions (\emph{cf}.~\fr{Eq:neumann}).}\label{Fig:singleedge}
\end{figure}

Using the mode expansion \fr{Eq:mode} and the expectation values \fr{Eq:thermalEVsas} we obtain the following expression for the two-point function \fr{Eq:corrfun}
\begin{align}
\label{Eq:correlation}
C_{ij}(x,y,t)&=\sum_{i_1,j_1=1}^2\sum_{n,m=1}^{\infty} [\mathcal{U}^\dag]_{i _1 i}[\mathcal{U}]_{j j_1}e^{i(\omega_{i_1}(n)-\omega_{j_1}(m))t}\phi^{i_1}(x,n)^*\phi^{j_1}(y,m) \tensor*[_{\mathbb{S}_0}]{ \braket{\Psi|a_{i_1}^\dag(n) a_{j_1}^{\phantom{\dag}}(m)|\Psi}}{_{\mathbb{S}_0}}\notag\\
&=\sum_{i_1,j_1,k=1}^{2}\sum_{n,m=1}^{\infty}\sum_{p=1}^{\mathcal{N}} [\mathcal{U}^\dag]_{i _1 i}[\mathcal{U}]_{j j_1}  [\mathcal{U}^\dag_0 \mathcal{U}]_{k i_1}  [\mathcal{U}^\dag \mathcal{U}_0]_{j_1 k} e^{i(\omega_{i_1}(n)-\omega_{j_1}(m))t}\phi^{i_1}(x,n)^*\phi^{j_1}(y,m) {B^{i_1k}_{np}}^*B^{j_1 k}_{mp}\,.
\end{align}
This expression holds for generic sudden changes of the scattering matrix. Specializing it to the case $\mathbb{S}_0=-\mathbb{I}$ and $\mathbb{S}_1=\mathbb{S}(\varepsilon,\theta)$ (\emph{cf}. Eq.~\fr{eq:cutandglue}) we obtain
\begin{align}
\label{Eq:CorrFunS01}
&C_{ij}(x,y,t)=[\mathcal{U}^\dag]_{1 i}[\mathcal{U}]_{j 1}\sum_{m=1}^{\mathcal{N}} \phi^{\textsc{dd}}(x,m)\phi^{\textsc{dd}}(y,m)+[\mathcal{U}^\dag]_{2 i}[\mathcal{U}]_{j 2} C(x,y,t)
\end{align}
where we introduced 
\be
\label{Eq:Cn1form1}
C(x,y,t)\equiv\sum_{n,m=1}^{\infty}\sum_{p=1}^{\mathcal{N}} e^{i(\omega_{\textsc{nd}}(n)-\omega_{\textsc{nd}}(m))t}\phi^{\textsc{dd}}(x,n)\phi^{\textsc{dd}}(y,m) B_{np}B_{mp}\,.
\ee
Here $B_{np}$ is the elementary overlap matrix introduced in Eq.~\fr{Eq:B}. We note that this building block is nothing but the two-point function that one would obtain in the ``single-edge version'' of the problem under examination. Namely, one considers fermions living on a single edge initially in the $\mathcal{N}$-particle ground state associated with Dirichlet boundary conditions on both ends, and then one suddenly changes the boundary condition at $x=0$ from Dirichlet to Neumann, see Fig.~\ref{Fig:singleedge}. By virtue of \fr{Eq:CorrFunS01} we see that for a ``cut and glue'' quench the time dependence of $C_{ij}(x,y,t)$ is completely determined by the ``single-edge'' two-point function.
Accordingly, for the density of particles we have 
\be
n_{i}(x,t)=\frac{1}{2}\left(1+(-1)^i\left(\frac{\varepsilon^2-1}{\varepsilon^2+1}\right)\right)\sum_{m=1}^{\mathcal{N}} \phi^{\textsc{dd}}(x,m)\phi^{\textsc{dd}}(x,m)+\frac{1}{2}\left(1-(-1)^i\left(\frac{\varepsilon^2-1}{\varepsilon^2+1}\right)\right)n(x,t)\,,\label{Eq:nS01}
\ee
where $n(x,t)\equiv\lim_{y\rightarrow x} C(x,y,t)$ and we used the explicit representation \fr{Eq:matrixU}. 

We now consider the ``thermodynamic limit'' -- $L\rightarrow\infty$, $\mathcal{N}\rightarrow\infty$ with fixed ``bulk density'' $n\equiv{\mathcal{N}}/{L}$. As shown in Appedix~\ref{app:intrep},  in this limit the building block \fr{Eq:Cn1form1} can be written in the following integral form
\be
 C(x,y,t)=\frac{2}{\pi}\int_{0}^{\pi n}\!\!\!\!\!{\rm d}p\, {\cal M}(x,t,p){\cal M}(y,t,p)^*\,,
\label{Eq:intrep}
\ee
where we introduced the function 
\begin{align}
\label{Eq:tildeMpIR}
{\cal M}(x,t,p)\equiv\frac{1}{2i}\left[e^{i p x}\textrm{erf}\left(e^{i\frac{\pi}{4}}\left(\frac{m_0 x + p t}{\sqrt{2 m_0 t}}\right)\right)-e^{-i p x}\textrm{erf}\left(e^{i\frac{\pi}{4}}\left(\frac{m_0 x - p t}{\sqrt{2 m_0 t}}\right)\right) \right]\,.
\end{align}
Here $\textrm{erf}(x)$ is the error function\cite{GRbook}. The representation \fr{Eq:intrep} is particularly useful to study the time-evolution of the two point function. The limit $t\rightarrow\infty$ for fixed $x$ can be found by noting
\be
\label{Eq:limit}
\lim_{t\rightarrow\infty}{\cal M}(x,t,p)=-i\cos\left(p x\right)\,.
\ee
This implies that, in the thermodynamic limit, correlation functions at infinite times after the local quench are given by the equilibrium correlation function of the final configuration. In other words, at large times (and fixed position) the time-evolved state can be replaced by the $\mathcal{N}$-particles ground state of the final configuration when computing expectation values of local observables. This is a direct consequence of the fact that the local quench does not change the energy density in the thermodynamic limit.

\begin{figure}[t]
\begin{tabular}{ll}
\includegraphics[width=0.45\textwidth]{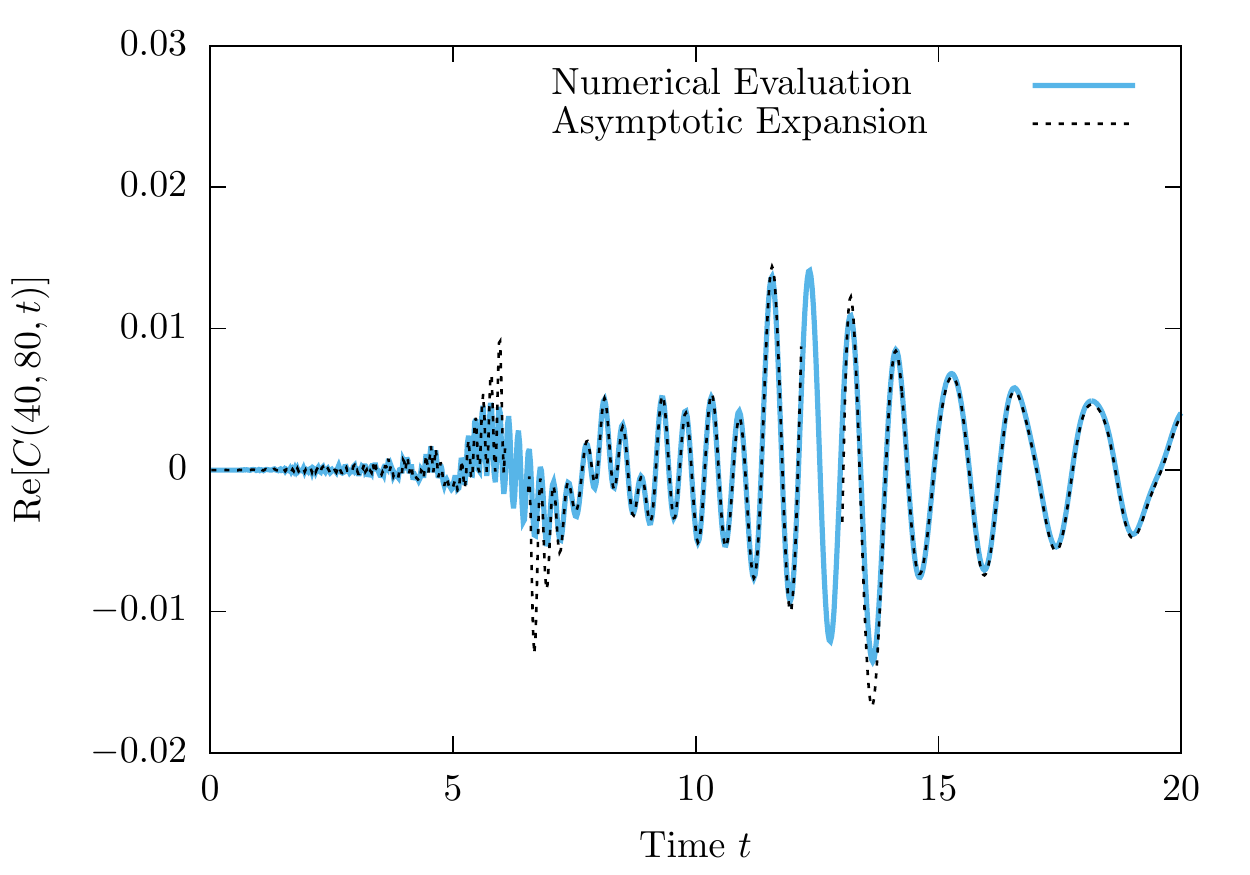} & \includegraphics[width=0.45\textwidth]{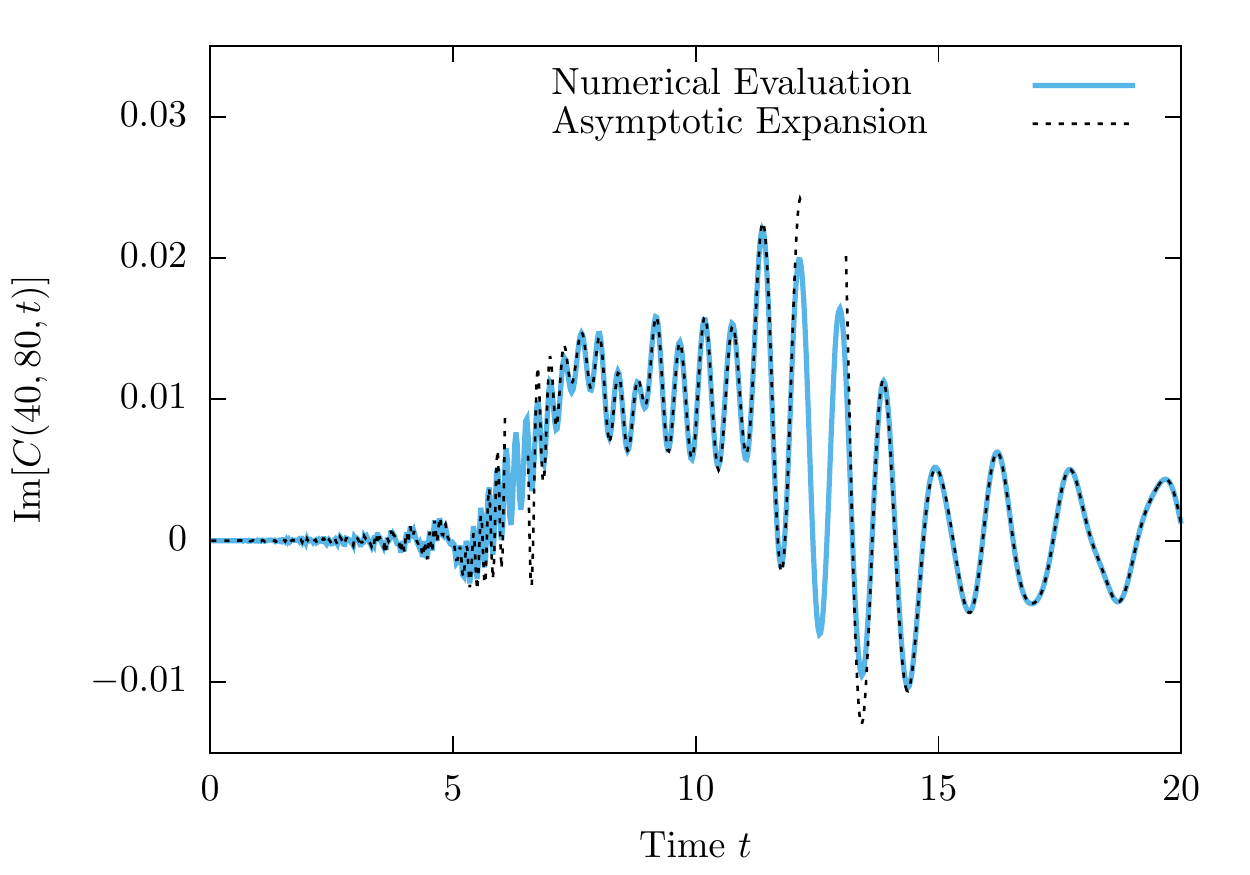} \\
\end{tabular}
\caption{Two-point function $C(x,y,t)$ as a function of time for $(x=40,y=80)$; we used $n=1$ and $m_0=1/2$. Dashed lines are the predictions of the asymptotic expansion \fr{Eq:asymptoticCorr} while full blue lines are the numerical evaluation of \fr{Eq:Cn1form1} for $L=4000$, the first $10 L$ terms are retained in the sums.}
\label{Fig:comp}
\end{figure}

\begin{figure}
\begin{tabular}{ll}
\includegraphics[width=0.45\textwidth]{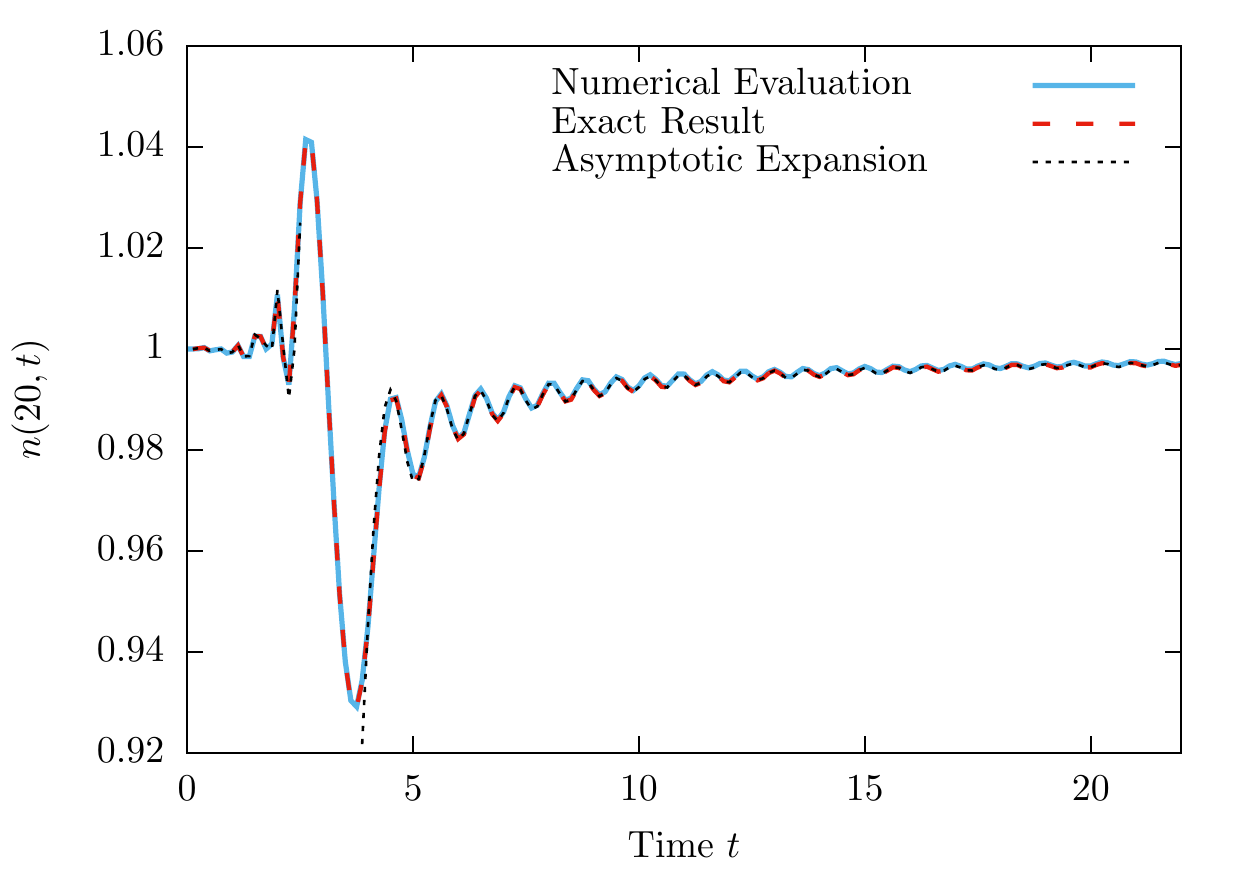} & \includegraphics[width=0.45\textwidth]{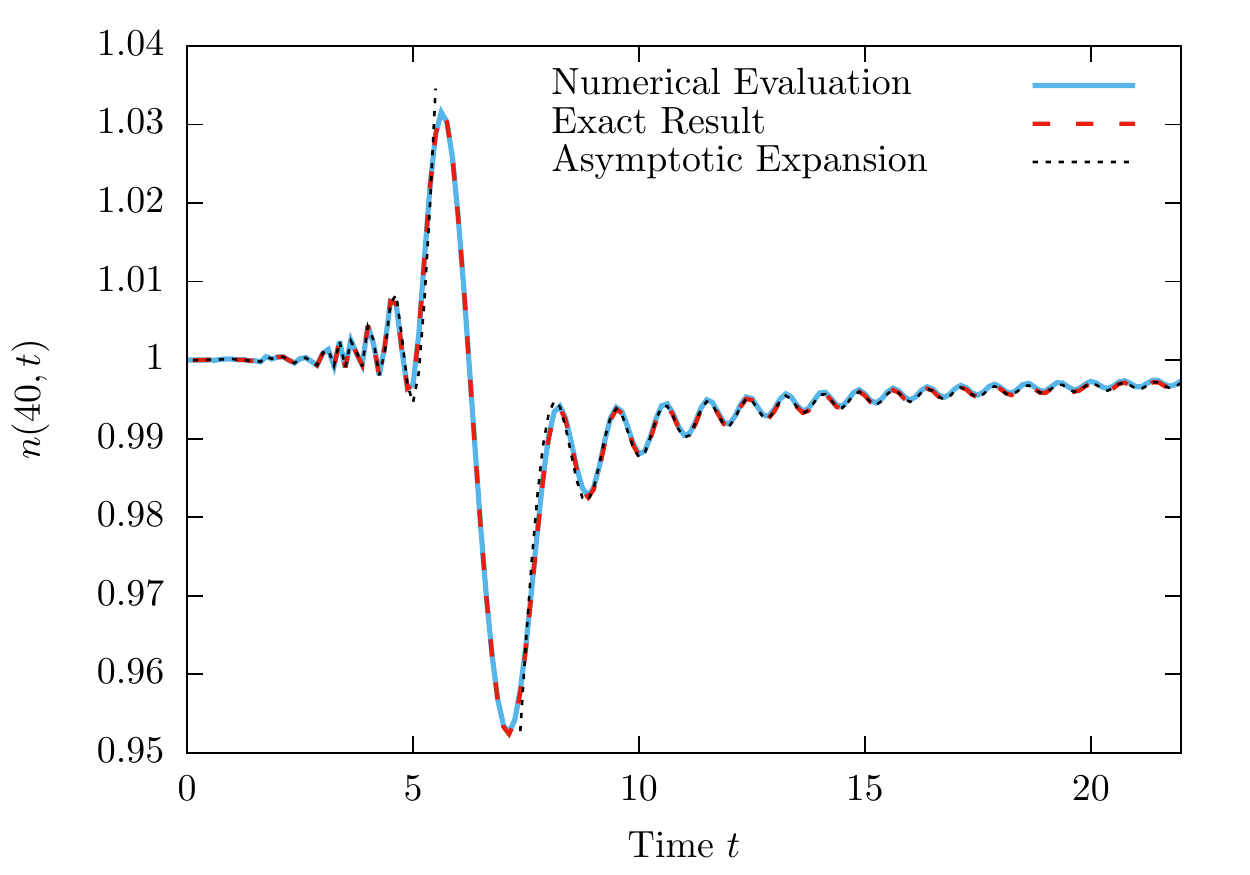} 
\end{tabular}
\caption{Density of particles $n(x,t)$ as a function of time for two different positions $x=20,40$; we used $n=1$ and $m_0=1/2$. Red dashed lines are the predictions of the exact expression \fr{Eq:exactdensity}, black dashed lines are the predictions of the asymptotic expansion \fr{Eq:asyresult} while full blue lines are the numerical evaluation of \fr{Eq:Cn1form1} for $x=y$ for $L=4000$, the first $10 L$ terms are retained in the sums.}
\label{Fig:compdens}
\end{figure}

For large but finite times, $n^2 t/m_0\gg1$, the integral representation \fr{Eq:intrep} can be asymptotically expanded.  We carry out this expansion for two fixed ``rays" $\xt=x/t$ and $\tilde\xt = y/t$ in Appendix~\ref{app:asympt}. The result reads as
\bea
&&C(x,y,t)\simeq\frac{e^{i \pi n (x-y)}-\text{sign}(v_{\rm F}- \tilde \xt)\text{sign}(v_{\rm F}  - \xt)e^{-i \pi n (x-y)}}{2\pi i (x-y)}\notag\\ 
&&\quad\quad\quad\quad\quad\quad+\frac{\text{sign}(v_{\rm F} - \tilde\xt)e^{i \pi n (x+y)}-\text{sign}(v_{\rm F} - \xt)e^{-i \pi n (x+y)}}{2\pi i (x+y)} \notag\\
&&\quad\quad\quad\quad\quad\quad
+\frac{2 e^{i t m_0 (\tilde \xt^2-\xt^2)/2}}{\pi^2 t (\tilde\xt^2 - \xt^2)} \left( \tilde \xt \log\left|\frac{v_{\rm F} - \tilde\xt}{v_{\rm F} +\tilde\xt}\right| -\xt \log\left|\frac{v_{\rm F} - \xt}{v_{\rm F} + \xt}\right|\right)\notag\\
&&\quad\quad\quad\quad\quad\quad-\sqrt{\frac{2}{\pi}}\frac{e^{i\frac{\pi}{4}} n}{(t m_0)^{3/2}}\left[\frac{\text{sign}(v_{\rm F} - \tilde \xt) (v_{\rm F} +\tilde \xt)e^{-i t m_0 (v_{\rm F}^2+\xt^2-2\tilde\xt v_{\rm F})/2}+(v_{\rm F} -\tilde \xt)e^{-i t m_0 (v_{\rm F}^2+\xt^2+2\tilde\xt v_{\rm F})/2}}{(v_{\rm F}^2-\tilde\xt^2)(v_{\rm F}^2-\xt^2)}\right]\notag\\
&&\quad\quad\quad\quad\quad\quad-\sqrt{\frac{2}{\pi}}\frac{e^{-i\frac{\pi}{4}} n}{(t m_0)^{3/2}}\left[\frac{\text{sign}(v_{\rm F}-\xt)(v_{\rm F} +\xt)e^{i t m_0 (v_{\rm F}^2+\tilde \xt^2-2 \xt v_{\rm F})/2}+(v_{\rm F} -\xt)e^{i t m_0(v_{\rm F}^2+\tilde \xt^2+2 \xt v_{\rm F})/2}}{(v_{\rm F}^2-\tilde\xt^2)(v_{\rm F}^2-\xt^2)}\right],\label{Eq:asymptoticCorr}
\eea
where we introduced the Fermi velocity
\be
\label{Eq:FV}
v_{\rm F} = \frac{\pi n}{m_0}\,.
\ee
The expansion \fr{Eq:asymptoticCorr} is up to $O(t^{-2})$ and is valid in the regions $|\xt-v_{\rm F}|\gg (m_0 t)^{-1/2}$ and $|\tilde \xt- v_{\rm F}| \gg (m_0 t)^{-1/2}$; its accuracy is confirmed by a direct comparison with the numerical evaluation of \fr{Eq:Cn1form1} as shown in Fig.~\ref{Fig:comp}. 
From \fr{Eq:asymptoticCorr} we can also find the asymptotic expansion of the density of particles by taking the limit
\be
n(x,t) \equiv \lim_{y\rightarrow x} C(x,y, t)\,.
\ee
The result reads as 
\begin{align}
&n(x,t)\simeq n+\textrm{sign}(v_{\rm F}-\xt) \frac{\sin(2\pi n x)}{2x\pi}-\frac{2n}{\pi t m_0 (v_{\rm F}^2-\xt^2)}+\frac{1}{x\pi^2}\log\Bigl|\frac{v_{\rm F} -\xt}{v_{\rm F} + \xt}\Bigr|\notag\\
&\quad-\sqrt{\frac{2}{\pi}}\frac{2 n}{(t m_0)^{3/2}}\left(\frac{\textrm{sign}(v_{\rm F}-\xt)\cos(t m_0 (v_{\rm F} -  \xt)^2/2-\pi/4)(v_{\rm F}+  \xt)+\cos(t m_0 (v_{\rm F} + \xt)^2/2-\pi/4)(v_{\rm F} - \xt)}{(v_{\rm F}^2 -\xt^2)^2}\right).\label{Eq:asyresult}
\end{align}
In fact, the density can be computed exactly by direct integration of \fr{Eq:intrep} 
\begin{eqnarray}
n(x,t)&=& \sqrt{\frac{m_0}{2 \pi^2 t}}\left(\frak F\left(( \xt + v_{\rm F})\sqrt{\frac{t m_0}{2}}\right)- \frak F\left((\xt - v_{\rm F})\sqrt{\frac{t m_0}{2}} \right)\right)\notag\\
&&-\frac{1}{2\pi x} \textrm{Im}\left [e^{2 i \pi n x} \textrm{erf}\left(e^{- i \frac{\pi}{4}}\sqrt{\frac{t m_0}{2}}( \xt - v_{\rm F})\right)\textrm{erf}\left(e^{i \frac{\pi}{4}}\sqrt{\frac{t m_0}{2}} (\xt + v_{\rm F})\right)\right ]\notag\\
&&-\frac{m_0}{2\pi^2 \xt}\textrm{Im}\left[ (\xt - v_{\rm F})^2  \,\tensor*[_2]{F}{_2}\left(1,1;\frac{3}{2},2;-i\frac{t m_0}{2} (\xt -  v_{\rm F})^2\right)\right] \notag\\
&&+\frac{m_0}{2\pi^2 \xt}\textrm{Im}\left[(\xt +  v_{\rm F})^2 \,\tensor*[_2]{F}{_2}\left(1,1;\frac{3}{2},2;-i \frac{t m_0}{2} ( \xt +  v_{\rm F} )^2\right)\right]\,,\label{Eq:exactdensity}
\end{eqnarray}
where $\tensor*[_2]{F}{_2}(a,b;c,d;z)$ is the generalized hypergeometric function\cite{GRbook} and we defined
\be
\frak F(z) \equiv \frac{1}{\sqrt{\pi}}\left(e^{i \frac{\pi}{4}} e^{i z^2} \textrm{erf}(e^{i \frac{\pi}{4}} z )+ e^{-i \frac{\pi}{4}} e^{-i z^2} \textrm{erf}(e^{-i \frac{\pi}{4}} z)\right)+z\,\textrm{erf}(e^{i \frac{\pi}{4}} z)\,\textrm{erf}(e^{-i \frac{\pi}{4}} z)\,.
\ee
A comparison of the exact expression \fr{Eq:exactdensity} and the asymptotic expansion \fr{Eq:asyresult} with the numerical evaluation of \fr{Eq:Cn1form1} for $x=y$ is presented in Fig.~\ref{Fig:compdens}. Going back to the expressions \fr{Eq:CorrFunS01} and \fr{Eq:nS01} for the correlations in our original problem, we see that \fr{Eq:exactdensity} allows one to determine exactly $C_{ij}(x,x,t)$ for $i,j=1,2$, while an asymptotic expansion of the correlation function $C_{ij}(x,y,t)$ for $x\neq y$ can be obtained from \fr{Eq:asymptoticCorr}.

\subsection{Approximate light cone behaviour}
Equations \fr{Eq:asymptoticCorr}, \fr{Eq:asyresult} and \fr{Eq:exactdensity} have a remarkable structure -- they display both features typical of light cone effects and some interesting corrections. 

Let us consider \fr{Eq:asymptoticCorr}, where we see that the leading contribution for $x,y,v_{\rm F}t\gg |x-y| $ -- reported in the first line of the r.h.s. -- changes in time only when $x=v_{\rm F} t$ or $y=v_{\rm F} t$. This is as if the information about the quench would travel at speed $v_{\rm F}$ -- a clear example light cone effect, with $v_{\rm F}$ playing the role of the velocity of light. Note that this light cone effect remains when the second term on the r.h.s. of  \fr{Eq:asymptoticCorr} is included. When $x,y>v_{\rm F}t$  the first two terms on the r.h.s. of \fr{Eq:asymptoticCorr} give the equilibrium value of the correlation function in the initial configuration  (with scattering matrix $\mathbb S_0$), while for $x,y<v_{\rm F} t$  they give the equilibrium value of the final configuration (with scattering matrix $\mathbb S_1$). In complete analogy, similar features are seen in the asymptotic expansion of the density of particles \fr{Eq:asyresult}; the first two contributions to the r.h.s. show light cone effects. Moreover, from the exact expression \fr{Eq:exactdensity}  we see that at $x/t\sim v_{\rm F}$ the density shows its strongest features (peak and dip). 
To quantify the strong features in the density, we extract the behaviour of $n(x,t)$ in the region $x\sim v_{\rm F} t$ from \fr{Eq:exactdensity} 
\begin{eqnarray}
n(x,t)&\simeq& \,\, n-\sqrt{\frac{m_0}{ 2 \pi^2 t}} (\frak F(z)-z)-\frac{1}{\pi^2 x}\textrm{Im}\left[ z^2  \,\tensor*[_2]{F}{_2}\left(1,1;\frac{3}{2},2;-i z^2\right)\right]\notag\\
&&- \frac{\sqrt{m_0}}{2 \pi^2 \sqrt{2 t} (\sqrt{2 \pi n v_{\rm F} t} + z)} -\frac{2 \log(2 (z + \sqrt{2 \pi n v_{\rm F} t}))+ \gamma}{2\pi^2 x} -\frac{ e^{2i \pi n x}\textrm{erf}(e^{-i \frac{\pi}{4}} z)- e^{-2i \pi n x} \textrm{erf}(e^{i \frac{\pi}{4}} z)}{4\pi x i}
\label{Eq:lightconeexpansion}
\end{eqnarray}
where $\gamma$ is the Euler-Mascheroni constant and we set 
\be
z\equiv\frac{x-v_{\rm F}t}{\sqrt{2t/m_0}}\,.
\label{Eq:z}
\ee 
We see that in the region $x\sim v_{\rm F} t$ the density minus its bulk value $n$ decays as $t^{-1/2}$, in other words it attains its leading contribution for $x,t\gg1$.  From Eq.~\fr{Eq:lightconeexpansion} we can also find how the width $w_{\rm lr}$ of light cone region scales with time (or equivalently with $x$, as $x\sim t$). Since most of the variation of the density in this region comes from $\frak F(z)-z$, it varies on scales $\Delta z \sim 1$; using Eq.~\fr{Eq:z} we then find $w_{\rm lr}\propto t^{1/2}$.

Together with the light cone effects described, Eqs. \fr{Eq:asymptoticCorr}, \fr{Eq:asyresult} and \fr{Eq:exactdensity} also show some corrections beyond light cone physics. These corrections are most easily identified from the asymptotic expansions \fr{Eq:asymptoticCorr} and \fr{Eq:asyresult}. Let us consider \fr{Eq:asymptoticCorr} at fixed $x<y$ and neglect the oscillatory contributions, since they decay with a larger power of $t$ (at fixed $x/t$). We see that, together with the ``light cone contributions'' analyzed above, there are also two logarithmic terms, which cause $C(x,y,t)$ to vary at times much shorter than the light cone arrival in $x$. Considering the asymptotic expansion \fr{Eq:asyresult} of $n(x,t)$ for fixed $x/t=\xt > v_{\rm F}$, we identify the leading contribution at large $x$ and $t$ as 
\be
n(x,t)-n(x,0)\simeq-\frac{2n}{\pi m_0 t (v_{\rm F}^2- \xt^2)}+\frac{1}{x\pi^2}\log\left[\frac{\xt-v_{\rm F}}{\xt+v_{\rm F}}\right]+O(x^{-3/2})
\label{Eq:corrections}
\ee
This result should be contrasted with what happens, \emph{e.g.}, in situations when Lieb-Robinson bounds~[\onlinecite{LR72}] apply, there ``superluminal" corrections are bounded by $e^{-x(\xt-v_{\rm F})/\lambda}$ ($\lambda$ is an appropriate length-scale). This means that the corrections are much larger in our case -- power law vs. exponential. The effects of the corrections can be easily detected in Fig.~\ref{Fig:compdens} -- there is a significant contribution approaching the light cone region $t\sim x/v_{\rm F}$ from smaller times, when the information traveling at velocity $v_{\rm F}$ has yet to reach the point $x$.

In summary, Eqs.~\fr{Eq:asymptoticCorr}\,--\,\fr{Eq:exactdensity} show ``approximate light cone effects" in a model without a finite maximal velocity. We call these light cone effects ``approximate'' because even though they share many of the properties with standard light cone effects in models with a \emph{finite} maximal velocity, they feature much more pronounced  ``superluminal'' corrections. Our goal now is to discuss the physical explanation of this behaviour, understanding both the origin of the light cone effect and the corrections. To do so we invoke a quasi-particle interpretation and compute the distribution of excitations produced by the local quench.

\subsubsection{Distribution of excitations}
\label{Sec:excitations}

Let us compute $\Delta N_{\rm ex}(p_m;i)$, the difference between the number of particles per mode $m$ in the edge $i$ in the state $\ket{\Psi}_{\mathbb{S}_0}$ and in the ground state of the post-quench configuration $\ket{\Psi}_{\mathbb{S}_1}$. This quantity is defined in terms of the fixed edge modes $\{b_i^{\phantom{\dag}}(m),b^\dag_i(m)\}$  (\emph{cf}.~\fr{Eq:fixededgemodes}) as follows
\be 
\label{Eq:excitations}
\Delta N_{\rm ex}(p_m; i) \equiv  \tensor*[_{\mathbb{S}_0}]{ \braket{\Psi|b_{i}^\dag(m) b_{i}^{\phantom{\dag}}(m)|\Psi}}{_{\mathbb{S}_0}}-\tensor*[_{\mathbb{S}_1}]{ \braket{\Psi|b_{i}^\dag(m) b_{i}^{\phantom{\dag}}(m)|\Psi}}{_{\mathbb{S}_1}}=\left|[\mathcal U]_{i2}\right|^2\left( \sum_{r=1}^{\mathcal N} B_{m r}^2-\theta_{\textsc{h}}(\mathcal{N}-m)\right)\,,
\ee
where $p_m\equiv\pi (m-1/2)/L$. In writing \fr{Eq:excitations} we used the expectation values \fr{Eq:thermalEVsas} and $\mathbb{S}_0=-\mathbb{I}$. The quantity $\Delta N_{\rm ex}(p_m; i) $ can be interpreted as
\be
\Delta N_{\rm ex}(p_m; i) = \tilde \delta_{\rm ex}(p_m; i) \Delta p_m\,,  
\ee
where $\Delta p_m=\pi/L$ and $\tilde \delta_{\rm ex}(p_m; i) $ is the distribution of excitations generated by the local quench on the $i$-th edge. In the semi-classical interpretation, these excitations are responsible for all the dynamics. Note that, depending on the mode $m$, $\tilde \delta_{\rm ex}(p_m; i)$ will also be negative, in that case a ``hole" excitation is created over the ground state. Finally, we note that the edge dependence of \fr{Eq:excitations} is trivial -- the total distribution of excitations 
\be
\tilde \delta_{\rm ex}(p_m) \equiv \sum_{i=1}^{2}\tilde \delta_{\rm ex}(p_m; i)\,, 
\ee
is distributed among the edges according to the factor $\left|[\mathcal U]_{i2}\right|^2$, so can we immediately pass from $\tilde \delta_{\rm ex}(p_m) $ to $\tilde \delta_{\rm ex}(p_m; i)$ and \emph{vice versa}.  

\begin{figure}
\includegraphics[width=0.6\textwidth]{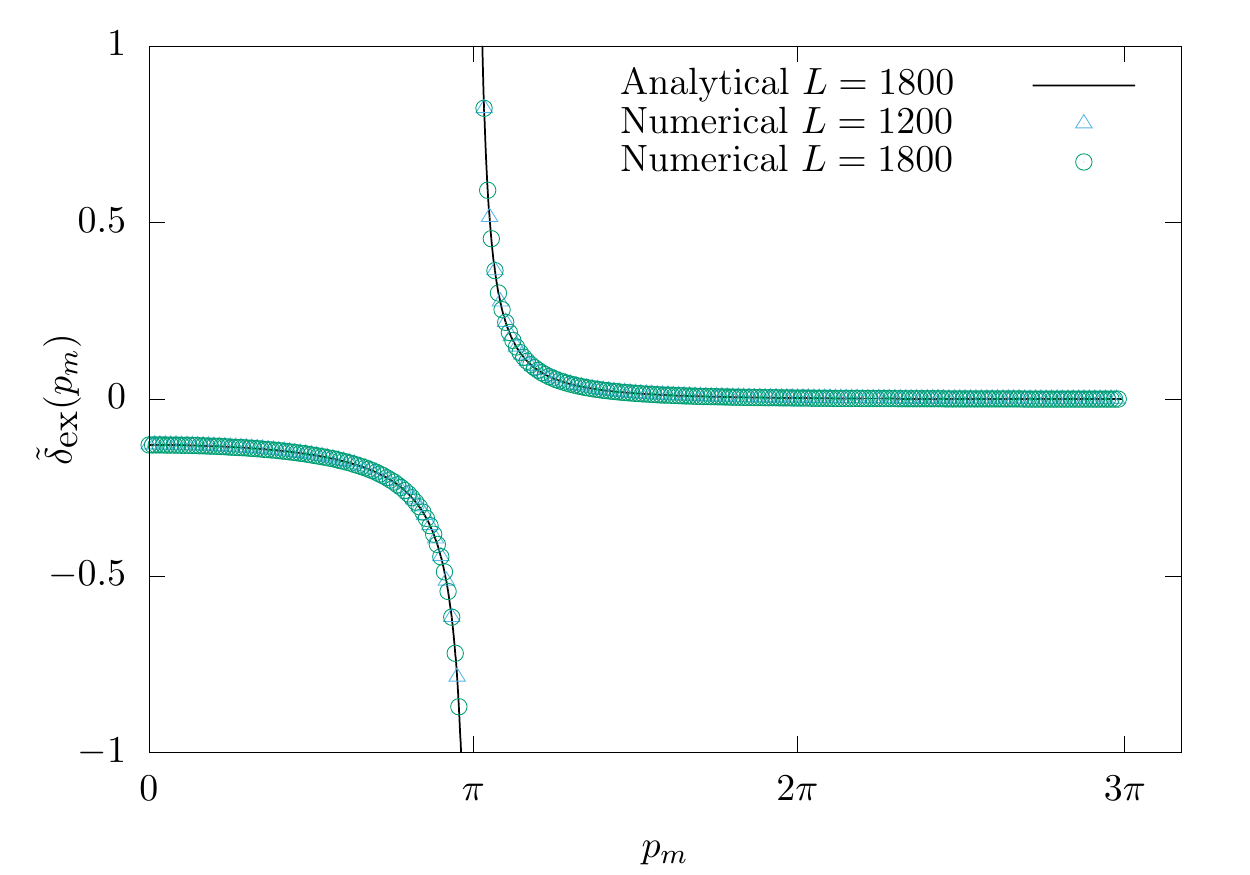}
\caption{Number of excitations over the ground state $\tilde\delta_{\rm ex}(p_m)$ as a function of $p_m=\frac{\pi (m-1/2)}{L}$.}
\label{fig:momentum}
\end{figure}

Let us now move to the computation of $\tilde \delta_{\rm ex}(p_m,i)$, to do that we need to evaluate 
\be
\label{Eq:seriesnp}
\sum_{r=1}^{\mathcal N} B_{m r}^2 =\frac{4}{\pi^2 } \sum_{r=1}^{\mathcal{N}} \frac{r^2}{(r^2-(m-{1}/2)^2)^2}=\frac{4}{L^2}\sum_{r=1}^{\mathcal{N}} \frac{(\frac{\pi r}{L})^2}{((\frac{\pi r}{L})^2-p_m^2)^2}\,,
\ee
where we used the explicit form \fr{Eq:B} of  $B_{nm}$. This sum can be turned into a contour integral as follows 
\be
\label{Eq:ISmodedist}
 \frac{4}{L^2}\sum_{r=1}^{\mathcal{N}} \frac{(\frac{\pi r}{L})^2}{((\frac{\pi r}{L})^2-p_m^2)^2}= \frac{i}{L} \underset{{z=p_m}}{\text{Res}}\left[ \frac{8 z^2 f_{\rm c b}(z)}{(z^2-p_m^2)^2}\right]\theta_{\textsc{h}}(\pi n- p_m)-\frac{1}{L}\oint\limits_{\mathcal{C}}\frac{{\rm d} z}{2 \pi } \frac{8 z^2 f_{\rm c }(z) }{(z^2-p_m^2)^2}\,,
\ee
where $\mathcal{C}$ encircles the interval $[0,\pi n]$ and we introduced the counting function  
\be
\label{Eq:contf}
 f_{\rm c }(z)=\frac{1}{1-e^{2 i L z}}\,.
\ee
In the thermodynamic limit $L\rightarrow\infty$ only the portion of $\mathcal{C}$ lying on the upper half complex plane contributes, the other is exponentially suppressed in $L$ (\emph{cf}. Eq.~\fr{Eq:contf}). Therefore, evaluating the residue, we have 
\be
\sum_{r=1}^{\mathcal N} B_{m r}^2-\theta_{\textsc{h}}(\mathcal{N}-m) = \theta_{\textsc{h}}(\pi n- p_m)-\theta_{\textsc{h}}(\pi n- p_m-\frac{\pi}{2L})+\frac{i}{p_m L}\theta_{\textsc{h}}(\pi n- p_m) +\frac{1}{L}\int_0^{\pi n} \frac{{\rm d} x}{2 \pi } \frac{8 x^2 }{(x^2-p_m^2 + i \epsilon)^2} + o\left(\frac{1}{L}\right)\!,
\ee
where $o(x)$ denotes little-O of $x$, \emph{i.e.} it goes to 0 \emph{faster} than $x$. The integral can be evaluated by deforming the integration contour, as done in Appendix~\ref{app:asympt} (\emph{cf}. Eq.~\fr{Eq:(v)kt}). Putting everything together we find
\be
\label{Eq:discretedist}
 \tilde\delta_{\rm ex}(p_m; i) = \left|[\mathcal U]_{i2}\right|^2\left( \frac{ L}{\pi}\left(\theta_{\textsc{h}}(\pi n- p_m)-\theta_{\textsc{h}}(\pi n- p_m-\frac{\pi}{2L})\right)+ \frac{1}{p_m \pi^2}\log\left|\frac{\pi n - p_m}{\pi n + p_m}\right|-\frac{1}{\pi}\frac{2n}{(\pi n)^2-p_m^2}\right).
\ee
Figure~\ref{fig:momentum} compares $\tilde\delta_{\rm ex}(p_m) $ computed from \fr{Eq:discretedist} to that found by numerical evaluation of  Eq.~\fr{Eq:excitations}. The agreement is excellent, with finite size effects being stronger near the singularity at $p_m\sim \pi n$. 
In the thermodynamic limit $\tilde \delta_{\rm ex}(p_m;i)$ is mapped into a distribution over a continuous variable, namely
\be
\label{Eq:tdlimit}
\lim_{L\rightarrow\infty}\sum_{m=1}^\infty f(p_m)  \tilde\delta_{\rm ex}(p_m;i)  \Delta p_m = \int_0^\infty\!\!f(p) \delta_{\rm ex}(p;i) {\rm d}p\,,
\ee 
for every smooth function $f(p)$. The distribution $\delta_{\rm ex}(p;i)$ can be found by turning the sum over $m$ in Eq.~\fr{Eq:tdlimit} into an integral (see Appendix~\ref{app:changeofvariables} for the detailed calculation), this procedure yields
\be
\label{Eq:distribution}
\delta_{\rm ex}(p;i) = \left|\mathcal U_{i2}\right|^2\left(\frac{1}{2}\delta(\pi n - p)+ \frac{1}{p \pi^2}\log\left|\frac{\pi n - p}{\pi n + p}\right|-\frac{1}{\pi}\left(\frac{2n}{\pi n+p}\right){\rm PV}\left(\frac{1}{\pi n - p}\right)\right)\,,\qquad p\in\mathbb{R}^+\,,
\ee
where PV denotes the principal value. We note that the presence of the delta function with coefficient $1/2$ ensures 
\be
\int_0^\infty \!\!{\rm d}p\, \delta_{\rm ex}(p;i)=0\,, \qquad \forall\,i\,,
\ee
which is consistent with Eq.~\fr{Eq:number} and with the direct summation of \fr{Eq:excitations}: there is no net production of particles, only particle-hole excitations are produced. Note that in \fr{Eq:distribution} particles have positive weight while holes have negative weight. The numbers $N_{\tt p}^i$ and $N_{\tt h}^i$ of particles and holes produced in the edge $i$ are, however, \emph{infinite} in the thermodynamic limit, as one can see by noting 
\be
N^i_{\tt p}=N^i_{\tt h}=\int_{\pi n}^\infty \!\!{\rm d}p\, \delta_{\rm ex}(p;i)=\infty\,.
\ee
This expression diverges due to the non integrable singularity at $p=\pi n$. Physically this means that an infinite number of particle-hole excitations are produced by taking the particles very close to the Fermi surface at $p_{\rm F}=m_0 v_{\rm F}$ and moving them right outside of it.  Particle-hole excitations of this kind are propagating exactly at the velocity $v_{\rm F}$ and have arbitrary small energy, so a macroscopic number of them can be created without changing the energy density.  

\subsubsection{Physical interpretation}
\label{subsec:SC}
From Eq.~\fr{Eq:distribution} we find the velocity distribution of the quasi-particles excitations by changing variables to $v= {p}/{m_0}$ 
\be
\label{Eq:vdistribution}
\delta^v_{\rm ex}(v; i) = \left| [\mathcal U]_{i2}\right|^2\left(\frac{m_0}{2}\delta(v_{\rm F} - v)+ \frac{1}{v \pi^2}\log\left|\frac{v_{\rm F} - v}{v_{\rm F} + v}\right|-\frac{1}{\pi m_0}\left(\frac{2n}{v_{\rm F}+v}\right){\rm PV}\left(\frac{1}{v_{\rm F} - v}\right)\right)\,.
\ee
As before, particles have positive weight while holes have negative weight. The distribution $\delta^v_{\rm ex}(v; i)$ gives a qualitative explanation to the physics described by Eqs.~\fr{Eq:asymptoticCorr}, \fr{Eq:asyresult}, and \fr{Eq:exactdensity}. The local quench produces a macroscopic number of excitations moving with velocity $v_{\rm F}$ ($\delta^v_{\rm ex}(v; i)$ has a non integrable singularity at $v_{\rm F}$). These are responsible for the light cone physics observed in the correlations. In particular when passing through a point they ``locally reorganize'' the state, changing the junction conditions felt by the observables. The emergence of a macroscopic number of low energy particle-hole excitations causes the ground states before and after the quench, $\ket{\Psi}_{\mathbb{S}_0}$ and $\ket{\Psi}_{\mathbb{S}_1}$ respectively, to become orthogonal in the thermodynamic limit. This is nothing other than the celebrated Anderson's orthogonality catastrophe.\cite{gogolinbook, orthogonalitycatastrophe} Interestingly, the light cone behaviour observed is not due to some inherent property of the time-evolving Hamiltonian, but instead is caused by the structure of the initial state. The discontinuous ``Fermi-sea-like'' mode occupation numbers of the initial state (\emph{cf}. Eqs.~\fr{Eq:thermalEVs}, \fr{Eq:thermalEVsas}, and \fr{Eq:ISmodedist}) persists to late times as the local quench can not change the macrostate. It is precisely at this discontinuity that a macroscopic number excitations can be produced with the finite energy injected by the local quench. We expect this physics to be robust under small softenings of the Fermi-sea discontinuity in the mode occupation of the initial state, for example when starting from a thermal state at finite temperature $T$ and chemical potential $\mu$. For low enough temperatures, $T\ll \mu$, light cone physics will still be observable, the width of the light cone region, however, will depend on $T$. At large temperature, instead, we expect that the light cone will be completely spread out. This situation is similar to that observed in quenches from inhomogeneous initial states in the the Lieb-Liniger model\cite{CADY:hydro}: an initial state constructed by joining two different thermal states ( at temperatures $T_L$ and $T_R$) shows Luttinger-Liquid (conformal) physics at long times when $T_R,T_L\ll \mu$. 

The distribution $\delta^v_{\rm ex}(v; i)$, however, shows that there are finite contributions from faster (particle) and slower (hole) excitations. In particular, particle excitations come from the right tail of $\delta^v_{\rm ex}(v; i)$ and generate the enhanced superluminal corrections to the correlation function discussed before (\emph{cf}. Eq.~\fr{Eq:corrections}). This semiclassical interpretation can be pushed further, and allows one to make quantitative predictions about the asymptotic behaviour of the density of particles. Let us embrace the semi classical interpretation and compute the density of particles in the thermodynamic limit $n_i(x,t)$ 
\be
\label{Eq:scdens}
n_i(x,t) = n^{\rm bg}_i(x,t) + n^{\rm ex}_ i(x,t)\,,
\ee
where $n^{\rm  bg}_ i(x,t)$ is the ``background density", while $n^{\rm ex}_ i(x,t)$ is the ``density of excitations''. The background density at distance $x$ and time $t$ is equal to the equilibrium density for the junction conditions ``felt'' in $x$. Since we assume that the information about the defect is carried by the particles moving at the Fermi velocity -- their number is macroscopic so they can ``reorganise the state'' --  the background density will be written as 
\be
\label{Eq:background}
n^{\rm bg}_i(x,t) = n_i(x)_{{\mathbb S}_0} + \theta_{\textsc{h}}( v_{\rm F}t -x) (n_i(x)_{{\mathbb S}_1}-n_i(x)_{{\mathbb S}_0})\,.
\ee 
Here $n_i(x)_{{\mathbb S}_0}$ and $n_i(x)_{{\mathbb S}_1}$ are the equilibrium densities in the initial and final configurations respectively. They can be respectively computed by taking $t\rightarrow 0,\infty$ in \fr{Eq:exactdensity} and plugging in \fr{Eq:nS01}.

Let us now focus on the case $x \neq v_{\rm F} t$ and determine $n^{\rm ex}_ i(x,t) $. The number of excitations $d N^{\rm ex}_i(x,t)$ in the interval $[x,x+dx]$ of the edge $i$ at time $t$ is the number of excitations produced by the quench at the position $x=0$ and time $t=0$ which propagate with a velocity between $\bar v_1= x/t$ and $\bar v_2= (x+dx)/t$. Thus, we have 
\be
\label{Eq:semiclassical}
d N^{\rm ex}_i(x,t) = n^{\rm ex}_i(x,t) dx = \delta^v_{\rm ex}\left(\frac{x}{t}; i\right) \frac{dx}{t}\,,\qquad\qquad\Rightarrow\qquad\qquad n^{\rm ex}_i(x,t)= \frac{1}{t}\delta^v_{\rm ex}\left(\frac{x}{t}; i\right)\,.
\ee
From \fr{Eq:scdens}, \fr{Eq:background}, \fr{Eq:semiclassical}, and the expression for the equilibrium densities $n_i(x)_{{\mathbb S}_1}$ and $n_i(x)_{{\mathbb S}_0}$ we find  
\begin{align}
\label{Eq:exc}
n_{i}(x,t)=& n-\frac{\sin(2\pi n x)}{2\pi x}
+\left(1-(-1)^i\left(\frac{\varepsilon^2-1}{\varepsilon^2+1}\right)\right)\frac{\sin(2\pi n x)}{2\pi x}\theta_{\textsc{h}}(x-v_{\rm F}t)\notag\\
&+\frac{1}{2}\left(1-(-1)^i\left(\frac{\varepsilon^2-1}{\varepsilon^2+1}\right)\right)\left(\frac{1}{x \pi^2}\log\left|\frac{v_{\rm F} t - x}{v_{\rm F} t + x}\right|-\frac{t}{\pi m_0}\left(\frac{2n}{v_{\rm F} t + x}\right){\rm PV}\left(\frac{1}{v_{\rm F} t - x}\right)\right)\,,
\end{align}
where we used $x/t\neq v_{\rm F}$ to disregard the contribution from the  Dirac delta function. This formula exactly reproduces the asymptotic expansion of the density (obtained by plugging \fr{Eq:asyresult} into \fr{Eq:nS01}) for $x,t\gg1$ up to $O(x^{-1})$. It is widely believed that the semiclassical quasi-particle interpretation gives the non trivial $O(x^0)$ contribution for global quenches and in transport problems, as shown in many remarkable examples\cite{CC, BF:defect,BCDF:transport, AC:entropy, VSDH:XX, CADY:hydro}. Here we see that in the case of local quenches still gives the first non-trivial contribution, which now is $O(x^{-1})$.

\subsection{Transmittance dependence of $n_i(x,t)$}
\begin{figure}
\begin{tabular}{ll}
\includegraphics[width=0.5\textwidth]{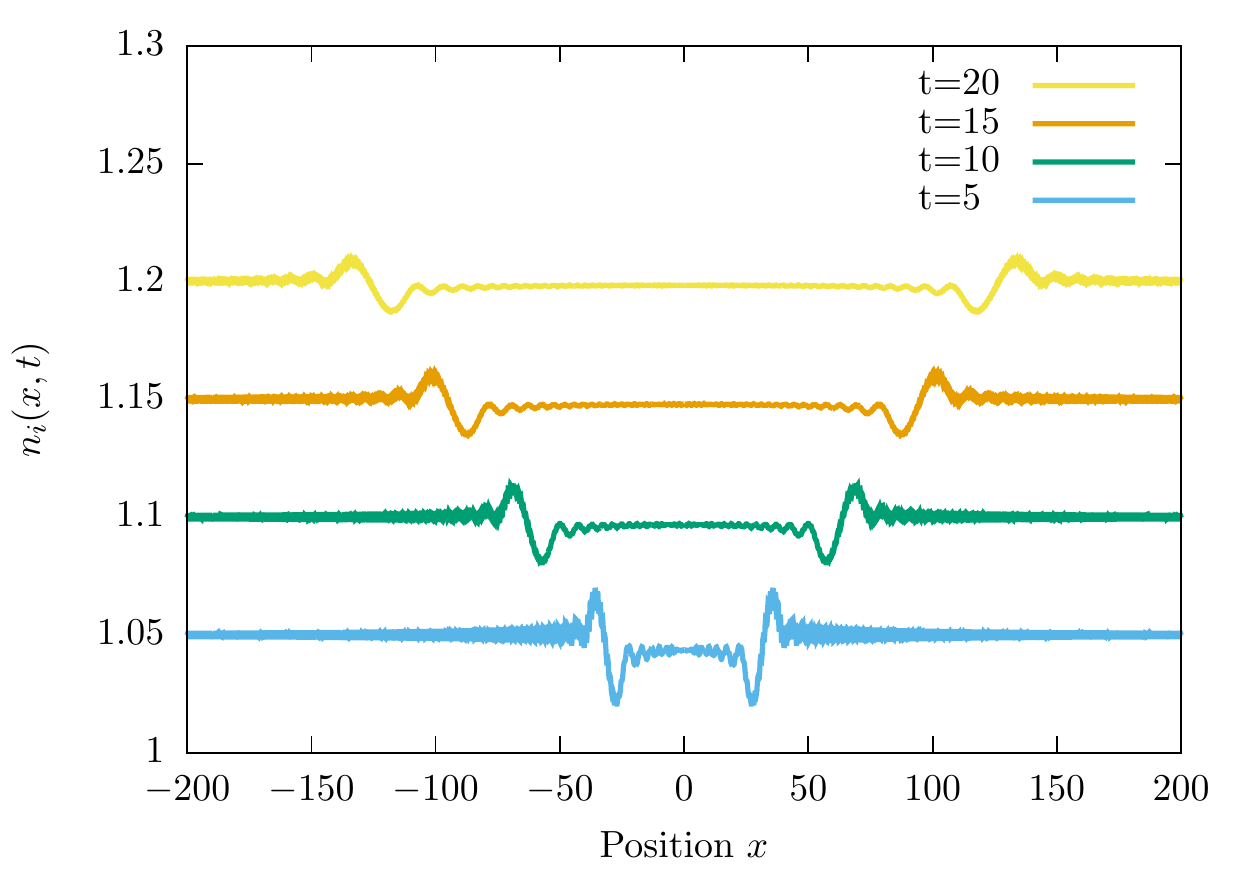} 
& \includegraphics[width=0.5\textwidth]{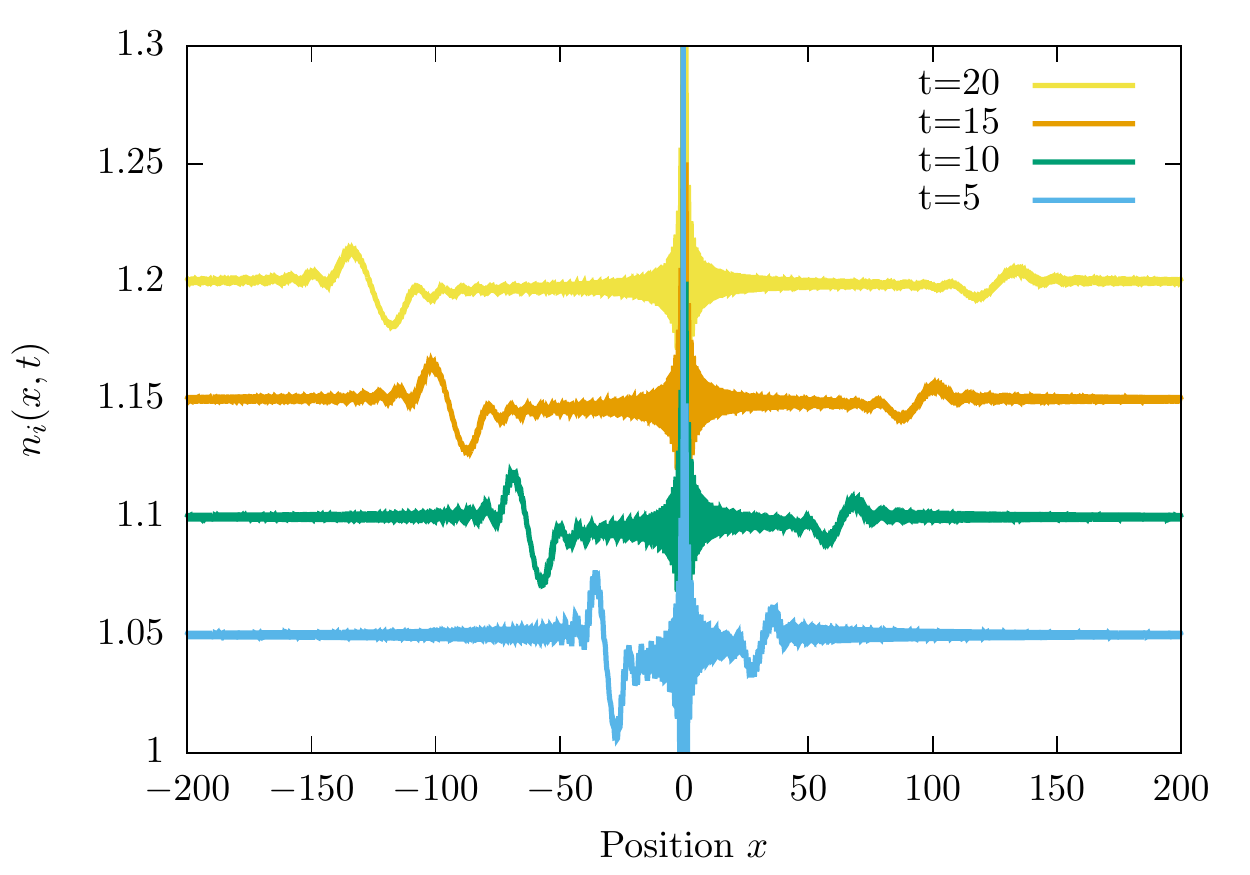} 
\end{tabular}
\caption{Profile of $n_{i}(x,t)$ (the edge $E_2$ lies on the negative real axis) at different times, for $\varepsilon=1$ (left) and $\varepsilon=0.6$ (right). We took $n=1$ and $m_0=1/2$. The curves have been obtained by plugging the exact expression \fr{Eq:exactdensity} into \fr{Eq:nS01}.}
\label{fig:twoedges}
\end{figure}

Let us move back to the exact solution \fr{Eq:exactdensity}. Inserting it into Eq.~\fr{Eq:nS01}, we can study the dependence of the density of particles $n_i(x,t)$ on the parameter $\varepsilon\in\mathbb{R}$, which characterizes the transmission and reflection probabilities at the junction. Some examples of density profiles for various times are reported in Fig.~\ref{fig:twoedges} for two different values of $\varepsilon$. The figure also highlights how the peak corresponding to the light cone is propagating through the system with velocity $v_{\rm F}$, and is slowly broadening (the width of the light cone region is $\propto t^{1/2}$).

For generic $\varepsilon$ both the initial-state and final-state values of the density of particles are characterized by the presence of Friedel oscillations, which decay as $x^{-1}$. The density is discontinuous at the origin; in particular, the value of the discontinuity $\Delta$ for $t\rightarrow\infty$ is equal to the difference of reflection amplitudes of the final scattering matrix (\emph{cf}.~\fr{Eq:TransRef})
\be
\Delta\equiv \lim_{t\rightarrow\infty} n_1(0,t)- n_2(0,t)=R_{11}-R_{22}=2
\frac{\varepsilon^2-1}{\varepsilon^2+1} \,.
\ee
The case $\varepsilon=1$ is special, corresponding to perfect transmission it gives $\Delta=0$. As a result, the Friedel oscillations vanish from the final state. We note that the limits $\varepsilon\rightarrow0,\infty$ correspond to vanishing final transmission amplitude between the two edges, but nonetheless the density of particles remains time-dependent in one of the two edges. This is because  in the quench 
\be
\mathbb{S}_0=-\mathbb{I}\longrightarrow\mathbb{S}_1=\lim_{\varepsilon\rightarrow 0,\infty}\mathbb{S}(\varepsilon,\theta)=\text{diag}(\pm1,\mp1)\,,
\ee
the junction conditions of one of the two edges are changed even if there is no mixing between the two edges.

\section{Entanglement entropy}
\label{sec:EE}
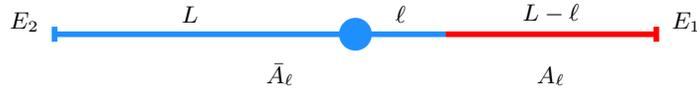
\begin{figure}[h]
\begin{tikzpicture}[xscale=4, yscale=4]
\draw[line width=0.75mm, -, lightblue] (0,0) -- (0.3,0) ;
\draw[line width=0.75mm, -, red] (0.3,0) -- (1,0) ;
\draw[line width=0.75mm, -, lightblue] (-1,0) -- (0,0) ;
\draw[line width=0.75mm, -, red] (1,-0.025) -- (1,0.025) ;
\draw[line width=0.75mm, -, lightblue] (-1,-0.025) -- (-1,0.025) ;
\draw node[below] at (1.1,0.1) {$E_1$};
\draw node[below] at (-1.1,0.1) {$E_2$};
\draw node[above] at (0.65,0.01) {$L-\ell$};
\draw node[above] at (-0.55,0.01) {$L$};
\draw node[above] at (0.15,0.01) {$\ell$};
\draw node[above] at (0.65,-0.2) {$A_\ell$};
\draw node[above] at (-0.25,-0.2) {$\bar A_\ell$};
\filldraw[lightblue] (0,0) circle (1.5pt);
\end{tikzpicture}
\caption{Partition Considered.}\label{Fig:partition}
\end{figure}

An alternative way of probing the quasi-particle content of the state after the quench is to study the time evolution of entanglement.\cite{CC, entanglement:review, CCD:review, ECP:review, L:review} We consider the subsystem $A_\ell\equiv\{(1,x)\!:\, \ell\leq x\leq L\}$, depicted in Fig.~\ref{Fig:partition}, and look at the behaviour of the entanglement between $A_\ell$ and the rest of the system. We measure the entanglement by computing the entanglement entropy,  the most accepted measure of bipartite entanglement for pure states.\cite{entanglement:review, CCD:review, ECP:review, L:review} It is defined as 
\be
S_\ell(t)=-{\textrm{tr}}\!\left[\rho_{A_\ell}(t)\log\rho_{A_\ell}(t)\right]\,.
\ee 
Here $\rho_A(t)=\textrm{tr}_{\bar A}\left[ e^{-i H_{\mathbb S_1}t}\tensor*[]{\ket{\Psi_0}}{_{\mathbb S_0}}\!\!\tensor*[_{\mathbb S_0}]{\bra{\Psi_0}}{}e^{i H_{\mathbb S_1}t}\right]$ is the density matrix of the system reduced to the subsystem $A$ ($\bar A$ denotes the complement of $A$). 

The time evolution of the entanglement entropy following a quantum quench has been intensively studied in the last decade. In particular, in the case of $1+1$ dimensional conformal field theory (CFT) many analytic results have been obtained.\cite{CC:review} These results include the calculation of $S_\ell(t)$ after the local quench of interest here, in case of perfect final transmission ($\varepsilon=1$).\cite{CC07, SD:localquenches} The conformal solution shows a very clear light cone effect -- the entropy starts to appreciably evolve in time only after $t^*=\ell/v$, where $v$ is the speed of light in the CFT. This is in full agreement with the quasi-particle interpretation -- pairs of correlated quasi-particles are emitted at the junction for $t=0$ and move at velocity $v$, then, for $t=t^*$, half of them reach the subsystem $A_\ell$ correlating it with the edge $E_2$ and therefore increasing the entanglement.

In the case of our non-relativistic model \fr{Eq:EOM} the quasi-particles are produced according to the velocity distribution \fr{Eq:vdistribution}. A macroscopic number of them move at the Fermi velocity $v_{\rm F}$, but other velocities also have non zero weight. It is then interesting to compare the time evolution of $S_\ell(t)$ in our system with the result one would have obtained considering only the excitations moving at the Fermi velocity, \emph{i.e.}, a CFT with $v=v_{\rm F}$ and central charge $c=1$. For our non-interacting field theory, the entanglement entropy can be computed following Ref.~[\onlinecite{SGEE}]. The result reads as 
\be
\label{Eq:EE}
S_\ell(t)=-\textrm{tr}\left[ \mathbb{A}_{\ell}(t) \log \mathbb{A}_{\ell}(t)+(\mathbb{I}-\mathbb{A}_{\ell}(t))\log(\mathbb{I}-\mathbb{A}_{\ell}(t)) \right]\,.
\ee
Here we introduced the $2\mathcal{N}\times 2\mathcal{N}$ matrix $\mathbb{A}_{\ell}(t)$, defined as
\be
[\mathbb{A}_{\ell}(t)]_{nm}=\int_\ell^L\! {\rm d} x\, \chi(x,t;n)^*\chi(x,t;m)\qquad\qquad n,m=1,...,2\mathcal{N}\,,
\label{Eq:matelA}
\ee
where 
\be
 \chi(x,t;n)=
\begin{cases}
\displaystyle
\sqrt{\frac{\varepsilon^2}{\varepsilon^2+1}}\sum_{p=1}^{\infty} B_{p n}e^{-i \omega_{\textsc{nd}}(p) t}\phi^{\textsc{nd}}(x,p) & 1 \leq n\leq \mathcal{N}\\
\displaystyle
\frac{1}{\sqrt{\varepsilon^2+1}}\phi^{\textsc{dd}}(x,n-\mathcal{N}) & \mathcal{N}+1\leq n\leq 2 \mathcal{N}\,.\\
\end{cases}
\label{Eq:chi}
\ee
\begin{figure}[t]
\includegraphics[width=0.7\textwidth]{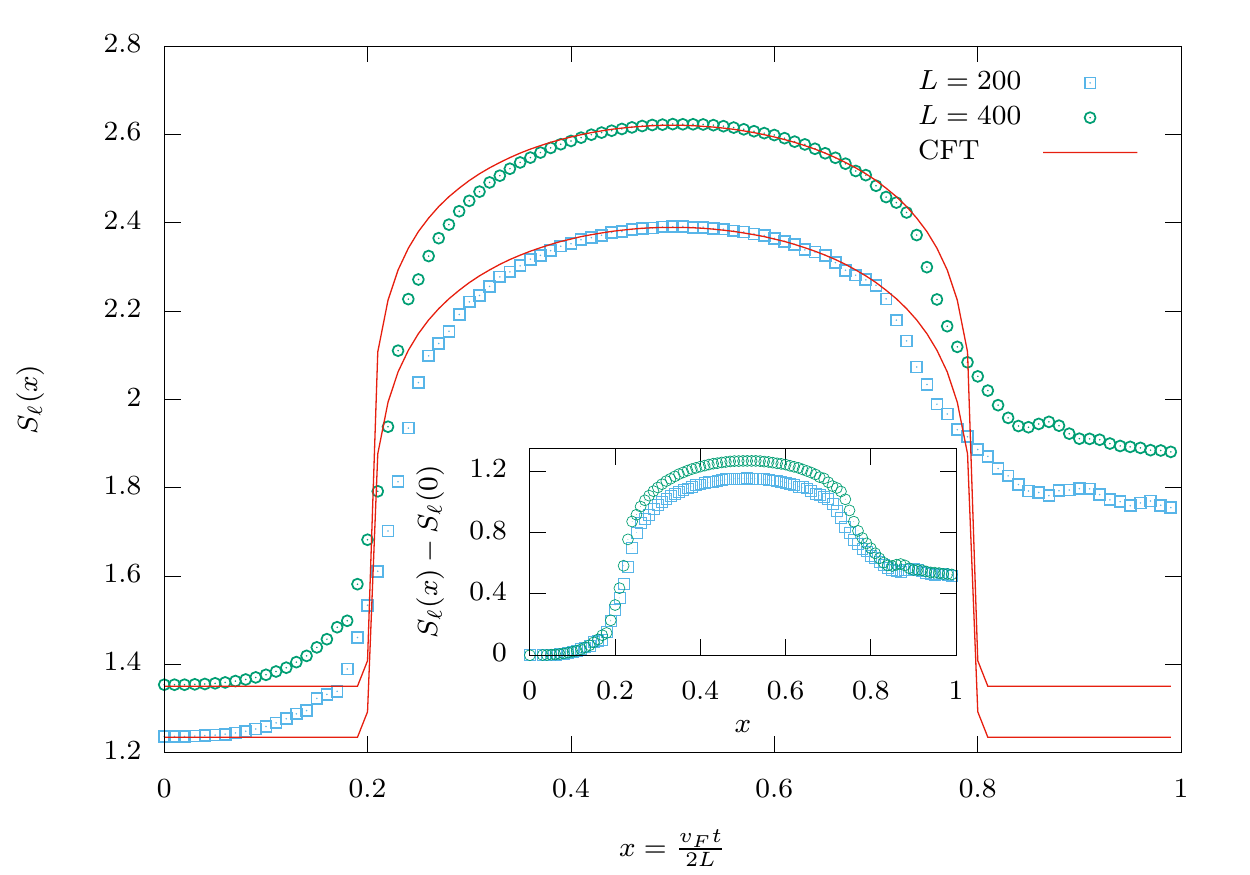}
\caption{Time evolution of the entanglement entropy $S_\ell(x)$ as a function of the rescaled variable $x=\frac{v_{\rm F}t}{2 L}$, for $\ell=0.4 L$. The CFT result of Ref.~[\onlinecite{SD:localquenches}] is compared with the behaviour found in our system of non relativistic fermions for various system sizes $2L=160,200$. We considered fixed density $n=1$ and perfect transmission $\varepsilon=1$ in the final configuration; points indicate the results for our model while red lines are the CFT result. The inset shows the behaviour of the entanglement entropy where the initial contribution $S_\ell(0)\sim \frac{1}{6}\log L$ has been subtracted.}
\label{fig:EE}
\end{figure}

The expression \fr{Eq:EE} reduces the problem of computing the entanglement entropy to that of diagonalizing a $2\mathcal{N}\times 2\mathcal{N}$ matrix. Using the explicit form of the matrix elements $[\mathbb{A}_\ell(t)]_{nm}$ reported in Appendix~\ref{app:matrixA}, we constructed and diagonalized $\mathbb{A}_\ell(t)$ numerically for various times, finding the time evolution of $S_\ell(t)$. Figure~\ref{fig:EE} reports the comparison between the CFT prediction of Ref.~[\onlinecite{SD:localquenches}] and the result; we set $\varepsilon=1$ (the entanglement entropy does not depend on $\theta$) and chosen the two non-universal constants in the CFT result to match the initial and peak value of the entanglement entropy we computed. We see that there is a substantial disagreement between the behaviour of $S_\ell(t)$ in our system and that in the CFT. At short times, $t<\ell/v_{\rm F}$, the ``conformal quasi-particles" (those moving with velocity $v_{\rm F}$) have not yet reached the subsystem $A_\ell$, consequently the CFT entanglement entropy remains constant. For non-relativistic fermions, on the other hand, the entanglement entropy starts to grow due to the presence of faster quasi-particles. For times $\ell/v_{\rm F}<t<L/v_{\rm F}$ the entanglement of non-relativistic fermions is lower than the conformal case, because slower quasi-particles are still outside the subsystem $A_\ell$ (we set the peaks to the same values). Similarly, for $L/v_{\rm F}<t<(2L-\ell)/v_{\rm F}$, faster quasi-particles leave the subsystem $A_\ell$, causing $S_\ell(t)$ for non-relativistic fermions to be lower than the CFT one. Finally, for $t>(2L-\ell)/v_{\rm F}$ the entanglement entropy of non-relativistic fermions remains larger than the initial value, unlike the CFT case. This is due to the slower quasi-particles remaining in the subsystem.

We note that the arrival of ``conformal quasi-particles" changes the scaling of the entanglement entropy with $L$, from $S_\ell(t)\sim S_{\ell}(0)\sim\frac{1}{6}\log L$ to $S_\ell(t)\sim\frac{1}{3}\log L$, as happens in the conformal case. This is clearly indicated by the inset of Fig.~\ref{fig:EE}: initially $S_\ell(x)-S_\ell(0)$ for different system sizes are lying on the top of each other while at time $t^*$ the curves for different system sizes separate. The fact that quasiparticles moving with velocity $v_{\rm F}$ change the scaling of the entanglement entropy is consistent with their number being macroscopic, as opposed to that of the quasi-particles with different velocities.

\begin{figure}[t]
\includegraphics[width=0.7\textwidth]{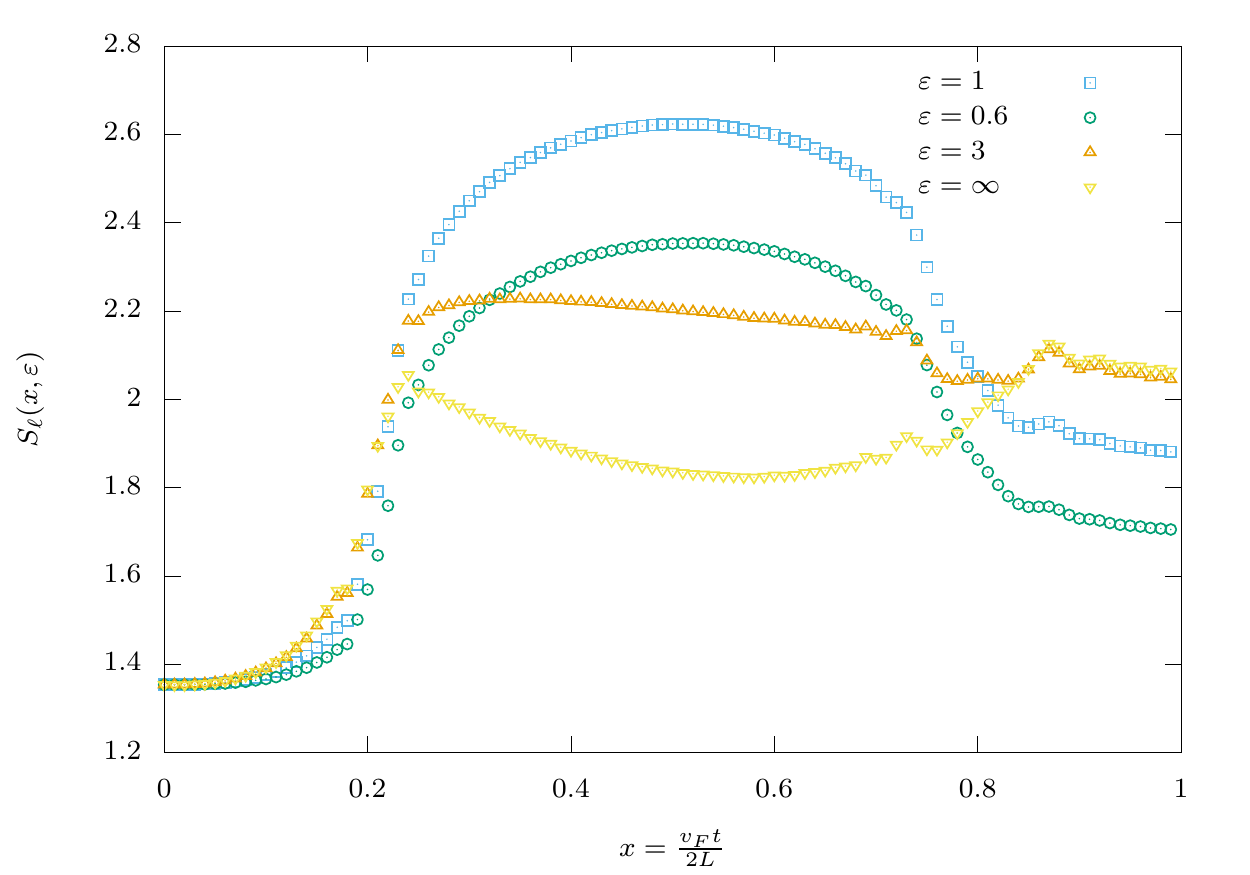}
\caption{Time evolution of the entanglement entropy $S_\ell(x,\varepsilon)$ as a function of the rescaled variable $x=\frac{v_{\rm F}t}{2 L}$, for $\ell=0.4 L$ and $L=200$. Different point types correspond to different values of $\varepsilon$, parametrizing the transmission at the defect.}
\label{fig:EEdef}
\end{figure}

We conclude this analysis by considering the behaviour of the entanglement entropy in the case of non-zero reflection amplitude, \emph{i.e.} $\varepsilon\neq1$. Here there is no CFT result to compare to. The time evolution of $S_\ell(t)$ can be again determined by diagonalising $\mathbb{A}_\ell(t)$ for various times; results are reported in Fig.~\ref{fig:EEdef}.  For $\varepsilon<1$ the EE peak decreases with $\varepsilon$ and for $\varepsilon=0$ the EE does not evolve: this clearly follows from \fr{Eq:chi}, as we are setting the time dependent part to zero. Physically, for $\varepsilon=0$ the two edges are disjoint in the final configuration  and the edge $E_1$ containing the subsystem $A_\ell$ does not change boundary conditions during the quench --  no dynamics are generated. In the quasi-particle interpretation this can be explained by saying that when $\varepsilon$ is decreased, less and less correlated quasi-particles reach the system $A_\ell$ as they are reflected by the defect. On the other hand, for $\varepsilon>1$ the form of $S_\ell(t)$ changes, see Fig.~\ref{fig:EEdef}. For  $v_{\rm F}t\sim\ell$, when the ``conformal'' quasi-particles arrive,  the entropy has a peak and then starts to decrease -- this is as if the conformal quasi-particles lower the entanglement when reaching the subsystem. The entropy remains time dependent also in the limit $\varepsilon\rightarrow\infty$. This is easy to explain -- even though in the final configuration the edges are decoupled, the conditions at $x=0$ for the edge containing $A_\ell$ have been changed from Dirichlet to Neumann, so non-trivial dynamics are generated.

\section{Generalizations}
\label{Sec:generalizations}%

\subsection{Harmonic Potential on the Wires}%
\label{subsec:harmonic}
So far, we have discussed ``cut and glue'' quenches in non-relativistic field theories with the simplest possible setting -- noninteracting fermions (in the bulk) living on two edges which are suddenly joined. The only complication introduced has been a scale-invariant defect localized at the junction in the final configuration, causing imperfect transmission of excitations. We now consider a generalisation of this setting, which could be of more direct applicability. Instead of finite edges of length $L$, we consider edges of infinite length with an additional harmonic trap which constrains the electrons within a finite region. This situation is somewhat closer to the standard setting of cold atomic experiments\cite{exp,explc}. We note that a CFT approach to deal with this inhomogeneous setting (in the ``clean case'' $\varepsilon=1$, $\theta=0$ \emph{cf}.~\fr{Eq:TransRef}) has recently been developed in Ref.~[\onlinecite{DSVC:inhomoCFT}]. 

Adding an external harmonic potential  $V(x)=\frac{1}{2}m_0\omega^2x^2$ the equation of motion for the field becomes 
\be
\label{Eq:EOMHarmonic}
\left(i\partial_t  +\frac{1}{2m_0}\partial^2_x-\frac{1}{2}m_0\omega^2x^2\right)\psi_i(x,t)=0\,.
\ee   
The conditions at the junction which guarantee unitary time evolution remain those described by Eq~\fr{Eq:Cvertex}. As before, we focus on the case where the one-body scattering matrix at the junction is scale invariant.

The computation of the equilibrium density $n_{i}(x,0^-)$ reveals that the harmonic trap forces the fermions to remain in the interval $x\in[0, \ell_{\rm eff}(\omega,\mathcal N)]$ with
\be 
\ell_{\rm eff}(\omega,\mathcal N)=\sqrt{\frac{4\mathcal N}{m_0 \omega}}\,.
\ee
As a result, the thermodynamic limit $\mathcal N\rightarrow \infty$, $\ell_{\rm eff}(\omega,\mathcal N)\rightarrow\infty$ with fixed bulk density $n=\mathcal N /\ell_{\rm eff}(\omega,\mathcal N)$ is reached by taking $\omega= 4 n^2 m_0^{-1} \mathcal N^{-1} $.

\subsubsection{Mode expansion of the field}

As before, the solution of the equation of motion \fr{Eq:EOMHarmonic} and the junction conditions \fr{Eq:Cvertex}, can be written in terms of a mode expansion (\emph{cf.} Eq.~\fr{Eq:mode})
\be
\label{Eq:modeH}
\psi_i(x,t)=\sum_{j=1}^{2}[\mathcal{U}]_{ij}\sum_{m=1}^{\infty}\phi^{j}_{\textsc{h}}(x,m)e^{-i \omega^{\textsc{h}}_{j}(m) t}a_j(m)\,.
\ee 
The difference with respect to the previous case is in the wave functions $\{\phi^j_{\textsc{h}}(x,m)\}$ and in the frequencies $\{\omega_j^{\textsc{h}}(m)\}$. Specifically, for a scattering matrix with $p$ negative eigenvalues, the functions $\{\phi^j_{\textsc{h}}(x,m)\}$  are the eigenfunctions of the single-particle Hamiltonian operator $\frac{-1}{2 m_0} \partial^2_x +\frac{1}{2} m_0 \omega^2 x^2$ fulfilling Dirichlet conditions for $1 \leq j \leq p$ and Neumann conditions for $p+1 \leq j \leq 2$. These functions are explicitly given by 
 \be
  \phi_{\textsc{h}}^j(x,m)=
 \begin{cases}
 \sqrt{2}\chi_{2 m - 1}(x)\,,  \qquad &1 \leq j \leq p\,,  \\
 \sqrt{2}\chi_{2 m- 2}(x)\,,   \qquad &p< j \leq 2\,,  \\
 \end{cases} 
\qquad
\qquad
 \omega^{\textsc{h}}_{j}(n)=
 \begin{cases}
 (2 n - \frac{1}{2}) \omega\,, \qquad  & 1 \leq j \leq p\,,  \\
 (2 n - \frac{3}{2})\omega\,, \qquad  & p< j \leq 2\,.  \\
 \end{cases} 
 \label{Eq:Hsingleparticle}
 \ee
Here we introduced the eigenfunctions of the one-dimensional harmonic oscillator 
 \be
 \chi_{n}(x) = \frac{1}{( 2^n n !)^{1/2}} \left( \frac{m_0 \omega}{\pi} \right)^{1/4} e^{-\frac{m_0 \omega x^2}{2}}H_n(\sqrt{m_0 \omega} x)\,,
 \ee
 where $H_n(x)$ are the Hermite polynomials.\cite{GRbook}

\subsubsection{Correlation Function After the Quench}
The two-point function after an instantaneous change of the scattering matrix $\mathbb{S}_0=-\mathbb I \longrightarrow  \mathbb S_1=\mathbb S(\varepsilon,\theta)$ in the ground state with 2$\mathcal N$ particles can be found as before. The result reads as 
\begin{align}
\label{Eq:correlationHarm}
C_{ij}(x,y,t)=\tensor*[_{\mathbb{S}_0}]{ \braket{\Psi|\psi_i^\dag(x,t)\psi_j(y,t)|\Psi}}{_{\mathbb{S}_0}} =&\,2 [\mathcal{U}^\dag]_{2 i}[\mathcal{U}]_{j 2}\sum_{p,q=1}^{\infty}\sum_{m=1}^{\mathcal{N}} e^{2i (p-q) \omega t} B^{\chi}_{pm}B^{\chi}_{qm}\chi_{2p-2}(x)\chi_{2q-2}(y)\notag\\
&+2[\mathcal{U}^\dag]_{1 i}[\mathcal{U}]_{j 1}\sum_{m=1}^{\mathcal{N}} \chi_{2m-1}(x)\chi_{2m-1}(y)\,,\qquad\qquad\qquad t>0\,.
\end{align}
Here we introduced the ``elementary overlap matrix''
\be
B^{\chi}_{nm}\equiv \int\limits_0^\infty \!{\rm d}x \,\,\chi_{2n-2}(x) \chi_{2 m -1}(x)= \frac{(-1)^{n + m} (2 m - 1)!! (2 n - 3)!!}{\sqrt{2\pi}
 \sqrt{(2 n - 2)!} \sqrt{(2 m - 1)!}(2 m - 2 n + 1)}\,.
\ee

\subsubsection{Time Evolution and light cone effects}
\begin{figure}
\includegraphics[width=0.5\textwidth]{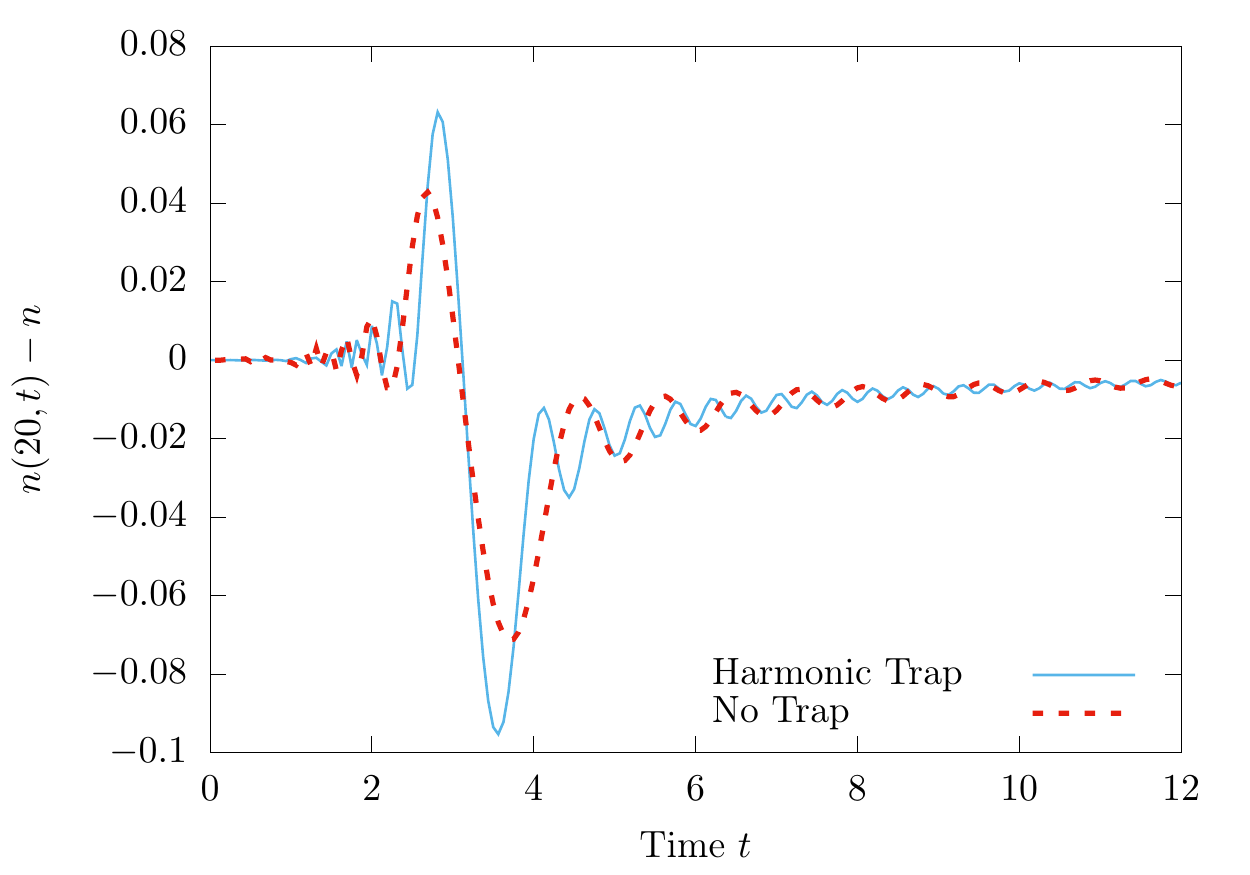}  
\caption{Time evolution of $n(x,t)-n$, comparing the result obtained in the absence of the trap (red dashed line) with the one in the presence of the trap (full blue line). The untrapped result is obtained through the exact formula \fr{Eq:exactdensity} while the trapped one via the numerical summation of \fr{Eq:Hdensity}, we took $\mathcal N=2000$ and retained the first $10 \mathcal N$ terms. The parameters are chosen such that the light-cone velocities are the same in both cases (we use the superscript ``{trap}" to indicate quantities in the presence of the Harmonic potential), in particular we took $n=1$, $m_0=1/2$,   $n^{\rm trap}=\pi/2 $, and $m_0^{\rm trap}=1$; these give $v_{\rm F}=v_{\rm F}^{\rm trap}=2 \pi$.}
\label{Fig:harm}
\end{figure}

Let us focus on the time evolution of the time dependent part  of the density (\emph{cf}. Eq.~\fr{Eq:nS01})
\be
n(x,t)=2 \sum_{p,q=1}^{\infty}\sum_{m=1}^{\mathcal{N}} e^{2i (p-q) \omega t} B^{\chi}_{pm}B^{\chi}_{qm}\chi_{2p-2}(x)\chi_{2q-2}(x)\,,
\label{Eq:Hdensity}
\ee
in the ``thermodynamic limit'' $\mathcal{N}\rightarrow\infty$ with $\omega\mathcal N = 4 n^2/m_0$. Figure~\ref{Fig:harm} reports the numerical evaluation of \fr{Eq:Hdensity}, compared with the previous result \fr{Eq:exactdensity} obtained for untrapped fermions. We see that the picture described above remains qualitatively the same; there is an approximate light cone effect with appreciable ``superluminal'' corrections.  Once again, this can be explained using a quasi-particle interpretation: deep in the trap, for large $\mathcal N$ and $\omega \propto \mathcal N^{-1}$ the eigenstates of the single-particle Hamiltonian characterized by the integer $m$ with $m \sim 2  \mathcal  N$ (\emph{cf}.~\fr{Eq:Hsingleparticle}) can be seen as quasi-particle states. These quasi-particles propagate with momentum $p_m=\sqrt{2 m \omega m_0}$ and energy $E_m=\omega m=p^2_m/(2 m_0)$. Their velocity is thus $v_m= d E_m/d p_m =\sqrt{2 m \omega/ m_0} $. The local quench creates a \emph{macroscopic} number of quasi-particle excitations moving at the ``Fermi velocity'' $v_{2 \mathcal N}= 4 n /m_0$. It also creates a finite number of excitations at higher or lower velocities, which are responsible for the corrections to the pure light cone behaviour. Numerical evaluation of \fr{Eq:Hdensity} confirms that $v_{2 \mathcal N}$ is indeed the velocity of the light cone, \emph{cf}. Fig.~\ref{Fig:harm}. 

\subsection{Star Graph}
\label{subsec:star}

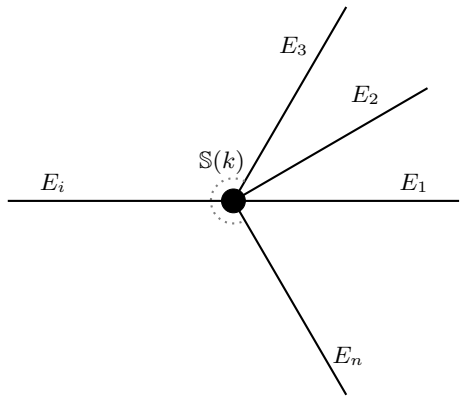
\begin{figure}
\begin{tikzpicture}[xscale=3, yscale=3]
\draw[-, thick] (0,0) -- (1,0) ;
\draw[-, thick] (0,0) -- (0.5,0.86) ;
\draw[-, thick] (0,0) -- (0.86,0.5) ;
\filldraw (0,0) circle (1.5pt);
\draw[dotted, thick,gray] (0.05,0.0866) arc (60:300:0.1);
\draw[-, thick] (0,0) -- (-1,0) ;
\node at (0.8,0) [above] {$E_1$};
\node at (0.688,0.47) [left] {$E_2$};
\node at (0.37,0.688)  [left] {$E_3$};
\node at (-0.8,0)  [above] {$E_i$};
\node at (0.4,-0.688) [right] {$E_n$};
\draw[-, thick] (0,0) -- (0.5,-0.86) ;
\draw node [above] at (-0.05,0.08) {$\mathbb{S}(k)$};
\end{tikzpicture}
\caption{A star graph.}\label{Fig:star}
\end{figure}

Throughout the paper we always focussed on the local quench which suddenly couples two edges. Our results can be easily extended to the more general case, where $n$ edges are suddenly coupled such that they all share a common end point in the final configuration, see Fig~\ref{Fig:star}. This configuration is  known as a \emph{star graph}\cite{SGQFT, SGQFTreview}; it can be experimentally realized in the context of solid state systems,\cite{carbonnanowires} and very recently substantial progress has been made in the direction of its realization with cold atomic systems.\cite{Tetal:stargraph} In the latter context a realization of the quench protocol under exam appears feasible.

In this case the scattering matrix is changed as 
\be
-\mathbb I_n\longrightarrow\mathbb S\,.
\ee
Here $\mathbb I_n$ is the $n\times n$ identity matrix and the final scattering matrix $\mathbb S$ is a $n\times n$ unitary matrix with real eigenvalues, of which $p>0$ are negative. All the formulae of Section~\ref{subsec:mode} extend to the star-graph case with the trivial substitution $2\rightarrow n$ for the number of edges. The correlation functions can then be easily computed, reading 
 \begin{align}
\label{Eq:CorrFunS0n}
&C_{ij}(x,y,t)=\sum_{k=1}^{p}[\mathcal{U}^\dag]_{k i}[\mathcal{U}]_{j k}\sum_{m=1}^{\mathcal{N}} \phi^{\textsc{dd}}(x,m)\phi^{\textsc{dd}}(y,m)+\sum_{k=p+1}^{n}[\mathcal{U}^\dag]_{k i}[\mathcal{U}]_{j k} C(x,y,t)\,,
\end{align}
where $\mathcal U^\dag=\mathcal U^{-1}$ diagonalizes $\mathbb S$ and $C(x,y,t)$ is the one given in \fr{Eq:Cn1form1}. The density of particles is then given by 
\be
n_{i}(x,t)=\frac{1}{2}\left(1+R_{ii}\right)\sum_{m=1}^{\mathcal{N}} \phi^{\textsc{dd}}(x,m)\phi^{\textsc{dd}}(x,m)+\frac{1}{2}\left(1-R_{ii}\right)n(x,t)\,,\label{Eq:nS0n}
\ee
where $R_{ii}=[\mathbb S]_{ii}$ is the reflection amplitude for a fermion on the $i$-th edge and $n(x,t)=\lim_{y\rightarrow x} C(x,y,t)$. We see that in the thermodynamic limit both \fr{Eq:CorrFunS0n} and \fr{Eq:nS0n} are completely determined once one knows the building blocks $C(x,y,t)$ and $n(x,t)$. Thus, the results of Sec.~\ref{Sec:twopoint} give respectively the exact density of particles $n_i(x,t)$ and the asymptotic expansion of the correlation function $C_{ij}(x,y,t)$ also when $n$ edges of noninteracting fermions are joined together to form a star graph, characterized by a scattering matrix $\mathbb S$.

\section{Conclusions}%
\label{Sec:conclusions}%

In this paper we studied the time evolution of correlations in a non relativistic quantum field theory, following a particular class of local quantum quenches: the sudden junction of two identical systems. At the joining, we allowed for the presence of a localized defect. We considered the example of noninteracting fermions in the bulk and found an analytical expression for the density of particles at any time after the quench. We also found an accurate asymptotic expansion of the equal-time two-point function for large times or large distances from the junction. 

We used these results to study the emergence of light cone effects in the absence of an inherent maximal velocity. In particular, we found that the leading part of the correlations at late times and large distances assume a CFT form, with the role of the velocity of light being played by the Fermi velocity $v_{\rm F}$ (\emph{cf}.~\fr{Eq:FV}). The light cone region scales as $t^{1/2}$, as opposed to the $t^{1/3}$ scaling found in lattice models for both  inhomogeneous quenches\cite{HRS,ER,VSDH:XX} and global quenches\cite{BEGR:long}. Correlations, however, also show ``superluminal corrections'' decaying in time as $t^{-1}$; to be compared to the exponentially decaying corrections occurring in the presence of Lieb-Robinson bounds. These anomalous corrections suggested naming the observed behaviour as ``approximate light cone effects".    We found approximate light cone effects also in the dynamics of the entanglement entropy between a subsystem away from the junction and the rest of the system. The behaviour is roughly described by the CFT result with $c=1$ and $v_{\rm F}$ playing the role of velocity of light, the corrections, however, are clearly visible.  

We explained these results by computing the distribution of excitations created by the quench and using the semiclassical quasi-particle picture of Ref.~[\onlinecite{CC}]. The quench creates a macroscopic number of particle-hole excitations very close to the Fermi surface and moving with velocity $v_{\rm F}$; these are responsible for the observed light cone effects in the correlation functions; it also creates a finite number of excitations with different velocities, leading to the visible corrections. This is due to the Fermi-sea-structure of the mode occupations in the initial state, which remains present in the final equilibrium state since the quench is local. It is close to this discontinuity that a macroscopic number of excitations can be produced with finite energy. The semiclassical interpretation allows us to go beyond the pure qualitative description. Using as input the distribution of excitations produced by the local quench, in Sec.~\ref{subsec:SC} we found the first non trivial contribution to the local density of particles -- for large times and distances from the junction and away from the light cone region -- only employing a simple semiclassical reasoning.

We began by studying the problem in the simplest possible setting -- an immediate question is: can we generalize our findings to more realistic situations, such as interacting systems in the presence of external trapping potentials? A first step in this direction has already been taken in Sec.~\ref{subsec:harmonic}, where we showed that the same qualitative picture emerges when one includes a harmonic trapping potential, the simplest way of modelling the traps used in real experiments. It should be also noted that the semiclassical picture adopted here has been shown to be extremely effective in the presence of integrable interactions~\cite{AC:entropy, BCDF:transport}, at least in the case of those solvable by thermodynamic Bethe ansatz. Therefore we expect the qualitative picture described here to hold also in this case. Discontinuities in the initial ``filling functions'' $\vartheta(\lambda)$\cite{Korepinbook} -- which generalise the occupation numbers to the interacting integrable case -- will correspond to the production of a macroscopic number of excitations and light cone effects in the correlations will be observable at the corresponding velocities. From the quantitative point of view, one might try to apply the argument of Sec.~\ref{subsec:SC} to find the leading contribution for the density of particles after the quench. To do that, however, a necessary ingredient is the exact expression for the distribution of excitations produced by the quench, which is non-trivial to find in interacting cases.

\acknowledgements
I am deeply indebted to Fabian Essler and Pasquale Calabrese for having suggested me the problem considered in this work and for many illuminating discussions. I kindly thank Lorenzo Piroli, Maurizio Fagotti, Neil Robinson, and Mihail Mintchev for extremely valuable comments on the manuscript and very useful suggestions. I thank Mihail Mintchev also for having introduced me to the model used in this work. This work has been supported by the ERC under Starting Grant 279391 EDEQS. Finally, I thank the Isaac Newton Institute for Mathematical Sciences, under grant EP/K032208/1, for hospitality during the earliest stages of this work. 

\appendix

\section{Mapping between mode operators}%
\label{app:modeopocc}

Both before and after the quench the field can be expressed in terms of a mode expansion \fr{Eq:mode}.  Introducing the index $0$ for the quantities before the quench we have 
\begin{align}
\label{Eq:modet>0}
&\psi_i(x,t)=\sum_{j=1}^{2}[\mathcal{U}]_{ij}\sum_{m=1}^{\infty}\phi^{j}(x,m)e^{-i \omega_{j}(m) t}a_j(m)\,, & &t>0\,,\\
\label{Eq:modet<0}
&\psi_i(x,t)=\sum_{j=1}^{2}[\mathcal{U}_0]_{ij}\sum_{m=1}^{\infty}\phi_0^{j}(x,m)e^{-i \omega_{0j}(m) t}a_{0j}(m)\,, & &t<0\,.
\end{align} 
In particular, from the continuity of the time evolution it follows  
\be
\label{Eq:modet0}
\sum_{j=1}^{2}[\mathcal{U}]_{ij}\sum_{m=1}^{\infty}\phi^{j}(x,m)a_{j}(m)=\sum_{j=1}^{2}[\mathcal{U}_0]_{ij}\sum_{m=1}^{\infty}\phi_0^{j}(x,m)a_{0j}(m) \,.
\ee
This implies that there is a simple linear (and unitary) relation between the mode operators before and after the quench, as generically true in non-interacting models. A peculiarity of this case is that there is no mixing between creation and annihilation operators, this implies that the number of particles remains unchanged during the quench and only particle-hole excitations are created. To find the explicit relation between $\{a_{0 j}(m)\}$ and $\{a_j(m)\}$ we consider  
\be
\sum_{j=1}^{2}\int_0^L{\rm d}x\,\,[\mathcal{U}^\dag]_{ij}\psi_j({x,0})\phi^i(x,n)^*\,,
\ee
and use the two different expressions of $\psi_j(x,0)$ given in \fr{Eq:modet0}. 
The result reads as
\be
\label{Eq:mappingapp}
a_i(n)=\sum_{k=1}^{2}\sum_{m=1}^{\infty} [\mathcal{U}^\dag \mathcal{U}_0]_{ik} B^{ik}_{nm}a_{0 k}(m)\,,
\ee
where we introduced the ``overlap-matrix'' given by  
\be
B^{ik}_{nm}=\int_0^L{\rm d}x\,\,\phi^i(x,n)^*\phi_0^k(x,m)=
\begin{cases}
\delta_{nm}\qquad &\textrm{if}\quad i\leq p_1\,,\quad k\leq p_0\\
B_{nm}\qquad &\textrm{if}\quad i> p_1\,,\quad k\leq p_0\\
B^*_{mn}\qquad &\textrm{if}\quad i\leq p_1\,,\quad k> p_0\\
\delta_{nm}\qquad &\textrm{if}\quad i> p_1\,,\quad k> p_0
\end{cases}\,.
\label{eq:overlapmatrixapp}
\ee
Here, the ``elementary overlap-matrix'' $B_{nm}$ reads as 
\be
\label{Eq:Bapp}
B_{nm}\equiv\int_0^L{\rm d}x\, \phi^{\textsc{nd}}(x,n)^* \phi^{\textsc{dd}}(x,m)=\frac{8}{\pi}\frac{m}{4m^2-(2n-{1})^2}\,.
\ee
The matrix $B_{nm}$ is an infinite orthogonal matrix, \emph{i.e.} it satisfies $\sum_{q=1}^\infty B_{nq}B_{mq}=\delta_{nm}$. From its definition \fr{Eq:Bapp} it follows that 
\bea
\sum_{m=1}^\infty B_{nm}^* \phi^{\textsc{dd}}(x,m)& = &  \phi^{\textsc{nd}}(x,n)\,, \\
\sum_{m=1}^\infty B_{mn} \phi^{\textsc{nd}}(x,m)& = &  \phi^{\textsc{dd}}(x,n)\,.
\label{eq:defB}
\eea

\section{Integral representation of $C(x,y,t)$ in the thermodynamic limit}%
\label{app:intrep}
Here we show that Eq.~\fr{Eq:intrep} gives an integral representation of the two point function in the thermodynamic limit. We start by noting that the expression \fr{Eq:Cn1form1} can be re-written as 
\be
\label{Eq:Cn1}
C(x,y,t)=\frac{2}{L}\sum_{p=1}^{\mathcal{N}}M_p(x,\tau)M_p(y,\tau)^*\,,
\ee
where we introduce
\be
\label{Eq:Mp}
M_p(x,\tau)=\sum_{n=1}^{\infty}B_{np}\cos\Bigl((n-\frac{1}{2})\frac{\pi x}{L}\Bigr)e^{i\tau[((n-\frac{1}{2})\frac{\pi }{L})^2-{\tilde p}^2]}\,,\qquad\qquad \tau\equiv \frac{t}{2m_0}\,, \qquad\qquad  \tilde p\equiv \frac{\pi p}{L}\,. 
\ee
The first step to construct an integral representation for \fr{Eq:Cn1} is to evaluate the sum \fr{Eq:Mp}. This sum can not be straightforwardly turned into an integral in the thermodynamic limit using the Euler-Maclaurin formula\cite{Wongbook}; this is because $B_{np}$ is not well defined for continuous variables (it features a singularity for $n=p-1/2$). A convenient trick is to take the derivative of \fr{Eq:Mp} with respect to $\tau$, this cancels the singularity and Euler-Maclaurin formula can be straightforwardly applied  
\begin{align}
\partial_\tau M_p(x,\tau)&=-\frac{2 i \tilde p}{L}\sum_{n=1}^{\infty}\cos\Bigl((n-\frac{1}{2})\frac{\pi x}{L}\Bigr)e^{i\tau[((n-\frac{1}{2})\frac{\pi }{L})^2-{\tilde p}^2]}=-2i \tilde p e^{-i\frac{x^2}{4\tau}-i\tau{\tilde p}^2}\int_{0}^{\infty}\!\frac{{\rm d}\eta}{\pi}\, e^{i\tau(\eta+i\epsilon^+)^2}+O\left(\frac{1}{L}\right)\,,\notag\\
&=\frac{\tilde p}{\sqrt{\pi \tau}} e^{-i\frac{\pi}{4}} e^{-i\frac{x^2}{4\tau}-i\tau{\tilde p}^2}+O\left(\frac{1}{L}\right)\,,\qquad \qquad \tau>0\,.\label{Eq:Mpprime}
\end{align}
Using the definition of $B_{np}$ (\emph{cf}. Eq.~\fr{eq:defB}) we can determine the value of \fr{Eq:Mp} for $\tau=0$
\be
M_p(x,0)=\sum_{n=1}^{\infty}B_{np}\cos\Bigl((n-\frac{1}{2})\frac{\pi x}{L}\Bigr)=\sin\left(\tilde p x\right)\,,
\label{Eq:Mpinitial}
\ee
Integrating \fr{Eq:Mpprime} with the initial condition \fr{Eq:Mpinitial} we find 
\begin{align}
\label{Eq:tildeMpIRap}
M_p(x,\tau)&=\sin\left(\tilde p x\right)+2 \tilde p e^{-i\frac{\pi}{4}}\sqrt{\frac{\tau}{\pi}} \int_{0}^{1} \!\!\!{\rm d}s\, e^{-i\tau \frac{x^2}{4\tau^2 s^2}-i\tau s^2 \tilde p^2}+O\left(\frac{1}{L}\right)\notag\\
&=\frac{1}{2i}\left(e^{i \tilde p x}\textrm{erf}\left(e^{i\frac{\pi}{4}}(m_0\xt+\tilde p)\sqrt{\tau}\right)-e^{-i \tilde p x}\textrm{erf}\left(e^{i\frac{\pi}{4}}(m_0\xt-\tilde p)\sqrt{\tau}\right) \right)+O\left(\frac{1}{L}\right)\,.
\end{align}
The expression \fr{Eq:tildeMpIRap} is a smooth function of $\tilde p\in \mathbb R$. Using this representation of $M_p(x,\tau)$, the sum \fr{Eq:Cn1} can be straightforwardly turned into an integral leading to the desired result \fr{Eq:intrep}.

\section{Asymptotic expansion of the correlation function}%
\label{app:asympt}
In this appendix we asymptotically expand the correlation function \fr{Eq:intrep} in the regions $|\xt-v_{\rm F}|\gg (m_0 t)^{-1/2}$ and $|\tilde\xt-v_{\rm F}|\gg (m_0 t)^{-1/2}$, \emph{i.e.} far enough from the light-cone. To this aim it is convenient to rewrite the correlation function as follows
\begin{align}
C(x,y,t) &=\frac{2}{\pi}\int_{0}^{\pi n}\!\!\!\!{\rm d}p\,\sin(p x)\sin(p y)\label{Eq:simple}\\
&+\frac{4e^{-i\frac{\pi}{4}}\sqrt{\tau}}{\pi^{3/2}}\iint\limits_{[0,1]\times[0,\pi n]}\!\!\!\!\!\!{\rm d}s{\rm d}p\,p\sin({p y}) e^{-i\tau[ \xt^2 s^{-2}+s^2{p}^2]}\label{Eq:double}\\
&+\frac{4e^{i\frac{\pi}{4}}\sqrt{\tau}}{\pi^{3/2}}\iint\limits_{[0,1]\times[0,\pi n]}\!\!\!\!\!\!{\rm d}s{\rm d}p\,p\sin({p x}) e^{i\tau[\tilde \xt^2 s^{-2}+s^2{p}^2]}\label{Eq:double2}\\
&+\frac{8\tau}{\pi^2}\iiint\limits_{[0,1]^2\times[0,\pi n]} \!\!\!\!\!\!{\rm d}s_1{\rm d}s_2 {\rm d}p\,p^2e^{-i\tau[\xt^2 s_1^{-2}+ s_1^2p^2]}e^{i\tau[\tilde \xt^2 s_2^{-2}+ s_2^2p^2]}\,,\label{Eq:triple}
\end{align}
Here for convenience we set $\tau ={t}/{2m_0}$ and $m_0=1$, the mass dependence can be easily restored by the replacement $\xt\rightarrow m_0\xt$ (analogously for $\tilde \xt$). We construct a separate asymptotic expansion for each of the terms \fr{Eq:double}\,--\,\fr{Eq:triple}. We start from \fr{Eq:double} and \fr{Eq:double2} which feature a double integral.

\subsubsection{Double integrals $\tilde\xt\neq\xt$}
Let us consider $\tilde \xt\neq \xt$
\begin{align}
I_1(\xt,\tilde \xt,\tau)\equiv\iint\limits_{[0,1]\times[0,\pi n]}\!\!\!\!\!\!{\rm d}s{\rm d}p\,p\sin({p y}) e^{-i\tau[ \xt^2 s^{-2}+s^2{p}^2]}&=\frac{1}{2i}\iint\limits_{[0,1]\times[-\pi n,\pi n]}\!\!\!\!\!\!{\rm d}s{\rm d}p\,p e^{-i\tau (s^2 p^2+\xt^2 s^{-2}-2\tilde\xt p)}\,.
\end{align}
We consider the case $0<\xt<\tilde \xt<\pi n$ there are no stationary points in the bulk of the integration domain, we then consider those on the boundary. They are given by
\begin{align}
I_1(\xt,\tilde \xt,\tau)=&\frac{1}{2\tau}\Biggl\{\int_0^1{\rm d}s\,\frac{2(\pi n s^2 -\tilde\xt)\pi n e^{-i\tau(s^2 (\pi n)^2+\xt^2 s^{-2}-2\tilde\xt \pi n)}}{4(\pi n s^2 -\tilde\xt)^2+4 s^2 ((\pi n)^2-\xt^2s^{-4})^2}\notag\\
&-\int_0^1{\rm d}s\,\frac{2(\pi n s^2 +\tilde\xt)\pi n e^{-i\tau(s^2 (\pi n)^2+\xt^2 s^{-2}+2\tilde\xt \pi n)}}{4(\pi n s^2 +\tilde\xt)^2+4 s^2 ((\pi n)^2-\xt^2s^{-4})^2}\notag\\
&+\int_{-\pi n}^{\pi n} \!{\rm d}p\, \frac{2(p^2 -\xt^2)p e^{-i\tau(p^2+\xt^2 -2\tilde\xt p)}}{4(p -\tilde\xt)^2+4 (p^2-\xt^2)^2}\Biggr\}\,+o\left(\frac{1}{\tau^{M}}\right)\,, \qquad M\in \mathbb{N}.
\label{Eq:I1boundarykt}
\end{align}
There are three stationary points of the function reduced on the boundary 
\begin{itemize}
\item[(i)]  $s=\sqrt{\frac{\xt}{\pi n}}$ in the first integral. 
\item[(ii)] $s=\sqrt{\frac{\xt}{\pi n}}$ in the second integral. 
\item[(iii)] $p=\tilde\xt$ in the third integral. 
\end{itemize}
The leading contribution given by those points are
\be
I_1(\xt,\tilde \xt,\tau)\Big|_{\textrm{(i)+(ii)+(iii)}}= \frac{\sqrt{\pi}e^{-i\frac{\pi}{4}}e^{2i\tau \pi n (\tilde\xt-\xt)}}{8 \tau^{3/2}(\xt-\tilde\xt)}-\frac{\sqrt{\pi}e^{-i\frac{\pi}{4}}e^{-2i\tau \pi n (\tilde\xt+\xt)}}{8 \tau^{3/2}(\xt+\tilde\xt)}+\frac{\sqrt{\pi}\tilde\xt e^{-i\frac{\pi}{4}}e^{i\tau(\tilde\xt^2-\xt^2)}}{4\tau^{3/2}(\tilde\xt^2-\xt^2)}+O(\tau^{-5/2})\,.
\ee
The next contributions in the expansion are found integrating by parts the one dimensional integrals, they are given by
\begin{align}
\label{Eq:boundarykt}
I_1(\xt,\tau)\Big|_{\textrm{1d-boundaries}}=\frac{i \pi n}{8\tau^2}\left[\frac{(\pi n +\tilde \xt)e^{-i\tau((\pi n)^2+\xt^2-2\tilde\xt \pi n)}-(\pi n -\tilde \xt)e^{-i\tau((\pi n)^2+\xt^2+2\tilde\xt \pi n)}}{((\pi n)^2-\tilde\xt^2)((\pi n)^2-\xt^2)}\right]
\end{align}
for other configurations of $\tilde \xt$ and $\xt$ some of the stationary points are outside of the integration domain. The general result reads
\be
\label{Eq:I1finalkt}
I_1(\xt, \tilde \xt,\tau)=a_{3/2}(\xt, \tilde \xt,\tau)\tau^{-3/2}+a_{2}(\xt, \tilde \xt,\tau)\tau^{-2} + O(\tau^{-5/2}).
\ee
Here we defined
\begin{align}
a_{3/2}(\xt, \tilde \xt,\tau)=&\left(\frac{\sqrt{\pi}e^{-i\frac{\pi}{4}}e^{2i\tau \pi n (\tilde\xt-\xt)}}{8 (\xt-\tilde\xt)}-\frac{\sqrt{\pi}e^{-i\frac{\pi}{4}}e^{-2i\tau \pi n (\tilde\xt+\xt)}}{8 (\xt+\tilde\xt)}\right)\theta_{\textsc{h}}(\pi n - \xt)\notag\\
&+\frac{\sqrt{\pi}\tilde\xt e^{-i\frac{\pi}{4}}e^{i\tau(\tilde\xt^2-\xt^2)}}{4(\tilde\xt^2-\xt^2)}\theta_{\textsc{h}}(\pi n -\tilde \xt)\,,\\
a_2(\xt, \tilde \xt,\tau)=&\frac{i \pi n}{8}\left[\frac{(\pi n +\tilde \xt)e^{-i\tau((\pi n)^2+\xt^2-2\tilde\xt \pi n)}-(\pi n -\tilde \xt)e^{-i\tau((\pi n)^2+\xt^2+2\tilde\xt \pi n)}}{((\pi n)^2-\tilde\xt^2)((\pi n)^2-\xt^2)}\right]\,.
\end{align}

\subsubsection{Triple integral for $\tilde k \neq k$}
Let us now consider the triple integral \fr{Eq:triple} for $\tilde \xt>\xt$
\be
I_2(\xt,\tilde \xt,\tau)\equiv\iiint\limits_{[0,1]^2\times[0,\pi n]} \!\!\!\!\!\!{\rm d}s_1{\rm d}s_2 {\rm d}p\,p^2e^{-i\tau[\xt^2 s_1^{-2}+ s_1^2p^2]}e^{i\tau[\tilde \xt^2 s_2^{-2}+ s_2^2p^2]}\,.
\ee
We consider separately the three cases 
\begin{itemize}
\item[(a)] $\tilde \xt>\xt>\pi n$. 
\item[(b)]  $\tilde \xt>\pi n>\xt$.
\item[(c)] $\pi n> \tilde \xt>\xt$. 
\end{itemize}
Let us start from the configuration (c). For $\tilde \xt \neq \xt$ there are no stationary points in the three dimensional domain, the leading contribution is thus found on the boundary 
\begin{align}
I_2(\xt,\tilde \xt,\tau)=&\frac{i}{\tau}\Biggl\{ \,\,\iint\limits_{[0,1]^2}{\rm d}s_1{\rm d}s_2\,\frac{2(\pi n)^3 (s_1^2 -s_2^2) e^{-i\tau\left[(\pi n)^2 (s_1^2-s_2^2) + (\xt^2 s_1^{-2}- \tilde\xt^2s_2^{-2})\right]}}{4(\pi n)^2 (s_1^2 -s_2^2)^2+4((\pi n)^2 s_1- \xt^2s_1^{-3})^2+4((\pi n)^2 s_2- \tilde\xt^2s_2^{-3})^2}\notag\\
&-\iint\limits_{[0,1]\times[0,\pi n]} \!\!\!{\rm d}s{\rm d}p\,\frac{(p^2 -\tilde\xt^2)p^2 e^{-i \tau \left[(p^2 (s^2-1) + ( \xt^2s^{-2}-\tilde\xt^2)\right]}}{2((p^2 -\tilde\xt^2)^2+ p^2(s^2-1)^2+(p^2 s- \xt^2s^{-3})^2)}\notag\\
&+\iint\limits_{[0,1]\times[0,\pi n]} \!\!\!{\rm d}s{\rm d}p\,\frac{(p^2 -\xt^2)p^2 e^{i\tau \left[p^2 (s^2-1) + ( \tilde\xt^2s^{-2}-\xt^2)\right]}}{2((p^2 -\xt^2)^2+ p^2(s^2-1)^2+(p^2 s-\tilde \xt^2s^{-3})^2)}\Biggr\}+o\left(\frac{1}{\tau^M}\right)\,.\qquad M\in \mathbb{N}\,.
\end{align}
The three integrals have stationary points in $(s_1=\sqrt{\xt/{\pi n}}, s_2=\sqrt{\tilde\xt/{\pi n}})$, $(p=\xt,s=1)$ and $(p=\tilde\xt,s=1)$ respectively. Their leading contributions read as 
\be
I_2(\xt,\tilde \xt,\tau)_{\textrm{boundarystationary}}=\frac{i\pi  e^{2 i \tau \pi n (\tilde\xt-\xt)}}{8 \tau^2 ( \xt -\tilde \xt)}+\frac{i \pi \xt e^{i \tau (\tilde \xt^2-\xt^2)}}{8 \tau^2 ( \tilde\xt^2 -  \xt^2)}+\frac{i \pi \tilde \xt e^{i \tau (\tilde \xt^2-\xt^2)}}{8 \tau^2 (\tilde \xt^2 -  \xt^2)}+O(\tau^{-3})
\ee
Finally, there are contributions coming from stationary points at the boundary of the two-dimensional integrals. There are five of them 
\begin{itemize}
\item[(i)] At the boundary $s_1=1$, there is a stationary point in $s_2=\sqrt{\tilde\xt/\pi n}$. \item[(ii)] At the boundary $s_2=1$, there is a stationary point in $s_1=\sqrt{\xt/\pi n}$. 
\item[(iii)] At the boundary $p=\pi n$, there is a stationary point at $s=\sqrt{\xt/\pi n}$.
\item[(iv)] At the boundary $p=\pi n$, there is a stationary point at $s=\sqrt{\tilde\xt/\pi n}$.
\item[(v)] The boundary $s=1$ is a continuous line of stationary points. 
\end{itemize}
Considering the contribution of the first four of them we obtain 
\begin{align}
I_2(\xt,\tilde \xt,\tau)\Bigr|_{\textrm{(i)}+\textrm{(ii)}+\textrm{(iii)}+\textrm{(iv)}}\sim-\frac{\pi^{3/2}ne^{i\tau (\tilde \xt^2 -\xt^2)}}{8 \tau^{5/2}}\left[\frac{e^{-i\tau (\pi n - \tilde \xt)^2}e^{i \frac{\pi}{4}}(n\pi+\tilde\xt)+e^{i\tau (\pi n - \xt)^2}e^{-i \frac{\pi}{4}}(n\pi+\xt)}{((\pi n)^2 - \xt^2)((\pi n)^2 - \tilde\xt^2)}\right]\,.
\end{align}
Considering the contribution of (v) we have  
\begin{align}
\label{Eq:(v)kt}
I_2(\xt,\tilde \xt,\tau)\Bigr|_{\textrm{(v)}}\!\!\!\sim& -\frac{e^{i \tau (\tilde \xt^2-\xt^2)}}{4\tau^2}\int_0^{\pi n} \!\!{\rm d}p\,\biggl\{ \frac{p^2}{((p^2 -\tilde\xt^2)^2+(p^2- \xt^2)^2)}\left[\frac{(p^2 -\tilde\xt^2)}{(p+\xt)}\textrm{PV}\frac{1}{(p-\xt)}+\frac{(p^2 -\xt^2)}{(p+\tilde\xt)}\textrm{PV}\frac{1}{(p-\tilde\xt )}\right]\biggr\}\notag\\
&=\frac{e^{i \tau (\tilde \xt^2-\xt^2)}}{8\tau^2}\biggl\{\int\limits_{0}^{\infty}\!\!{\rm d}p\,\frac{2i(p-i\pi n)^2}{((p-i\pi n)^2+\xt^2)((p-i\pi n)^2+\tilde\xt^2)}-\int\limits_{0}^{\infty}\!\!{\rm d}p\,\frac{2ip^2}{(p^2+\xt^2)(p^2+\tilde\xt^2)}+\frac{ i\pi}{ (\xt+\tilde\xt)}\biggl\}\,.
\end{align}
where we used the residues theorem and deformed the contours according to the Cauchy's theorem. Performing the elementary integrals we obtain 
\be
\label{Eq:(v)resultkt}
I_2(\xt,\tilde \xt,\tau)\Bigr|_{\textrm{(v)}}\sim \frac{e^{i \tau (\tilde \xt^2-\xt^2)}}{8\tau^2(\tilde\xt^2 - \xt^2)} \left( \tilde \xt \log\left|\frac{\pi n -\tilde\xt}{\pi n + \tilde\xt}\right| -\xt \log\left|\frac{\pi n - \xt}{\pi n + \xt}\right|\right)+O(\tau^{-3})\,.
\ee
The cases (a) and (b) can be easily recovered by taking into account only the stationary points remaining within the integration domain. The result reads as 
\be
\label{Eq:I2finalkt}
I_2(\xt,\tilde \xt,\tau)=b_{2}(\xt,\tilde \xt,\tau)\tau^{-2}+b_{5/2}(\xt,\tilde \xt,\tau)\tau^{-5/2}+O(\tau^{-3})\,.
\ee
Here we defined
\begin{align}
b_{2}(\xt,\tilde\xt,\tau)&= \frac{e^{i \tau (\tilde \xt^2-\xt^2)}}{8(\tilde\xt^2 - \xt^2)} \left( \tilde \xt \log\left|\frac{\pi n -\tilde\xt}{\pi n + \tilde\xt}\right| -\xt \log\left|\frac{\pi n - \xt}{\pi n + \xt}\right|\right)+ \frac{i\pi  e^{2 i \tau \pi n (\tilde\xt-\xt)}}{8  ( \xt -\tilde \xt)}\theta_{\textsc{h}}(\pi n -\xt)\theta_{\textsc{h}}(\pi n -\tilde\xt)\notag\\
&+\frac{i \pi \xt e^{i \tau (\tilde \xt^2-\xt^2)}}{8( \tilde\xt^2 - \xt^2)}\theta_{\textsc{h}}(\pi n -\xt)+\frac{i \pi \tilde \xt e^{i \tau (\tilde \xt^2-\xt^2)}}{8( \tilde\xt^2 -  \xt^2)}\theta_{\textsc{h}}(\pi n -\tilde\xt)\,,\\
b_{5/2}(\xt,\tilde\xt,\tau) &=-\frac{\pi^{3/2}ne^{i\tau (\tilde \xt^2 -\xt^2)}}{8}\left[\frac{e^{-i\tau (\pi n - \tilde \xt)^2}e^{i \frac{\pi}{4}}(n\pi+\tilde\xt)\theta_{\textsc{h}}(\pi n -\tilde\xt)+e^{i\tau (\pi n - \xt)^2}e^{-i \frac{\pi}{4}}(n\pi+\xt)\theta_{\textsc{h}}(\pi n -\xt)}{((\pi n)^2 - \xt^2)((\pi n)^2 - \tilde\xt^2)}\right]\,.
\end{align}

\section{Distribution of excitations in the thermodynamic limit}%
\label{app:changeofvariables}

In Section~\ref{Sec:excitations} we determined the number of excitations of mode $m$ on the edge $i$, it is given by $\Delta N_{\rm ex}(p_m;i)= \tilde\delta_{\rm ex}(p_m;i) \Delta p_m$, where $\Delta p_m = \pi/L$. Here we determine the distribution corresponding to $\tilde \delta_{\rm ex}(p_m;i)$ in the thermodynamic limit, namely a distribution $\delta_{\rm ex}(p;i)$ such that 
\be
\label{Eq:}
\lim_{L\rightarrow\infty}\sum_{m=1}^\infty f(p_m)\tilde\delta_{\rm ex}(p_m;i) \Delta p_m = \int_0^\infty\!\! f(p)\delta_{\rm ex}(p;i) {\rm d}p\,,
\ee
where $f(p)$ is an arbitrary smooth function. Let us start considering the terms in \fr{Eq:discretedist} involving step functions
\be
\sum_{m=1}^\infty\left(\theta_{\textsc{h}}(\pi n- p_m)-\theta_{\textsc{h}}(\pi n- p_m-\frac{\pi}{2L}) \right)f(p_m)=L \int^0_{-{\pi}/{2L}} \frac{{\rm d}p}{\pi} f(\pi n + p)+O\left(\frac{1}{L}\right)=\frac{1}{2}f(\pi n)+O\left(\frac{1}{L}\right)\,.
\ee
In the first step we approximated $\theta_{\textsc{h}}(x)$ with smooth functions and employed Euler-Maclaurin summation formula. Using the residues theorem we have  
\begin{align}
|[\mathcal U]_{i2}|^{-2}\sum_{m=1}^\infty  f(p_m)\tilde\delta_{\rm ex}(p_m;i) \Delta p_m =& \frac{1}{2}f\left(\pi n\right)+\frac{i}{\pi}f(\pi n)+\int\limits_{\mathcal{C}_1}\!\!\frac{{\rm d}z}{\pi}\,  \left\{\frac{2}{z \pi}\arctan\left(\frac{z}{\pi}\right)-\frac{2n}{(\pi n)^2-z^2}\right\}f_{\rm cf}(z)f(z)\notag\\
&+\int\limits_{\mathcal{C}_2}\!\!\frac{{\rm d}z}{\pi}\,  \left\{\frac{2}{z \pi}\arctan\left(\frac{\pi}{z}\right)-\frac{2n}{(\pi n)^2-z^2}\right\}f_{\rm cf}(z)f(z)+O\left(\frac{1}{L}\right)\,,
\end{align}
where $\mathcal{C}_1$ and $\mathcal{C}_2$ encircle the intervals $[0,\pi n[$ and $[\pi n,\infty[$ respectively and we introduced
\be
f_{\rm cf}(z)=\frac{1}{1+e^{2 i L z}}\,.
\ee
In the thermodynamic limit only the branches of the paths $\mathcal{C}_1$ and $\mathcal{C}_2$ lying on the upper half plane contribute and we have 
\begin{align}
\hspace{-0.15cm}\lim_{L\rightarrow\infty}\frac{\pi }{L |[\mathcal U]_{i2}|^{2}}\sum_{m=1}^\infty f(p_m)\tilde\delta_{\rm ex}(p_m; i) &= \frac{1}{2}f(\pi n)+\frac{i}{\pi}f(\pi n)+\int_{0}^{\infty}\!\!\!\!{{\rm d}p}  \left\{\frac{f(p)}{p \pi^2}\log\left|\frac{\pi n - p}{\pi n + p}\right|-\frac{2nf(p)}{\pi (\pi n + p)(\pi n-p-i\epsilon)}\right\}\notag\\
&=\int_{0}^{\infty}\!\!\!\!{{\rm d}p}\left\{\frac{1}{p \pi^2}\log\left|\frac{\pi n - p}{\pi n + p}\right|-\frac{2n}{\pi (\pi n + p)}\textrm{PV}\left(\frac{1}{\pi n-p}\right)+\frac{1}{2}\delta(\pi n -p)\right\}f(p)\,,
\end{align}  
where we used 
\be
\lim_{\epsilon\rightarrow 0^+}\frac{1}{x-i \epsilon}= \textrm{PV}\frac{1}{x}+i\pi\delta(x)\,.
\ee
From the arbitrariness of $f(p)$ we conclude
\be
|[\mathcal U]_{i2}|^{-2}\delta_{\rm ex}(p; i)=\frac{1}{p \pi^2}\log\left|\frac{\pi n - p}{\pi n + p}\right|-\frac{2n}{\pi (\pi n + p)}\textrm{PV}\left(\frac{1}{\pi n-p}\right)+\frac{1}{2}\delta(\pi n -p)\,.
\ee

\section{Explicit matrix elements of $\mathbb{A}_\ell(t)$}%
\label{app:matrixA}
Here we report the explicit matrix elements of $\mathbb{A}_\ell(t)$, which we used to numerically construct the matrix 
\begin{align}
&[\mathbb{A}_{\ell}(t)]_{nm}= \frac{\varepsilon^2}{\varepsilon^2+1}\left( \delta_{nm}-\sum_{k,p=1}^{\infty}e^{i(\omega_{\textsc{nd}}(k)-\omega_{\textsc{nd}}(p))t} B_{pn} B_{km}\left\{\frac{\sin(\frac{\pi \ell}{L}(p-k))}{\pi (p-k)}+\frac{\sin(\frac{\pi \ell}{L}(p+k-1))}{\pi (p+k-1)}\right\}\right) && 1\leq n,m\leq \mathcal{N}\notag\\
&[\mathbb{A}_{\ell}(t)]_{nm}= \frac{\varepsilon}{\varepsilon^2+1} \sum_{p=1}^{\infty}e^{-i\omega_{\textsc{nd}}(p) t} B_{pn}\left\{\frac{\cos(\frac{\pi \ell}{L}(m-\mathcal{N}-p+\frac{1}{2}))}{\pi (m-\mathcal{N}-p+\frac{1}{2})}+\frac{\cos(\frac{\pi \ell}{L}(m -\mathcal{N}+p-\frac{1}{2}))}{\pi (m-\mathcal{N}+p-\frac{1}{2})}\right\} && 1\leq n, m-\mathcal{N}\leq \mathcal{N} \notag\\
&[\mathbb{A}_{\ell}(t)]_{nm}= \frac{\varepsilon}{\varepsilon^2+1} \sum_{p=1}^{\infty}e^{i\omega_{\textsc{nd}}(p) t} B_{pm}\left\{\frac{\cos(\frac{\pi \ell}{L}(n-\mathcal{N}-p+\frac{1}{2}))}{\pi (n-\mathcal{N}-p+\frac{1}{2})}+\frac{\cos(\frac{\pi \ell}{L}(n -\mathcal{N}+p-\frac{1}{2}))}{\pi (n-\mathcal{N}+p-\frac{1}{2})}\right\} && 1\leq m, n-\mathcal{N}\leq \mathcal{N} \notag\\
&[\mathbb{A}_{\ell}(t)]_{nm}=\frac{1}{\varepsilon^2+1}\left( \delta_{nm}-\left\{\frac{\sin(\frac{\pi \ell}{L}(n-m))}{\pi (n-m)}-\frac{\sin(\frac{\pi \ell}{L}(n+m-2\mathcal{N}))}{\pi (n+m-2\mathcal{N})}\right\}\right) && \mathcal{N}+1\leq n, m\leq 2 \mathcal{N}\,. \notag
\end{align}
These matrix elements are obtained by plugging the explicit form of $\chi(x,t,n)$ (\emph{cf}.~\fr{Eq:chi}) into \fr{Eq:matelA}.


\begin{thebibliography}{99}

\bibitem{exp} M. Greiner, O. Mandel, T.W. H\"ansch, and I. Bloch, 
\href{\doi10.1038/nature00968}{Nature {\bf 419}, 51-54 (2002)}; T. Kinoshita, T. Wenger,  and D. S. Weiss, 
\href{\doi10.1038/nature04693}{Nature {\bf 440}, 900 (2006)};
 S. Hofferberth, I. Lesanovsky, B. Fischer, T. Schumm, and J. Schmiedmayer, 
\href{\doi10.1038/nature06149}{Nature {\bf 449}, 324-327 (2007)}; 
L. Hackermuller, U. Schneider, M. Moreno-Cardoner, T. Kitagawa, S. Will, T. Best, E. Demler, E. Altman, I. Bloch and B. Paredes,
\href{\doi10.1126/science.1184565}{Science {\bf 327}, 1621 (2010)};
S. Trotzky, Y.-A. Chen, A. Flesch, I. P. McCulloch, U. Schollw\"ock, J. Eisert, and I. Bloch, 
\href{\doi10.1038/nphys2232}{Nature Phys. {\bf 8}, 325 (2012)}; M. Gring, M. Kuhnert, T. Langen, T. Kitagawa, B. Rauer, M. Schreitl, I. Mazets, D. A. Smith, E. Demler, and J. Schmiedmayer,
\href{\doi10.1126/science.1224953}{Science {\bf 337}, 1318 (2012)};
U. Schneider, L. Hackerm\"uller, J. P. Ronzheimer, S. Will, S. Braun, T. Best, I. Bloch, E. Demler, S. Mandt, D. Rasch, and A. Rosch, 
\href{\doi10.1038/nphys2205}{Nature Phys. {\bf 8}, 213 (2012)}; 
F. Meinert, M.J. Mark, E. Kirilov, K. Lauber, P. Weinmann, A.J. Daley, and H.-C. N\"agerl, 
\href{http://dx.doi.org/10.1103/PhysRevLett.111.053003}{Phys. Rev. Lett. {\bf 111}, 053003 (2013)}; T. Fukuhara, A. Kantian, M. Endres, M. Cheneau, P. Schau{\ss},	 S. Hild,	 D. Bellem,	 U. Schollw\"ock, T. Giamarchi, C. Gross, I. Bloch, and S. Kuhr,  
\href{\doi10.1038/nphys2561}{Nature Physics {\bf 9}, 235 (2013)}; T. Fukuhara, P. Schau{\ss}, M. Endres, S. Hild, M. Cheneau, I. Bloch, and C. Gross, 
\href{\doi10.1038/nature12541}{Nature {\bf 502}, 76 (2013)};
J.P. Ronzheimer, M. Schreiber, S. Braun, S.S. Hodgman, S. Langer, I.P. McCulloch, F. Heidrich-Meisner, I. Bloch and U. Schneider,
\href{http://dx.doi.org/10.1103/PhysRevLett.110.205301}{Phys. Rev. Lett. {\bf 110}, 205301 (2013)}.


\bibitem{explc} M. Cheneau, P. Barmettler, D. Poletti, M. Endres, P. Schauss, T. Fukuhara, C. Gross, I. Bloch, C. Kollath, and S. Kuhr, 
\href{http://dx.doi.org/10.1038/nature10748}{Nature {\bf 481}, 484 (2012)}; T. Langen,  R. Geiger,  M. Kuhnert,  B. Rauer,  and J. Schmiedmayer,
\href{http://dx.doi.org/10.1038/nphys2739}{Nature Physics {\bf 9}, 640 (2013)}; P. Jurcevic, B. P. Lanyon, P. Hauke, C. Hempel, P. Zoller, R. Blatt, and C. F. Roos,
\href{http://dx.doi.org/10.1038/nature13461}{Nature {\bf 511}, 202 (2014)}; P. Richerme, Z.-X. Gong, A. Lee, C. Senko, J. Smith, M. Moss-Feig, S. Michalakis, A. V. Gorshkov, and C. Monroe, 
\href{http://dx.doi.org/10.1038/nature13450}{Nature {\bf 511}, 198 (2014)}.

\bibitem{EF:review}
F.~H.~L. Essler and M. Fagotti, 
\href{http://iopscience.iop.org/article/10.1088/1742-5468/2016/06/064002}{J. Stat. Mech. (2016) 064002}.


\bibitem{CC:review}
P. Calabrese and J. Cardy, 
\href{http://iopscience.iop.org/article/10.1088/1742-5468/2016/06/064003}{J. Stat. Mech. (2016) 064003}.

\bibitem{CaCh16}
M.~A. Cazalilla and M.-C. Chung, 
\href{http://iopscience.iop.org/article/10.1088/1742-5468/2016/06/064004}{J. Stat. Mech. (2016) 064004}.


\bibitem{BD:review} 
D.~Bernard and B.~Doyon, 
\href{http://dx.doi.org/10.1088/1742-5468/2016/06/064005}{J. Stat. Mech. (2016) 064005}.

\bibitem{C16}
J.-S. Caux, 
\href{http://iopscience.iop.org/article/10.1088/1742-5468/2016/06/064006}{J. Stat. Mech. (2016) 064006}.

\bibitem{VR:review}
L.~Vidmar and M.~Rigol, 
\href{http://dx.doi.org/10.1088/1742-5468/2016/06/064007}{J. Stat. Mech. (2016) 064007}.


\bibitem{ProsenReview}
 E.~Ilievski, M.~Medenjak, T.~Prosen, and L.~Zadnik,
\href{http://iopscience.iop.org/article/10.1088/1742-5468/2016/06/064008}{J. Stat. Mech. (2016) 064008}.


\bibitem{LangenReview}
 T.~Langen, T.~Gasenzer and J.~Schmiedmayer
\href{http://iopscience.iop.org/article/10.1088/1742-5468/2016/06/064009}{J. Stat. Mech. (2016) 064009}.



\bibitem{VM:review} 
R.~Vasseur and J.~E.~Moore, 
\href{http://dx.doi.org/10.1088/1742-5468/2016/06/064010}{J. Stat. Mech. (2016) 064010}. 





\bibitem{DeLucaArxiv16}
A.~De~Luca and G.~Mussardo,
\href{http://iopscience.iop.org/article/10.1088/1742-5468/2016/06/064011}{J. Stat. Mech. (2016) 0640011}.






\bibitem{RigolPRL07}
M.~Rigol, V.~Dunjko, V.~Yurovsky, and M.~Olshanii, 
\href{http://dx.doi.org/10.1103/PhysRevLett.98.050405} {Phys. Rev. Lett. {\bf 98}, 050405 (2007)}.


\bibitem{D91}
J. M. Deutsch, 
\href{http://dx.doi.org/10.1103/PhysRevA.43.2046}{Phys. Rev. A {\bf 43}, 2046 (1991)}.

\bibitem{S94}
M. Srednicki, 
 \href{http://dx.doi.org/10.1103/PhysRevE.50.888}{Phys. Rev. E {\bf 50}, 888 (1994)}.


\bibitem{R08}  
M. Rigol, V. Dunjko, and M. Olshanii, 
\href{\doi10.1038/nature06838}{Nature {\bf 452}, 854 (2008)}.

\bibitem{RS12}
M. Rigol and M. Srednicki, 
 \href{http://dx.doi.org/10.1103/PhysRevLett.100.100601}{Phys. Rev. Lett. {\bf 108}, 110601 (2012)}.


\bibitem{RigolPRA06}
M.~Rigol, A.~Muramatsu, and M.~Olshanii,
\href{http://dx.doi.org/10.1103/PhysRevA.74.053616} {Phys. Rev. A {\bf 74}, 053616 (2006)}.

\bibitem{CazalillaPRL06}
M.~A.~Cazalilla, 
\href{http://dx.doi.org/10.1103/PhysRevLett.97.156403} {Phys. Rev. Lett. {\bf 97}, 156403 (2006)}.

\bibitem{CalabreseJStatMech07}
P.~Calabrese and  J.~Cardy,  
\href{http://dx.doi.org/10.1088/1742-5468/2007/06/P06008}{J. Stat. Mech. (2007) P06008}.

\bibitem{CramerPRL08}
M.~Cramer, C.~M.~Dawson, J.~Eisert, and T.~J.~Osborne, 
\href{http://dx.doi.org/10.1103/PhysRevLett.100.030602}{Phys. Rev. Lett. {\bf 100}, 030602 (2008)}.

\bibitem{BarthelPRL08}
T.~Barthel and U.~Schollw\"ock, 
\href{http://dx.doi.org/10.1103/PhysRevLett.100.100601}{Phys. Rev. Lett. {\bf 100}, 100601 (2008)}.


\bibitem{FiorettoNJP10}
D.~Fioretto and G.~Mussardo, 
\href{http://dx.doi.org/10.1088/1367-2630/12/5/055015}{New J. Phys.  {\bf 12}, 055015 (2010)}.

\bibitem{P:meanvalues}  
B.~Pozsgay, 
\href{http://dx.doi.org/10.1088/1742-5468/2011/01/P01011}{J. Stat. Mech. (2011) P01011}.

\bibitem{DBZ:work} 
B.~D\'ora, \'A.~B\'acsi, and G.~Zar\'and, 
\href{http://dx.doi.org/10.1103/PhysRevB.86.161109}{Phys. Rev. B {\bf 86}, 161109(R) (2012)}.

\bibitem{CEF:TFIC} P. Calabrese, F.H.L. Essler, and M. Fagotti, 
\href{http://dx.doi.org/10.1103/PhysRevLett.106.227203}{Phys. Rev. Lett. {\bf 106}, 227203 (2011)}; 
\href{http://dx.doi.org/10.1088/1742-5468/2012/07/P07016}{J. Stat. Mech. (2012) P07016}; 
\href{http://dx.doi.org/10.1088/1742-5468/2012/07/P07022}{J. Stat. Mech. (2012) P07022}.

\bibitem{CauxPRL12}
J.-S.~Caux and R.~M.~Konik, 
\href{http://dx.doi.org/10.1103/PhysRevLett.109.175301}{Phys. Rev. Lett. {\bf 109}, 175301 (2012)}.

\bibitem{EsslerPRL12}
F.~H.~L.~Essler, S.~Evangelisti, and M.~Fagotti, 
\href{http://dx.doi.org/10.1103/PhysRevLett.109.247206}{Phys. Rev. Lett.  {\bf 109}, 247206 (2012)}.

\bibitem{FE}
M.~Fagotti and F.~H.~L.~Essler, 
\href{http://dx.doi.org/10.1103/PhysRevB.87.245107}{Phys.~Rev.~B {\bf 87}, 245107 (2013)}.

\bibitem{ColluraPRL13}
M.~Collura, S.~Sotiriadis, and P.~Calabrese, 
\href{http://dx.doi.org/10.1103/PhysRevLett.110.245301}{Phys. Rev. Lett. {\bf 110}, 245301 (2013)}; 
\href{http://dx.doi.org/10.1088/1742-5468/2013/09/P09025}{J. Stat. Mech. (2013) P09025}.

\bibitem{QAPRL} 
J.-S.~Caux and F.~H.~L.~Essler, 
\href{http://dx.doi.org/10.1103/PhysRevLett.110.257203}{Phys. Rev. Lett. {\bf 110}, 257203 (2013)}.

\bibitem{MussardoPRL13}
G.~Mussardo, 
\href{http://dx.doi.org/10.1103/PhysRevLett.111.100401} {Phys. Rev. Lett. {\bf 111}, 100401 (2013)}.

\bibitem{PozsgayJStatMech13}
B.~Pozsgay, 
\href{http://dx.doi.org/10.1088/1742-5468/2013/07/P07003} {J. Stat. Mech. (2013) {P07003}}.

\bibitem{KSCCI}
M. Kormos, A. Shashi, Y.-Z. Chou, J.-S. Caux and A. Imambekov,
\href{http://dx.doi.org/10.1103/PhysRevB.88.205131}{Phys. Rev. B {\bf 88}, 205131 (2013)}.

\bibitem{FagottiJStatMech13}
M.~Fagotti and F.~H.~L.~Essler, 
\href{http://dx.doi.org/10.1088/1742-5468/2013/07/P07012}{J. Stat. Mech. (2013) P07012}.

\bibitem{BKCexcited} 
L.~Bucciantini, M.~Kormos, and P.~Calabrese, 
\href{http://dx.doi.org/10.1088/1751-8113/47/17/175002}{J. Phys. A {\bf 47}, 175002 (2014)}. 

\bibitem{FagottiPRB14}
M.~Fagotti, M.~Collura, F.~H.~L.~Essler, and P.~Calabrese,  
\href{http://dx.doi.org/10.1103/PhysRevB.89.125101} {Phys. Rev. B {\bf 89}, 125101 (2014)}.

\bibitem{WoutersPRL14}
B.~Wouters, J.~De~Nardis, M.~Brockmann, D.~Fioretto, M.~Rigol, and J.-S.~Caux, 
\href{http://dx.doi.org/10.1103/PhysRevLett.113.117202}{Phys. Rev. Lett.  {\bf 113}, 117202 (2014)}.

\bibitem{PozsgayPRL14}
B.~Pozsgay, M.~Mesty\'{a}n, M.~A.~Werner, M.~Kormos, G.~Zar\'{a}nd, and G.~Tak\'{a}cs,  
\href{http://dx.doi.org/10.1103/PhysRevLett.113.117203}{Phys. Rev. Lett.  {\bf 113}, 117203 (2014)}.


\bibitem{KCC:freetohardbosons}
M.~Kormos, M.~Collura, and P.~Calabrese,  
\href{http://dx.doi.org/10.1103/PhysRevA.89.013609}{Phys. Rev. A {\bf 89}, 013609 (2014)}.

\bibitem{DeNardisPRA14}
J.~De~Nardis, B.~Wouters, M.~Brockmann, and J.-S.~Caux, 
\href{http://dx.doi.org/10.1103/PhysRevA.89.033601}{Phys. Rev. A  {\bf 89}, 033601 (2014)}.

\bibitem{SotiriadisJStatMech14}
S.~Sotiriadis and P.~Calabrese, 
\href{http://dx.doi.org/10.1088/1742-5468/2014/07/P07024}{J. Stat. Mech. (2014) {P07024}}.

\bibitem{GoldsteinArxiv14}
G.~Goldstein and N.~Andrei, 
\href{http://dx.doi.org/10.1103/PhysRevA.90.043625}{Phys. Rev. A {\bf 90}, 043625 (2014)}.

\bibitem{Betal:LL} 
M.~Brockmann, B.~Wouters, D.~Fioretto, J.~De.~Nardis, R.~Vlijm, and J.-S.~Caux, 
\href{http://dx.doi.org/10.1088/1742-5468/2014/12/P12009}{J. Stat. Mech. (2014) P12009}. 


\bibitem{qbosons}
B.~Pozsgay, 
\href{http://dx.doi.org/10.1088/1742-5468/2014/10/P10045}{J. Stat. Mech. (2014) P10045}.

\bibitem{MPTW:XXZ} 
M.~Mesty\'an, B.~Pozsgay, G.~Tak\'acs, and M.~A.~Werner, 
\href{http://dx.doi.org/10.1088/1742-5468/2015/04/P04001}{J. Stat. Mech. (2015) P04001}. 


\bibitem{EsslerArxiv14}
F.~H.~L.~Essler, G.~Mussardo, and M.~Panfil, 
\href{http://dx.doi.org/10.1103/PhysRevA.91.051602}{Phys. Rev. A {\bf 91}, 051602(R) (2015)}.

\bibitem{IlievskiPRL15}
E.~Ilievski, J.~De~Nardis, B.~Wouters, J.-S.~Caux, F.~H.~L.~Essler, and T.~Prosen,
\href{http://link.aps.org/doi/10.1103/PhysRevLett.115.157201}{Phys. Rev. Lett. {\bf 115}, 157201 (2015)}. 

\bibitem{S:memory} 
S.~Sotiriadis, 
\href{https://doi.org/10.1103/PhysRevA.94.031605}{Phys. Rev. A {\bf 94}, 031605(R) (2016)}.

\bibitem{Ilievskietal} 
E.~Ilievski, E.~Quinn, J.~De~Nardis, and M.~Brockmann, 
\href{http://dx.doi.org/10.1088/1742-5468/2016/06/063101}{J. Stat. Mech. (2016) 063101}.


\bibitem{Pirolibound} 
L.~Piroli, P.~Calabrese, and F.~H.~L.~Essler, 
\href{http://dx.doi.org/10.1103/PhysRevLett.116.070408}{Phys. Rev. Lett. {\bf 116}, 070408 (2016)}; 
\href{http://dx.doi.org/10.21468/SciPostPhys.1.1.001}{SciPost Phys. {\bf 1}, 001 (2016)}.

\bibitem{RigolPRL16}
M.~Rigol,
\href{http://link.aps.org/doi/10.1103/PhysRevLett.116.100601}{Phys. Rev. Lett. {\bf 116}, 100601 (2016)}. 

\bibitem{BPC:sinhG} 
B.~Bertini, L.~Piroli, and P.~Calabrese, 
\href{http://dx.doi.org/10.1088/1742-5468/2016/06/063102}{J. Stat. Mech. (2016) 063102}.

\bibitem{PVC:XXZ} 
L.~Piroli, E.~Vernier, P.~Calabrese, 
\href{http://dx.doi.org/10.1103/PhysRevB.94.054313}{Phys. Rev. B {\bf 94}, 054313 (2016)}.

\bibitem{BS:qlGGE} 
A.~Bastianello, S.~Sotiriadis, 
\href{http://arxiv.org/abs/1608.00924}{arXiv:1608.00924 (2016)}.

\bibitem{SGcharges}
E.~Vernier, A.~C.~Cubero,
\href{http://arxiv.org/abs/1609.03220}{arXiv:1609.03220 (2016)}.





\bibitem{RS-14}
M. A. Rajabpour and S. Sotiriadis, 
 \href{http://dx.doi.org/10.1103/PhysRevA.89.033620}{Phys. Rev. A {\bf 89}, 033620 (2014)}.

\bibitem{MCKC-14} P.~P.~Mazza, M.~Collura, M~Kormos, and P.~Calabrese,  
\href{http://dx.doi.org/10.1088/1742-5468/2014/11/P11016}{ J. Stat. Mech. (2014) P11016}.

\bibitem{ES:IFT}
D.~Schuricht and F.~H.~L.~Essler,
\href{http://dx.doi.org/10.1088/1742-5468/2012/04/P04017}{J. Stat. Mech. (2012) P04017}. 


\bibitem{B:sG} 
B. Bertini, D. Schuricht, and F.H.L. Essler, 
\href{http://dx.doi.org/10.1088/1742-5468/2014/10/P10035}{J. Stat. Mech. (2014) P10035}.


\bibitem{NC-14}
J. De Nardis and J.-S. Caux, 
\href{http://dx.doi.org/10.1088/1742-5468/2014/12/P12012}{J. Stat. Mech. (2014) P12012}.

\bibitem{D:nogo}
G.~Delfino,  \href{http://dx.doi.org/10.1088/1751-8113/47/40/402001}{J. Phys. A: Math. Theor. {\bf 47} 402001 (2014)}. 

\bibitem{DPC:relaxationdynamics}
J.~De Nardis, L.~Piroli, J.-S.~Caux, \href{http://dx.doi.org/10.1088/1751-8113/48/43/43FT01}{J. Phys. A: Math. Theor. {\bf 48}, 43FT01 (2015)}.

\bibitem{RS:long} M.A. Rajabpour, S. Sotiriadis, 
\href{http://dx.doi.org/10.1103/PhysRevB.91.045131}{ Phys. Rev. B  {\bf 91}, 045131 (2015)}.



\bibitem{CCE:XXZquench}
M.~Collura, P.~Calabrese, and F.~H.~L.~Essler, \href{http://dx.doi.org/10.1103/PhysRevB.92.125131}{Phys. Rev. B {\bf 92}, 125131 (2015)}.

\bibitem{RCK:QS}
N.~J.~Robinson, J.-S.~Caux, R.~M.~Konik, \href{http://dx.doi.org/10.1103/PhysRevLett.116.145302}{Phys. Rev. Lett. {\bf 116}, 145302 (2016)}.

\bibitem{KZ:semiclassicalSG}
M.~Kormos, G.~Zar\`and, \href{http://dx.doi.org/10.1103/PhysRevE.93.062101}{Phys. Rev. E {\bf 93}, 062101 (2016)}.

\bibitem{C:planarQQ}
A.~C.~Cubero, J. Stat. Mech. (2016) \href{http://dx.doi.org/10.1088/1742-5468/2016/08/083107}{083107}.
 
\bibitem{RCK:2componentLL}
N.~J.~Robinson, J.-S.~ Caux, R.~M.~Konik, \href{https://arxiv.org/abs/1602.05532}{arXiv:1602.05532}.

\bibitem{AC:entropy}
V.~Alba and P.~Calabrese, 
\href{http://arxiv.org/abs/1608.00614}{arXiv:1608.00614 (2016)}.


\bibitem{MKZ:uSC}
C.~P.~Moca, M.~Kormos, G.~Zarand, \href{http://arxiv.org/abs/1609.00974}{arXiv:1609.00974}.

\bibitem{MK:prethermalization}
M. Moeckel and S. Kehrein, 
\href{http://dx.doi.org/10.1103/PhysRevLett.100.175702}{Phys. Rev. Lett. {\bf 100}, 175702 (2008)};  
\href{http://dx.doi.org/10.1016/j.aop.2009.03.009}{Ann. Phys. {\bf 324}, 2146 (2009)}.

\bibitem{RoschPRL08}
A.~Rosch, D.~Rasch, B.~Binz, and M.~Vojta, 
\href{http://dx.doi.org/10.1103/PhysRevLett.101.265301}{Phys. Rev. Lett. {\bf 101}, 265301 (2008)}.

\bibitem{KollarPRB11}
M.~Kollar, F. A.~Wolf, and M.~Eckstein,
\href{http://dx.doi.org/10.1103/PhysRevB.84.054304}{Phys. Rev. B {\bf 84}, 054304 (2011)}.

\bibitem{worm13}
M.~van~den~Worm, B.~C.~Sawyer, J.~J.~Bollinger, and M.~Kastner, 
\href{http://dx.doi.org/10.1088/1367-2630/15/8/083007}{New J. Phys. {\bf 15}, 083007 (2013)}.

\bibitem{MarcuzziPRL13}
M.~Marcuzzi, J.~Marino, A.~Gambassi, and A.~Silva, 
\href{http://dx.doi.org/10.1103/PhysRevLett.111.197203}{Phys. Rev. Lett. {\bf 111}, 197203 (2013)}.

\bibitem{EsslerPRB14}
F.~H.~L.~Essler, S.~Kehrein, S.~R.~Manmana, and N.~J.~Robinson, 
\href{http://dx.doi.org/10.1103/PhysRevB.89.165104}{Phys. Rev. B {\bf 89}, 165104 (2014)}.

\bibitem{NIC14}
N.~Nessi, A.~Iucci and M.~A.~Cazalilla, 
\href{http://dx.doi.org/10.1103/PhysRevLett.113.210402}{Phys. Rev. Lett. {\bf 113}, 210402 (2014)}.

\bibitem{Fagotti14}
M. Fagotti,
\href{http://dx.doi.org/10.1088/1742-5468/2014/03/P03016}{J. Stat. Mech. (2014) P03016}.

\bibitem{konik14}
G.~P.~Brandino, J.-S.~Caux, and R.~M.~Konik, 
\href{http://dx.doi.org/10.1103/PhysRevX.5.041043}{Phys. Rev. X {\bf 5}, 041043 (2015)}. 

\bibitem{BF15}
B. Bertini and M. Fagotti, 
\href{http://dx.doi.org/10.1088/1742-5468/2015/07/P07012}{J. Stat. Mech. (2015) P07012}.

\bibitem{CTGM:pret} 
A.~Chiocchetta, M.~Tavora, A.~Gambassi, and A.~Mitra, 
\href{http://dx.doi.org/10.1103/PhysRevB.91.220302}{Phys. Rev. B {\bf 91}, 220302(R) (2015)}; \href{http://dx.doi.org/10.1103/PhysRevB.92.219901}{Phys. Rev. {\bf B} 92, 219901(E) (2015).}

\bibitem{knap15}
M. Babadi, E.~Demler, and M.~Knap, 
\href{http://dx.doi.org/10.1103/PhysRevX.5.041005}{Phys. Rev. X {\bf 5}, 041005 (2015)}.

\bibitem{SmacchiaPRB15}
P.~Smacchia, M.~Knap, E.~Demler, and A.~Silva, 
\href{http://link.aps.org/doi/10.1103/PhysRevB.91.205136}{Phys. Rev. B {\bf 91}, 205136 (2015).}

\bibitem{BEGR:PRL}
B. Bertini, F.~H.~L.~Essler, S.~Groha, and N.~J.~Robinson, 
\href{http://journals.aps.org/prl/abstract/10.1103/PhysRevLett.115.180601}{Phys. Rev. Lett. {\bf 115}, 180601 (2015)}.

\bibitem{FC15}
M. Fagotti and M.~Collura, 
\href{http://arxiv.org/abs/1507.02678}{arXiv:1507.02678 (2015)}.

\bibitem{MenegozJStatMech15}
G.~Menegoz and A.~Silva,
\href{http://stacks.iop.org/1742-5468/2015/i=5/a=P05035}{J. Stat. Mech. (2015) P05035}.

\bibitem{KaminishiNatPhys15}
E.~Kaminishi, T.~Mori, T.~Ikeda, N.~Tatsuhiko, and M.~Ueda,
\href{http://dx.doi.org/10.1038/nphys3478}{Nat. Phys. {\bf 11}, 1050 (2015)}.

\bibitem{BEGR:long}
B. Bertini, F.~H.~L.~Essler, S.~Groha, and N.~J.~Robinson, 
\href{http://arxiv.org/abs/1608.01664}{arXiv:1608.01664 (2016)}.




\bibitem{CC} P. Calabrese and J. Cardy, 
J. Stat. Mech. (2005) \href{http://dx.doi.org/10.1088/1742-5468/2005/04/P04010}{P04010};  
\href{http://dx.doi.org/10.1103/PhysRevLett.96.136801}{Phys. Rev. Lett. {\bf 96}, 136801 (2006)}.



\bibitem{DMCF:entropyheisenberg} G. De Chiara, S. Montangero, P. Calabrese, and R. Fazio, 
\href{http://dx.doi.org/10.1088/1742-5468/2006/03/P03001}{J. Stat. Mech. (2006) P03001}.

\bibitem{CF:EE} M.  Fagotti  and  P.  Calabrese, 
 \href{http://dx.doi.org/10.1103/PhysRevA.78.010306}{Phys. Rev. A {\bf 78}, 010306(R) (2008)}.

\bibitem{LK:lcbosehubbard} A. L\"auchli and C. Kollath, 
\href{http://dx.doi.org/10.1088/1742-5468/2008/05/P05018}{J. Stat. Mech. (2008) P05018}.




\bibitem{MWNM:fermioncorrelation} S.  R. Manmana,  S.  Wessel,  R. M.  Noack,  and  A. Muramatsu, 
\href{http://dx.doi.org/10.1103/PhysRevB.79.155104}{Phys. Rev. B {\bf 79}, 155104 (2009)}.

\bibitem{CBSSF:lcbosons} G. Carleo, F. Becca, L. Sanchez-Palencia, S. Sorella, and M. Fabrizio, 
\href{http://dx.doi.org/10.1103/PhysRevA.89.031602}{Phys. Rev. A {\bf 89}, 031602 (2014)}.

\bibitem{BEL:PRL} L. Bonnes, F.H.L. Essler and A. L\"auchli, 
\href{http://dx.doi.org/10.1103/PhysRevLett.113.187203}{Phys. Rev. Lett. {\bf 113}, 187203 (2014).}



\bibitem{ABGM:magneticimpurityXY} D.B. Abraham, E. Barouch, G. Gallavotti and A. Martin-L\"of, 
 \href{http://dx.doi.org/10.1103/PhysRevLett.25.1449}{Phys. Rev. Lett. {\bf 25}, 1449 (1970)};  
 \href{http://dx.doi.org/10.1002/sapm1971502121}{Stud. Appl. Math. {\bf 50}, 121 (1971)}.

\bibitem{KSZ:spinchargeseparation} C. Kollath, U. Schollwock and W. Zwerger, 
 \href{http://dx.doi.org/10.1103/PhysRevLett.95.176401}{Phys. Rev. Lett. {\bf 95}, 176401 (2005)}.
\bibitem{CC07} P. Calabrese and J. Cardy, 
\href{http://dx.doi.org/10.1088/1742-5468/2007/10/P10004}{J. Stat. Mech. (2007) P10004}.
\bibitem{SD:localquenches} J.-M. Stephan and J. Dubail, 
\href{http://dx.doi.org/10.1088/1742-5468/2011/08/P08019}{J. Stat. Mech. (2011) P08019}.

\bibitem{EP:localquench}
I. Peschel, V. Eisler, \href{http://dx.doi.org/10.1088/1742-5468/2007/06/P06005}{J. Stat. Mech. (2007) P06005}.

\bibitem{EP:latticeversion}
I. Peschel, V. Eisler, \href{http://dx.doi.org/10.1088/1751-8113/45/15/155301}{J. Phys. A: Math. Theor. {\bf 45} (2012) 155301}.


\bibitem{GREE:boundstateslocalquenches} M. Ganahl, E. Rabel, F.H.L. Essler and H.-G. Evertz, 
\href{http://dx.doi.org/10.1103/PhysRevLett.108.077206}{Phys. Rev. Lett.{\bf 108}, 077206 (2012)}.

\bibitem{CP:localquenchXXZ} M.~Collura and P.~Calabrese, \href{http://dx.doi.org/10.1088/1751-8113/46/17/175001}{J. Phys. A: Math. Theor. {\bf 46} 175001}. 


\bibitem{F:LocalSwitch}
M. Fagotti,  
\href{http://arxiv.org/abs/1508.04401}{arXiv:1508.04401 (2015)}.

\bibitem{VSDH:XX}
J.~Viti, J.-M.~St\'ephan, J.~Dubail, M.~Haque, \href{https://doi.org/10.1209/0295-5075/115/40011
}{EPL {\bf 115} (2016) 40011}.	

\bibitem{BF:defect} 
B.~Bertini and M.~Fagotti, 
\href{https://doi.org/10.1103/PhysRevLett.117.130402}{Phys. Rev. Lett. {\bf 117}, 130402 (2016)}.

\bibitem{CADY:hydro} 
O.~A.~Castro-Alvaredo, B.~Doyon, and T.~Yoshimura, 
\href{http://arxiv.org/abs/1605.07331}{arXiv:1605.07331 (2016)}.


\bibitem{BCDF:transport} 
B.~Bertini, M.~Collura, J.~De~Nardis, and M.~Fagotti, 
\href{http://arxiv.org/abs/1605.09790}{arXiv:1605.09790 (2016)}.

\bibitem{F:Currents}
M.~Fagotti, \href{http://arxiv.org/abs/1608.02869}{arXiv:1608.02869 (2016)}.


\bibitem{C:furtherresults} 
J.~Cardy, 
\href{http://dx.doi.org/10.1088/1742-5468/2016/02/023103}{J. Stat. Mech. (2016) 023103}. 

\bibitem{confinementnoneq}
M.~Kormos, M.~Collura, G.~Tak\'acs, and P.~Calabrese, 
\href{http://arxiv.org/abs/1604.03571}{arXiv:1604.03571 (2016)}.


\bibitem{RMCKT:confinementnoneq} 
T.~Rakovszky, M.~Mesty\'an, M.~Collura, M.~Kormos, and G.~Tak\'acs, 
\href{http://dx.doi.org/10.1016/j.nuclphysb.2016.08.024}{Nucl. Phys. B {\bf 911}, 805-845 (2016)}.

\bibitem{LR72}
E. H. Lieb and D. W. Robinson, 
\href{http://dx.doi.org/10.1007/BF01645779}{Commun. Math. Phys. {\bf 28}, 251
(1972)}.


\bibitem{KS:graph} V. Kostrykin and R. Schrader, \href{\doi10.1002/1521-3978(200008)48:8<703::AID-PROP703>3.0.CO;2-O}{Fortschr. Phys. {\bf 48}, 703 (2000)}.

\bibitem{H:graph} M. Harmer, \href{http://dx.doi.org/10.1088/0305-4470/33/49/302}{J. Phys. A {\bf 33}, 9015 (2000)}.


\bibitem{SGQFT} B. Bellazzini and M. Mintchev, \href{http://dx.doi.org/10.1088/0305-4470/39/35/011}{J. Phys. A: Math. Gen. {\bf 39} 11101 (2006)}.


\bibitem{SGQFTreview} B. Bellazzini, M. Burrello, M. Mintchev and P. Sorba,  \href{http://dx.doi.org/10.1090/pspum/077 }{Proc. Symp. Pure Math. {\bf 77} 639 (2008)}.

\bibitem{SGbos} B. Bellazzini, M. Mintchev and P. Sorba, \href{http://dx.doi.org/10.1088/1751-8113/40/10/017}{J. Phys. A: Math. Theor. {\bf 40}   2485 (2007)}.

\bibitem{AnyonsSG} B. Bellazzini, P. Calabrese and M. Mintchev, \href{http://dx.doi.org/10.1103/PhysRevB.79.085122}{Phys. Rev. B {\bf 79}, 085122 (2009)}. 

\bibitem{SGNESS} M. Mintchev, \href{http://dx.doi.org/10.1088/1751-8113/44/41/415201}{J. Phys. A: Math. Theor. {\bf 44} 415201 (2011)}.

\bibitem{SGEE}
P. Calabrese, M. Mintchev, E. Vicari, \href{http://dx.doi.org/10.1103/PhysRevLett.107.020601}{Phys. Rev. Lett. {\bf 107}, 020601 (2011)}; \href{http://dx.doi.org/10.1088/1751-8113/45/10/105206}{J. Phys. A: Math. Theor. {\bf 45} (2012) 105206}.  


\bibitem{MSS} M.~Mintchev, L.~Santoni and P. Sorba, \href{http://dx.doi.org/10.1088/1751-8113/48/5/055003}{J. Phys. A {\bf 48}, 055003 (2015)}.





\bibitem{note}
The factor $\sqrt{L}$ takes care of the infinite volume normalization. 


\bibitem{orthogonalitycatastrophe} 
P. W. Anderson, \href{http://dx.doi.org/10.1103/PhysRevLett.18.1049}{Phys. Rev. Lett. {\bf 18}, 1049 (1967)}.


\bibitem{ND:solution}
P.~Nozi\`eres, and C.~de~Dominicis, \href{https://doi.org/10.1103/PhysRev.178.1097}{Phys. Rev. {\bf 178}, 1097 (1969)}.

\bibitem{SS:bos}
K. D. Schotte and U. Schotte, \href{https://doi.org/10.1103/PhysRev.182.479}{Phys. Rev. {\bf 182}, 479 (1969)}.

\bibitem{gogolinbook}
A.~O.~Gogolin, A.~A.~Nersesyan, and A.~M.~Tsvelik, \emph{Bosonization and strongly correlated systems}, (Cambridge University Press, 2004).


\bibitem{GRbook}
I.~S.~Gradshteyn and I.~M.~Ryzhik, \emph{Table of integrals, series, and products} (Academic press, 2014).


\bibitem{entanglement:review}
L.~Amico, R.~Fazio, A.~Osterloh, and V.~Vedral, \href{https://doi.org/10.1103/RevModPhys.80.517}{Rev. Mod. Phys. {\bf 80}, 517 (2008)}.

\bibitem{CCD:review}
P.~Calabrese, J.~Cardy, and B.~Doyon, \href{http://dx.doi.org/10.1088/1751-8121/42/50/500301}{J. Phys. A: Math. Theor. {\bf 42} 500301 (2009)}.


\bibitem{ECP:review}
J. Eisert, M. Cramer, and M. B. Plenio, \href{https://doi.org/10.1103/RevModPhys.82.277}{Rev. Mod. Phys. {\bf 82}, 277 (2010)}.


\bibitem{L:review}
N.~Laflorencie, \href{http://dx.doi.org/10.1016/j.physrep.2016.06.008}{Physics Report {\bf 643}, 1-59 (2016)}.


\bibitem{DSVC:inhomoCFT}
J.~Dubail, J.-M.~St\'ephan, J.~Viti, P.~Calabrese, 
\href{http://arxiv.org/abs/1606.04401}{arXiv:1606.04401 (2016)}.

  
\bibitem{carbonnanowires} M. S. Fuhrer, J. Nyg{\aa}rd, L. Shih, M. Forero, Y.-G.~Yoon, M. S. C. Mazzoni, H.~J.~Choi, J.~Ihm, S.~G.~Louie, A. Zettl, P.~L. McEuen, \href{\doi 10.1126/science.288.5465.494}{Science {\bf 288}, 494-497 (2000)}.
  
\bibitem{Tetal:stargraph}
F. Buccheri, G. D. Bruce, A. Trombettoni, D. Cassettari, H. Babujian, V. E. Korepin, P. Sodano,
\href{\doi10.1088/1367-2630/18/7/075012}{New J. Phys. {\bf 18}, 075012 (2016)}.

\bibitem{HRS} 
V.~Hunyadi, Z.~R\'acz, and L.~Sasv\'ari, 
\href{http://dx.doi.org/10.1103/PhysRevE.69.066103}{Phys. Rev. E  {\bf69}, 066103 (2004)}.  

\bibitem{ER} 
V.~Eisler and Z.~R\'acz, 
\href{http://dx.doi.org/10.1103/PhysRevLett.110.060602}{Phys. Rev. Lett. {\bf 110}, 060602 (2013)}. 

\bibitem{Korepinbook}
V.E. Korepin, A.G. Izergin, and N.M. Bogoliubov, {\emph{Quantum Inverse
  Scattering Method, Correlation Functions and Algebraic Bethe Ansatz}}
  (Cambridge University Press, 1993).


\bibitem{Wongbook}
R. Wong, \emph{Asymptotic Approximations of Integrals} (SIAM ed., 2001).



\end{thebibliography}
\end{document}